\documentclass[aps,prd,superscriptaddress,showpacs,nofootinbib,floatfix,twocolumn]{revtex4}

\usepackage{graphicx}
\usepackage{subfigure}
\usepackage{dcolumn}
\usepackage{comment}
\usepackage{bm}

\def\fig#1{Fig.~\ref{#1}}
\def\eq#1{Eq.~(\ref{#1})}
\def\tab#1{Tab.~\ref{#1}}

\def\jtfinal{\ensuremath{\sqrtrms{\jt{}}=585 \pm 6(\rm stat)\pm 15(\rm sys) \mevc}}
\def\ktfinal{\ensuremath{ \sqrtrms{\kt{}} = 2.68 \pm 0.07(\rm stat) \pm 0.34(\rm sys) \gevc }}
\def\ptnfinal{\ensuremath{\mean{\pt{}}_{pair} = 3.36 \pm 0.09(\rm stat) \pm 0.43(\rm sys) \gevc }}

\def\gev{\mbox{~GeV}}
\def\gevc{\mbox{~GeV/$c$}}

\def\mevc{\mbox{~MeV/$c$}}

\def\la{\left< }
\def\ra{\right> }
\def\mean#1{\ensuremath{\la#1\ra}}
\def\meanabs#1{\ensuremath{\la|#1|\ra}}
\def\meankv#1{\ensuremath{\la#1^2\ra}}
\def\rms#1{\meankv{#1}}
\def\sqrtrms#1{\ensuremath{\sqrt{\meankv{#1}}}}

\def\eg{{\it e.g.}}

\newcommand{\sign} {\ensuremath{\sigma_{\rm N}}}
\newcommand{\siga} {\ensuremath{\sigma_{\rm A}}}
\newcommand{\yn} {\ensuremath{Y_{\rm N}}}
\newcommand{\yf} {\ensuremath{Y_{\rm F}}}

\newcommand{\D} {\ensuremath{D^q_\pi}}
\newcommand{\fq}{\ensuremath{f_q(\hat{p}_T)}}

\newcommand{\sq}{\ensuremath{\Sigma_q(\hat{p}_T)}}

\newcommand{\fkt}{\ensuremath{f^\prime_q(\hat{p}_{T\rm t})}}

\newcommand{\skt}{\ensuremath{\Sigma^\prime_q(\hat{p}_{T\rm t})}}

\newcommand{\condta}{\ensuremath{\Big|_{\pt{t},\pt{a}}}}

\newcommand{\auau} {\ensuremath{Au+Au}}
\newcommand{\dau} {\ensuremath{d+Au}}
\newcommand{\pp} {\ensuremath{p+p}}
\newcommand{\ee} {\ensuremath{e^+ e^-}}

\def\ptq#1{\ensuremath{\hat{p}_{T\rm #1}}} 
\def\ptqkv#1{\ensuremath{\hat{p}^2_{T\rm #1}}} 
\def\vptq#1{\ensuremath{\vec{\hat{p}}_{T\rm #1}}} 

\def\pt#1{\ensuremath{p_{T\rm #1}}} 
\def\ptkv#1{\ensuremath{p^2_{T\rm #1}}} 
\def\vpt#1{\ensuremath{\vec{p}_{T\rm #1}}} 

\def\kt#1{\ensuremath{k_{T\rm #1}}} 
\def\ktkv#1{\ensuremath{k^2_{T\rm #1}}} 
\def\vkt#1{\ensuremath{\vec{k}_{T\rm #1}}} 

\def\jt#1{\ensuremath{j_{T\rm #1}}} 
 
\def\vjt#1{\ensuremath{\vec{j}_{T\rm #1}}}

\newcommand{\ptt}{\ensuremath{p_{T\rm t}}}
 
\newcommand{\pta}{\ensuremath{p_{T\rm a}}}
 
\def\pn{\ensuremath{\hat{p}_{\rm n}}} 
\def\pnkv{\ensuremath{\hat{p}^2_{\rm n}}}

\newcommand{\pout} {\ensuremath{p_{\rm out}}}

\newcommand{\mz} {\mean{z}}
\newcommand{\zt} {\ensuremath{z_{\rm t}}}
\newcommand{\mzt} {\mean{\zt}}
\newcommand{\za} {\ensuremath{z_{\rm a}}}
\newcommand{\mza} {\mean{\za}}
\newcommand{\xe} {\ensuremath{x_{\rm E}}}
\newcommand{\xh} {\ensuremath{x_{\rm h}}}
\newcommand{\xhq} {\ensuremath{\hat{x}_{\rm h}}}
\newcommand{\zkt} {\ensuremath{ \mean{\zt}\sqrtrms{\kt{}} }}
\newcommand{\xzkt} {\ensuremath{ \xhq^{-1}\mean{\zt}\sqrtrms{\kt{}} }}

\newcommand{\snn} {\ensuremath{\sqrt{s_{NN}}}}
\newcommand{\s} {\ensuremath{\sqrt{s}}}
\newcommand{\piz} {\ensuremath{\pi^0}}
\newcommand{\pizh}{\ensuremath{\piz-h^\pm}}

\def\bgi{\begin{itemize}}
\def\endi{\end{itemize}}
\def\bge{\begin{equation}}
\def\ende{\end{equation}}
\def\bgc{\begin{center}}
\def\endc{\end{center}}

\begin{document}

\title{Jet Properties from Dihadron Correlations in p+p Collisions at \s~=~200~\gev}

\newcommand{\abilene}{Abilene Christian University, Abilene, TX 79699, U.S.}
\newcommand{\acadsin}{Institute of Physics, Academia Sinica, Taipei 11529, Taiwan}
\newcommand{\banaras}{Department of Physics, Banaras Hindu University, Varanasi 221005, India}
\newcommand{\barc}{Bhabha Atomic Research Centre, Bombay 400 085, India}
\newcommand{\bnl}{Brookhaven National Laboratory, Upton, NY 11973-5000, U.S.}
\newcommand{\caucr}{University of California - Riverside, Riverside, CA 92521, U.S.}
\newcommand{\ciae}{China Institute of Atomic Energy (CIAE), Beijing, People's Republic of China}
\newcommand{\cns}{Center for Nuclear Study, Graduate School of Science, University of Tokyo, 7-3-1 Hongo, Bunkyo, Tokyo 113-0033, Japan}
\newcommand{\colorado}{University of Colorado, Boulder, CO 80309, U.S.}
\newcommand{\columbia}{Columbia University, New York, NY 10027 and Nevis Laboratories, Irvington, NY 10533, U.S.}
\newcommand{\dapnia}{Dapnia, CEA Saclay, F-91191, Gif-sur-Yvette, France}
\newcommand{\debrecen}{Debrecen University, H-4010 Debrecen, Egyetem t{\'e}r 1, Hungary}
\newcommand{\elte}{ELTE, E{\"o}tv{\"o}s Lor{\'a}nd University, H - 1117 Budapest, P{\'a}zm{\'a}ny P. s. 1/A, Hungary}
\newcommand{\fsu}{Florida State University, Tallahassee, FL 32306, U.S.}
\newcommand{\gsu}{Georgia State University, Atlanta, GA 30303, U.S.}
\newcommand{\hiroshima}{Hiroshima University, Kagamiyama, Higashi-Hiroshima 739-8526, Japan}
\newcommand{\ihepprot}{IHEP Protvino, State Research Center of Russian Federation, Institute for High Energy Physics, Protvino, 142281, Russia}
\newcommand{\illuiuc}{University of Illinois at Urbana-Champaign, Urbana, IL 61801, U.S.}
\newcommand{\isu}{Iowa State University, Ames, IA 50011, U.S.}
\newcommand{\jinrdubna}{Joint Institute for Nuclear Research, 141980 Dubna, Moscow Region, Russia}
\newcommand{\kek}{KEK, High Energy Accelerator Research Organization, Tsukuba, Ibaraki 305-0801, Japan}
\newcommand{\kfki}{KFKI Research Institute for Particle and Nuclear Physics of the Hungarian Academy of Sciences (MTA KFKI RMKI), H-1525 Budapest 114, POBox 49, Budapest, Hungary}
\newcommand{\korea}{Korea University, Seoul, 136-701, Korea}
\newcommand{\kurchatov}{Russian Research Center ``Kurchatov Institute", Moscow, Russia}
\newcommand{\kyoto}{Kyoto University, Kyoto 606-8502, Japan}
\newcommand{\labllr}{Laboratoire Leprince-Ringuet, Ecole Polytechnique, CNRS-IN2P3, Route de Saclay, F-91128, Palaiseau, France}
\newcommand{\lawllnl}{Lawrence Livermore National Laboratory, Livermore, CA 94550, U.S.}
\newcommand{\losalamos}{Los Alamos National Laboratory, Los Alamos, NM 87545, U.S.}
\newcommand{\lpc}{LPC, Universit{\'e} Blaise Pascal, CNRS-IN2P3, Clermont-Fd, 63177 Aubiere Cedex, France}
\newcommand{\lund}{Department of Physics, Lund University, Box 118, SE-221 00 Lund, Sweden}
\newcommand{\muenster}{Institut f\"ur Kernphysik, University of Muenster, D-48149 Muenster, Germany}
\newcommand{\myongji}{Myongji University, Yongin, Kyonggido 449-728, Korea}
\newcommand{\nagasaki}{Nagasaki Institute of Applied Science, Nagasaki-shi, Nagasaki 851-0193, Japan}
\newcommand{\newmex}{University of New Mexico, Albuquerque, NM 87131, U.S. }
\newcommand{\nmsu}{New Mexico State University, Las Cruces, NM 88003, U.S.}
\newcommand{\ornl}{Oak Ridge National Laboratory, Oak Ridge, TN 37831, U.S.}
\newcommand{\orsay}{IPN-Orsay, Universite Paris Sud, CNRS-IN2P3, BP1, F-91406, Orsay, France}
\newcommand{\peking}{Peking University, Beijing, People's Republic of China}
\newcommand{\pnpi}{PNPI, Petersburg Nuclear Physics Institute, Gatchina, Leningrad region, 188300, Russia}
\newcommand{\riken}{RIKEN (The Institute of Physical and Chemical Research), Wako, Saitama 351-0198, JAPAN}
\newcommand{\rikjrbrc}{RIKEN BNL Research Center, Brookhaven National Laboratory, Upton, NY 11973-5000, U.S.}
\newcommand{\saopaulo}{Universidade de S{\~a}o Paulo, Instituto de F\'{\i}sica, Caixa Postal 66318, S{\~a}o Paulo CEP05315-970, Brazil}
\newcommand{\seoulnat}{System Electronics Laboratory, Seoul National University, Seoul, South Korea}
\newcommand{\stonybrkc}{Chemistry Department, Stony Brook University, SUNY, Stony Brook, NY 11794-3400, U.S.}
\newcommand{\stonycrkp}{Department of Physics and Astronomy, Stony Brook University, SUNY, Stony Brook, NY 11794, U.S.}
\newcommand{\subatech}{SUBATECH (Ecole des Mines de Nantes, CNRS-IN2P3, Universit{\'e} de Nantes) BP 20722 - 44307, Nantes, France}
\newcommand{\tenn}{University of Tennessee, Knoxville, TN 37996, U.S.}
\newcommand{\titech}{Department of Physics, Tokyo Institute of Technology, Oh-okayama, Meguro, Tokyo 152-8551, Japan}
\newcommand{\tsukuba}{Institute of Physics, University of Tsukuba, Tsukuba, Ibaraki 305, Japan}
\newcommand{\vandy}{Vanderbilt University, Nashville, TN 37235, U.S.}
\newcommand{\waseda}{Waseda University, Advanced Research Institute for Science and Engineering, 17 Kikui-cho, Shinjuku-ku, Tokyo 162-0044, Japan}
\newcommand{\weizmann}{Weizmann Institute, Rehovot 76100, Israel}
\newcommand{\yonsei}{Yonsei University, IPAP, Seoul 120-749, Korea}
\newcommand{\deceased}{\dagger}
\affiliation{\abilene}
\affiliation{\acadsin}
\affiliation{\banaras}
\affiliation{\barc}
\affiliation{\bnl}
\affiliation{\caucr}
\affiliation{\ciae}
\affiliation{\cns}
\affiliation{\colorado}
\affiliation{\columbia}
\affiliation{\dapnia}
\affiliation{\debrecen}
\affiliation{\elte}
\affiliation{\fsu}
\affiliation{\gsu}
\affiliation{\hiroshima}
\affiliation{\ihepprot}
\affiliation{\illuiuc}
\affiliation{\isu}
\affiliation{\jinrdubna}
\affiliation{\kek}
\affiliation{\kfki}
\affiliation{\korea}
\affiliation{\kurchatov}
\affiliation{\kyoto}
\affiliation{\labllr}
\affiliation{\lawllnl}
\affiliation{\losalamos}
\affiliation{\lpc}
\affiliation{\lund}
\affiliation{\muenster}
\affiliation{\myongji}
\affiliation{\nagasaki}
\affiliation{\newmex}
\affiliation{\nmsu}
\affiliation{\ornl}
\affiliation{\orsay}
\affiliation{\peking}
\affiliation{\pnpi}
\affiliation{\riken}
\affiliation{\rikjrbrc}
\affiliation{\saopaulo}
\affiliation{\seoulnat}
\affiliation{\stonybrkc}
\affiliation{\stonycrkp}
\affiliation{\subatech}
\affiliation{\tenn}
\affiliation{\titech}
\affiliation{\tsukuba}
\affiliation{\vandy}
\affiliation{\waseda}
\affiliation{\weizmann}
\affiliation{\yonsei}
\author{S.S.~Adler}	\affiliation{\bnl}
\author{S.~Afanasiev}	\affiliation{\jinrdubna}
\author{C.~Aidala}	\affiliation{\columbia}
\author{N.N.~Ajitanand}	\affiliation{\stonybrkc}
\author{Y.~Akiba}	\affiliation{\kek} \affiliation{\riken}
\author{A.~Al-Jamel}	\affiliation{\nmsu}
\author{J.~Alexander}	\affiliation{\stonybrkc}
\author{K.~Aoki}	\affiliation{\kyoto}
\author{L.~Aphecetche}	\affiliation{\subatech}
\author{R.~Armendariz}	\affiliation{\nmsu}
% ...
\collaboration{PHENIX Collaboration} \noaffiliation

\begin{abstract}
The properties of jets produced in \pp\ collisions at \s=200~GeV are
measured using the method of two particle correlations. The trigger
particle is a leading particle from a large transverse momentum jet
while the associated particle comes from either the same jet or the
away-side jet. Analysis of the angular width of the near-side peak in
the correlation function determines the jet fragmentation transverse
momentum \jt{}.  The extracted value, \jtfinal, is constant with
respect to the
trigger particle transverse momentum, and comparable to the previous
lower \s\ measurements.  The width of the away-side peak is shown to
be a convolution of \jt{}\ with the fragmentation variable, $z$, and
the partonic transverse momentum, \kt{}.  The \mz\ is determined
through a combined analysis of the measured \piz\ inclusive and
associated spectra 
%%mjt042106by determining the 
using jet fragmentation functions measured in $e^+ e^-$ collisions. The
final extracted values of \kt{}\ are then determined to also be
independent of the trigger particle transverse momentum, over the
range measured, with value of \ktfinal.

\end{abstract}

\maketitle

\section{Introduction}

The goal of this paper is to explore the systematics of jet production
and fragmentation in \pp\ collisions at \s=200~\gev\ by the method of
two-particle azimuthal correlations. Knowledge of the
jet-fragmentation process is useful not only as a reference
measurement for a similar analysis in \auau\ collisions, but can be
used as a
stringent test of perturbative QCD (pQCD) 
calculations beyond leading order.

The two-particle azimuthal correlations method worked well at ISR
energies (\s=63~\gev) and below
\cite{CCORjt,Darriulat_poutxe,CCHK_jet_structure}, where it is
difficult to directly reconstruct jets, but has not been attempted at
higher values of \s. This method is also suitable for jet-analysis in
heavy ion data where the large particle multiplicity severely
interferes with direct jet reconstruction.

With the beginning of RHIC operation, heavy-ion physics entered a new
regime, where pQCD phenomena can be fully explored. High-energy
partons materializing into hadronic jets can be used as sensitive
probes of the early stage of heavy ion collisions.  Measurements
carried out during the first three years of RHIC operation at
\snn=130 and 200~\gev\ exhibit many new and interesting features. 
The high-\pt\ particle yield was found to be strongly suppressed in
\auau\ central collisions \cite{PHENIX_supp_130}. Furthermore, the
non-suppression of the high-\pt\ particle yield in \dau\ induced
collisions \cite{PHENIX_supp_dAu} confirmed that the suppression can be fully
attributed to the final state interaction of high-energy partons with an 
extremely opaque nuclear medium formed in \auau\ collisions at RHIC.

Other striking features found in RHIC data are the large asymmetry of
particle azimuthal distributions which is attributed to sizable
elliptic flow \cite{PHENIX_v2,STAR_v2} and the observation of the apparent
disappearance of the back-to-back jet correlation in central \auau\
collisions \cite{STAR_b2b_suppression}.

Many of the above mentioned observations can be explained by a large
opacity of the medium produced in central \auau\ collisions which causes the 
scattered partons to lose energy via coherent (Landau-Pomeranchuk-Migdal \cite{LPM_1956}) 
gluon bremsstrahlung \cite{quenching_WangMiklos,Wang_jetQuenching98_D(z)pars}.
It is expected that the medium effect will cause the apparent
modification of fundamental properties of hard-scattering like
broadening of intrinsic parton transverse momentum \kt{}\
\cite{Urs_ktBroadening,Ivan_dAu} and modification of jet fragmentation
\cite{Wang_fragModif1}. Thus the measurement of jet fragmentation
properties and intrinsic parton transverse momentum \kt{}\ for \pp\
collisions presented here provides a baseline for comparison to the
results in heavy ion collisions, helping to disentangle the complex
processes of propagation and possible fragmentation of partons within
the excited nuclear medium.

This paper is organized as follows:  Section \ref{sec:jetCorrel} discusses the
method of two-particle correlations and the relations between jet
properties and the angular correlation between parton fragments.  The
details of the PHENIX experiment relevant to this analysis are
outlined in section \ref{sec:experimental}.  Section
\ref{sec:rawResults} deals with the analysis of the correlation functions
extracted from the \pp\ data and an evaluation of the \mean{\jt{}}\ and
\mean{\kt{}}\ quantities.  
%%%mjt042106
The combined analysis of the inclusive and associated
{\pt{}}-distributions is discussed in section \ref{sec:inclusFrag} and the sensitivity of the associated {\pt{}}-distributions to the fragmentation function is discussed in section \ref{sec:mz_cond}. 
Section \ref{sec:finRes} presents the resulting values of the partonic
transverse momenta \kt{}\ corrected for the mean momentum fraction
\mzt.  Section \ref{sec:summary} summarizes the results from this
paper.

%%%%%%%%%%%%%%%%%%%%%%%%%%%%%%%%%%%%%%%%%\input{ppg_jet_correl}
%%%%%%%%%%%%%%%%%%%%%%%%%%%%%%%%%%%%%%%%%%%%%%%%%%%%%%%%%%%%%%%
\section{Jet angular correlations}
\label{sec:jetCorrel}
%%%%%%%%%%%%%%%%%%%%%%%%%%%%%%%%%%%%%%%%%%%%%%%%%%%%%%%%%%%%%%%

Jets are produced in the hard scattering of two partons
\cite{Feynman1,Feynman2,Feynman3,Feynman4}.
The overall \pp\ hard-scattering cross section in ``leading logarithm" pQCD
is the sum over parton reactions $a+b\rightarrow c +d$
(e.g. $g+q\rightarrow g+q$) at parton-parton center-of-mass (c.m.) energy $\sqrt{\hat{s}}$, 
\begin{equation}
\frac{d^3\sigma}{dx_1 dx_2 d\cos\theta^*}=
\frac{1}{s}\sum_{ab} f_a(x_1) f_b(x_2)
\frac{\pi\alpha_s^2(Q^2)}{2x_1 x_2} \Sigma^{ab}(\cos\theta^*)
\label{eq:QCDabscat}
\end{equation}
where $f_a(x_1)$, $f_b(x_2)$, are parton distribution functions, the
differential probabilities for partons $a$ and $b$ to carry momentum
fractions $x_1$ and $x_2$ of their respective protons (e.g. $u(x_2)$),
and where $\theta^*$ is the scattering angle in the parton-parton
c.m. system.  The parton-parton c.m. energy squared is $\hat{s}=x_1
x_2 s$, where $\sqrt{s}$ is the c.m. energy of the \pp\
collision. The parton-parton c.m. system moves with rapidity $y=(1/2)
\ln (x_1/x_2)$ in the \pp\ c.m. system.  
%%%mjt1beginning:

Equation~\ref{eq:QCDabscat} gives the \pt{}\ spectrum of outgoing
parton $c$ (emitted at $\theta^*$), which then fragments into hadrons,
e.g. a $\pi^0$.  The fragmentation function $D^{\pi^0}_{c}(z,\mu^2)$ is
the probability for a $\pi^0$ to carry a fraction $z=p^{\pi^0}/p^{c}$
of the momentum of outgoing parton $c$. Equation~\ref{eq:QCDabscat}
must be summed over all subprocesses leading to a $\pi^0$ in the final
state.  The parameter $\mu^2$ is an unphysical ``factorization" scale
introduced to account for collinear singularities in the structure and
fragmentation functions~\cite{Owens:1987mp,Bunce:2000uv}, which will
be ignored for the purposes of this paper.

%%%%%%%%%%%%%%%%%%%%%%%%%%%%%%%%%%%%%%%%%%%%%%%%%%%%%%%%%%%% Fig. 1
\begin{figure}[tbh]
\includegraphics[width=1.0\linewidth]{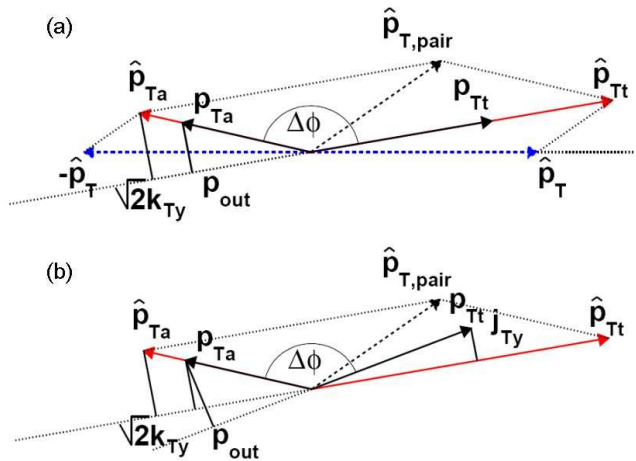}
%\includegraphics[width=1.0\linewidth]{jets_j0j0}
%\includegraphics[width=1.0\linewidth]{jets_j1j0}
%\vspace*{0.5 in}
\caption []{\label{fig:correl-schematic} (color online)
(a) Schematic view of a hard scattering event in the plane
perpendicular to the beam. Two scattered partons with transverse
momenta \ptq{}\ in the 
partons' center of mass frame are seen in the laboratory
frame to have a momenta \ptq{t}\ and \ptq{a}. The net pair transverse
momentum \ptq{pair}\ corresponds to the sum of %%%two 
%%%mjt043006the trigger and associated jet.  
the \vkt{}-vectors of the two colliding partons. %%
The trigger and associated jet 
fragments producing high-\pt{}\ particles are labeled as \ptt\ and \pta.
The projection of \vkt{}\ perpendicular to \ptq{t}\ is labeled as
\kt{y}.  The transverse momentum component of the away-side particle
\vpt{a}\ perpendicular to trigger particle \vpt{t}\ is labeled as
\pout.  (b) The same schematics as in {\sl (a)}, but the jet
fragmentation transverse momentum component \jt{y}\ of the trigger jet
is also shown.  }
\end{figure}

In this formulation, $f_a(x_1)$, $f_b(x_2)$ and $D^{\pi^0}_c (z)$ 
% input parameters
represent the ``long-distance phenomena" to be determined by experiment;
while the characteristic subprocess angular distributions,
{\bf $\Sigma^{ab}(\cos\theta^*)$},
and the coupling constant,
$\alpha_s(Q^2)=\frac{12\pi}{25 \ln(Q^2/\Lambda^2)}$,
are fundamental predictions of QCD~\cite{Cutler:1978qm,Cutler:1977mw,Combridge:1977dm}
for the short-distance, large-$Q^2$, phenomena.
%Actually, the scale $\Lambda$ is not predicted; and
The momentum scale $Q^2\sim\ptkv{}$ for the scattering subprocess,
while $Q^2\sim\hat{s}$ for a Compton or annihilation subprocess, but
the exact meaning of $Q^2$ tends to be treated as a parameter rather
than a dynamical quantity.

Figure~\ref{fig:correl-schematic} shows a schematic view of a 
hard-scattering event.  The transverse momentum of the outgoing 
scattered parton is:
\begin{equation}
\pt{}=\pt{}^*={ \sqrt{\hat{s}} \over 2 } \; \sin\theta^* \qquad .
\label{eq:cpT}
\end{equation}
%%mjt1end:
The two scattered partons
propagate nearly back-to-back in azimuth from the collision point and
fragment into the jet-like spray of final state particles (see
\fig{fig:correl-schematic}(a) where only one fragment of each parton is
shown). 

	It was originally thought that parton collisions were collinear with
the \pp\ collision axis so that the two emerging partons would have
the same magnitude of transverse momenta pointing opposite in
azimuth. However, it was found~\cite{CCHK_jet_structure} that each of
the partons carries initial transverse momentum $\vec{k}_{T}$,
originally described as ``intrinsic'' \cite{Feynman5}. This results in a
momentum imbalance (the partons' \pt\ are not equal) and an
acoplanarity (the transverse momentum of one jet does not lie in the
plane determined by the transverse momentum of the second jet and the
beam axes).  The jets are non- collinear having a net transverse
momentum $\rms{\pt{}}_{\rm pair} = 2\cdot\rms{\kt{}}$.

It is important to emphasize that the \mean{\kt{}}\ denotes the effective
magnitude of the apparent transverse momentum of each colliding
parton. The net transverse momentum of the outgoing parton-pair is
$\sqrt{2}\cdot\mean{\kt{}}$.  The naive expectation for the pure intrinsic parton
transverse momentum based on nucleon constituent quark mass is about
$\approx$~300~\mevc\ \cite{Feynman5,Dokshitzer_basics_of_pQCD}.
However, the measurement of net transverse momenta of diphotons, dileptons
or dijets over a wide range of center-of-mass energies gives
\mean{\kt{}}\ as large as 5\gevc\ \cite{Apanasevich_kt_E609}. 
%%%mjt2begin:
%Clearly, these
%values can not be attributed to the intrinsic transverse momentum
%given by the uncertainty relation and the re-summation technique
%\cite{Werner_resummation} has to be invoked.
It is common to think of the net transverse momentum of a
dilepton or dijet pair as composed of 3~components:
\begin{equation}
{\rms{\pt{}}_{\rm pair} \over 2}=\rms{\kt{}}=
\rms{\kt{}}_{\rm intrinsic} +\rms{\kt{}}_{\rm soft} + \rms{\kt{}}_{\rm NLO}\;\;,
\end{equation}
where the intrinsic part refers to the possible ``fermi motion" of the
confined quarks or gluons inside a proton, the NLO part refers to the
power law tail at large values of $p_{T_{pair}}$ due to the radiation
of an initial state or final state hard gluon, which is divergent as
the momentum of the radiated gluon goes to zero, and the soft part
refers to the actual Gaussian-like distribution observed as
$p_{T_{pair}}\rightarrow 0$, which is explained by
resummation \cite{Werner_resummation}.
%%mjt2end:

In the discussion below we will assume that the two components of the
vector \vkt{}, \kt{x}\ and \kt{y}\ are Gaussian distributed
with equal standard deviations $\sigma_{\rm 1parton,1d}$, in which
case \ktkv{}=\ktkv{x}+\ktkv{y}\ is distributed according to
a 2-dimensional (2D) Gaussian
\cite{Apanasevich_kt_E609}. For the net transverse momentum of the jet pair,
$\rms{\pt{}}_{\rm pair}=\sigma^2_{\rm 2partons,2d}=2\sigma^2_{\rm 1parton,2d}$.  
Note that the principal difference between the 1 and 2 dimensional Gaussians 
is that \mean{\kt{x}}=\mean{\kt{y}}=0, while 
$\mean{\kt{}}\neq 0$ since \vkt{}\ is a 2D radius vector. 
%%mjt3begin:

The two components of \kt{}\ result in different experimentally
measurable effects. \kt{y}\ leads to the acoplanarity of the dijet
pair while \kt{x}\ makes the momenta of the jets unequal which
results in the smearing of the steeply falling \pt{}\ spectrum. This
causes the measured inclusive jet or single particle cross section to
be larger than the pQCD value given by Eq.~\ref{eq:QCDabscat}. This
was observed in the original discovery of high \pt{}\ particle
production at the CERN ISR in 1972~\cite{Busser:1973hs} and led to
much confusion until the existence and effects of \kt{}\ were
understood.

	Before the advent of QCD, the invariant cross section for the
hard-scattering of the electrically charged partons of deeply
inelastic scattering was predicted for \pp\ collisions to follow a
general scaling form:~\cite{Berman:1971xz,Blankenbecler:1972cd}
\begin{equation} 
E \frac{d^3\sigma}{d^3p}=\frac{1}{\pt{}^{n}} F({x_T}) =
\frac{1}{\sqrt{s}^{n}} G({x_T}) \qquad, \label{eq:bbg}
\end{equation}
where $x_T=2\pt{}/\sqrt{s}$. The cross section has two factors, a
function $F({x_T})$ ($G({x_T})$) which `scales', i.e. depends only on
the ratio of momenta, and a dimensioned factor, ${1/\pt{}^{n}}$
($1/\sqrt{s}^{\, n}$), where $n$ equals 4 for QED, and for LO-QCD
(Eq.~\ref{eq:bbg}), analogous to the $1/q^4$ form of Rutherford
Scattering. The structure and fragmentation functions are all in the
$F(x_T)$ ($G(x_T)$) term. The original high \pt{}\ measurements at CERN 
%%%mjt101405 added references to CCR and Cronin
\cite{Busser:1973hs} and Fermilab \cite{Cronin}, showed beautiful $x_T$ scaling, but with a value of
$n=8$ instead of $n=4$, for values of $3\leq\pt{}\leq 7$ GeV/c. Later
measurements at larger \pt{}\ showed the correct scaling in agreement
with pQCD and it was realized that the value $n=8$ at lower values of
\pt{}\ and $\sqrt{s}$ was produced by the \kt{x}\ smearing 
%%%mjt101405 added references
\cite{Feynman2,Feynman3}. More
recently, the deviation of \piz\ and direct photon inclusive cross
sections measurements from pQCD predictions has been used to derive
the values of $\kt{}$ required to bring the measured and smeared pQCD
predictions into agreement.~\cite{Apanasevich_kt_E609}.

A more direct method to determine \kt{y}\ is to measure the
acoplanarity of the dijet pair. Such measurements were originally
performed at the CERN-ISR using two
particle correlations
%%%mjt101405mjt added references
~\cite{CCORjt,CCHK_jet_structure,Darriulat_poutxe,Feynman5}. The same method will be used in
the present work.

%%%mjt042106
Hard-scattering in \pp\ collisions at $\sqrt{s}=200$ GeV is detected by triggering on a $\pi^0$ with transverse momentum $p_{T_t}\geq 3$ GeV/c; and the properties of jets are measured using the method of two particle correlations. The trigger $\pi^0$ is a leading particle from a large transverse momentum jet while the associated particle comes from either the same jet or the
away-side jet. 
%%mjt3end:
We will analyze an outgoing dijet pair, with trigger jet transverse momentum
magnitude
\ptq{t}\ which fragments to a trigger particle with transverse momentum \vpt{t}, and an 
away-side jet transverse momentum magnitude of
\ptq{a}\ which fragments to a particle with 
transverse momentum \vpt{a}. The average transverse momentum component of the 
away-side particle \vpt{a}\ perpendicular to trigger particle \vpt{t}\ in the 
azimuthal plane is labeled as \pout. If the magnitude of the jet transverse 
fragmentation momentum \jt{}\ (\fig{fig:correl-schematic}(a) is 
neglected, the magnitude of $\sqrt{2}\kt{y}$ can be related to \pout:  
$\sqrt{2} \kt{y} = \pout\,\ptq{a}/\pta\equiv\pout/\za$. Thus the measurement 
of \pout\ and the knowledge of the fragmentation variable (\za) 
determines the magnitude of the parton's transverse momentum \kt{}.

%%mjt4begin: 
	The smearing of the steeply falling parton \ptq{}\ spectrum by the
\kt{x}\ distribution tends to make the trigger jet transverse momentum
\ptq{t}\ larger than the away jet transverse momentum \ptq{a}\ . The
component of the net transverse momentum of the parton pair along the
trigger direction is smeared by $\sqrt{2}\,\kt{x}$ such that:
\begin{equation}
\rms{(\ptq{t}-\ptq{ax})}=2\,\rms{\kt{x}}=\rms{\kt{}} \qquad.
\end{equation}
For a flat \ptq{}\ spectrum, the smearing would average to zero so
that there would be no net shift in the transverse momentum spectrum:
\begin{equation}
\mean{\ptq{t}-\ptq{}} = \mean{\ptq{}-\ptq{ax}} = 0 \qquad. 
\end{equation}
However, due to the steeply falling \ptq{}\ spectrum, the \kt{x}\ smearing
results in a net imbalance of the jet-pair towards the trigger direction.
In the limit when \kt{}\ is collinear with the trigger jet and with the
requirement of the Lorentz invariance of $\hat{s}$ 
(\ptqkv{}=\ptq{t}\ptq{a}) 
it is easy to see that
\begin{equation}
\mean{\ptq{t}-\ptq{}} = \mean{{\ptq{t}\over\ptq{}}(\ptq{}-\ptq{a})}
\simeq {1\over 2}{\mean{\ptq{t}-\ptq{a}}}> 0 \;. 
\end{equation}
We denote the imbalance of  \ptq{a}\ and \ptq{t}\ by the quantity 
\begin{equation}
\xhq={\mean{\ptq{a}}/\mean{\ptq{t}}}. 
\label{eq:defhatxh}
\end{equation}
%%%mjt101405I restored the above equation since you use \xhq below.

Jet fragments have a momentum \vjt\ perpendicular to the partonic
transverse momentum (\fig{fig:correl-schematic}(b). This
vector is again a two-dimensional vector with one component
perpendicular to the jet transverse axis, \vptq, in the transverse
plane and the other component perpendicular to the jet transverse axis
in the longitudinal plane (defined by the beam and jet axes).  The
component of \vjt\ projected onto the azimuthal plane is labeled as
\jt{y}. The magnitude of \mean{\jt{y}}, the mean value of \jt{}\ projected into
the plane perpendicular to the jet thrust (see App.\ref{app:means}),
measured at lower energies \cite{CCORjt} has been found to be \pt{}\
independent and $\approx$~400\mevc, consistent with measurements in
\ee\ collisions~\cite{Darriulat:1980nk,e+e-jet_width}.
%%mjt101405added ref above plus some english below.

This analysis uses two-particle azimuthal correlation functions to
measure the average relative angles between a trigger \piz\ and an
associated charged hadron. The angular width of the near- and
away-side peak in the correlation function is used to extract the
value of \rms{\jt{}}\ and \xzkt. An analysis of the associated yields is
used to extract the fragmentation function which provides the \mzt\ and
\mza\ values used for \rms{\kt{}}\ extraction. 
The details on the PHENIX experiment relevant to this analysis follow.

%%%%%%%%%%%%%%%%%%%%%%%%%%%%%%%%%%%%%%%%%\input{ppg_exp}
%%%%%%%%%%%%%%%%%%%%%%%%%%%%%%%%%%%%%%%%%%%%%%%%%%%%%%%%%%%%%%%
\section{Experimental Details}
\label{sec:experimental}
%%%%%%%%%%%%%%%%%%%%%%%%%%%%%%%%%%%%%%%%%%%%%%%%%%%%%%%%%%%%%%%

\begin{figure}[tbh]
\includegraphics[width=1.0\linewidth]{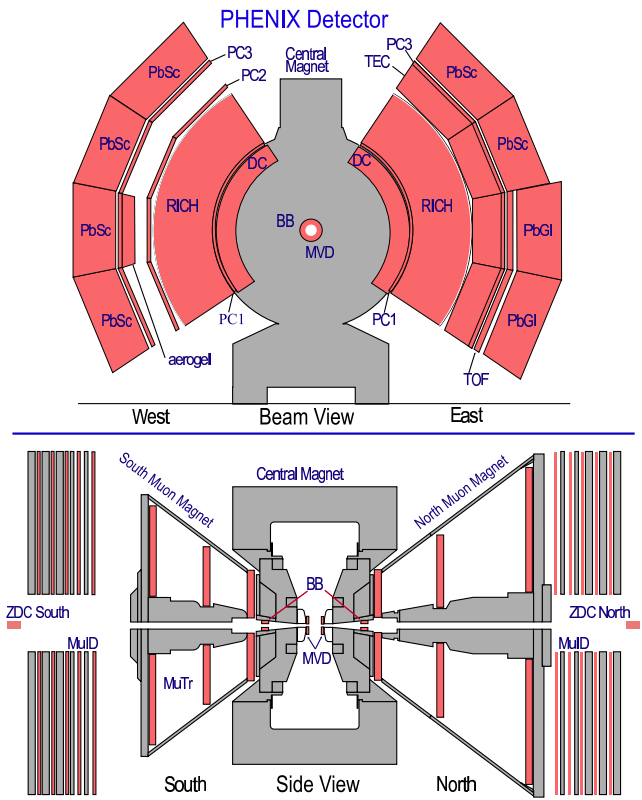}
\caption{ \label{fig:Phenix} (color online)
The PHENIX experimental layout for the Au+Au run
in 2003. The top panel shows the PHENIX central arm spectrometers
viewed along the beam axis. The bottom panel shows a side view of the
PHENIX muon arm spectrometers and the position of the global detectors
(BBC and ZDC). }
\end{figure}

The PHENIX experiment consists of four spectrometer arms - two
around mid-rapidity (the central arms) and two at forward rapidity
(the muon arms) - along with a set of global detectors. The layout of the
PHENIX experiment during RHIC Run-3 is shown in \fig{fig:Phenix}.

%%%%%%%%%%%%%%%%%%%%%%%%%%%%%%%%%%%%%%%%%%%%%%%%%%%%%%%%%%%
%% there has to be a bounding box for not undestood reasons
%%%%%%%%%%%%%%%%%%%%%%%%%%%%%%%%%%%%%%%%%%%%%%%%%%%%%%%%%%%

Each central arm covers the pseudorapidity range $|\eta|<0.35$ and 90
degrees in azimuthal angle $\phi$. In each of the central arms,
charged particles are tracked by a drift chamber (DC) positioned from
2.0 to 2.4m radially outward from the beam axis and 2 or 3 layers of
pixel pad chambers (PC1, PC2, PC3 located at 2.4m, 4.2m, 5m in the 
radial direction, respectively).  Particle identification is provided by
ring imaging \v{C}erenkov counters (RICH), a time of flight scintillator
wall (TOF), and two types of electromagnetic calorimeters 
(EMCal), lead
scintillator (PbSc) and lead glass (PbGl).  The magnetic field for
the central arm spectrometers is axially symmetric around the beam
axis. Its component parallel to the beam axis has an approximately
Gaussian dependence on the radial distance from the beam axis,
dropping from 0.48 T at the center to 0.096 T (0.048 T) at the inner
(outer) radius of the DC. A pair of Zero-Degree Calorimeters (ZDC) and
a pair of Beam-Beam Counters (BBC) were used for global event
characterization. Further details about the design and performance of
PHENIX can be found in \cite{PHENIX_NIM_overview,PHENIX_NIM_EMC,PHENIX_NIM_CA}.

%%%%%%%%%%%%%%%%%%%%%%%%%%%%%%%%%%%%%%%%%%%%%%%%%%%%%%%%%%%% Fig. 2

A \pp\ data sample corresponding to an integrated luminosity 0.35
pb$^{-1}$ at $\sqrt{s}=200$ GeV has been used for the present
analysis. It contains a minimum bias (MB) sample of 121M events and a
high-\pt\ triggered sample of 50M events. The MB trigger is obtained
from the charge multiplicity in the two BBCs situated at large
pseudo-rapidity ($\eta\approx\pm (3.0-3.9)$). The BBCs were also used
to determine the collision vertex, which is limited to a $\pm$30cm
range in this analysis.  The high-\pt\ trigger requests an additional
discrimination on sums of the analog signals from
non-overlapping, 2x2 groups of adjacent EMCal towers situated at
mid-rapidity ($|\eta|<0.35$) equivalent to an energy deposition of 750
MeV \cite{PHENIX_pi0ppPRL}.  The analysis has been performed
separately on the two data sets and no trigger selection bias was
found within the quoted errors.

%%%%%%%%%%%%%%%%%%%%%%%%%%%%%%%%%%%%%%%%%%%%%%%%%%%%%%%%%%%% Fig. 3
\begin{figure}[tbh]%\bgc 
\includegraphics[width=1.0\linewidth]{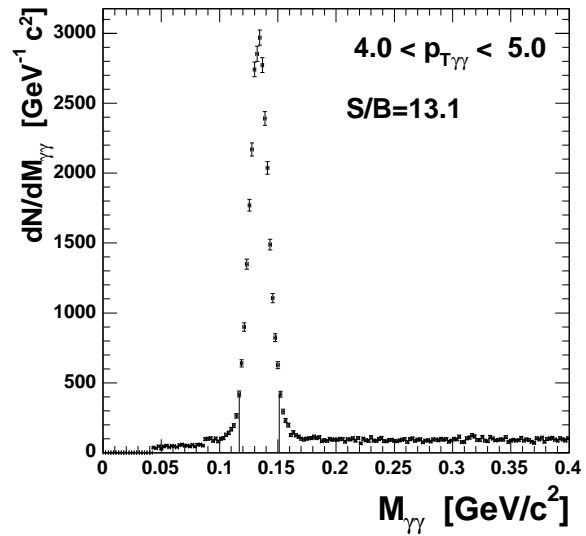}
\caption{\label{fig:mass}
The measured $\gamma\gamma$ invariant mass distribution for pair \pt\ in
$4<p_{T\gamma\gamma}<5$\gevc. The peak is fitted with a Gaussian. The
signal/background ratio within 2$\sigma$ of the mean ranges from
$\approx$~6 at {\pt{}} of 3\gevc\ up to $\approx$~15 at 8\gevc.
}
%\endc
\end{figure}

Neutral pions, which are used as trigger particles, are detected by the
reconstruction of their $\gamma\gamma$ decay channel.
Photons are detected in the EMCal, which has a timing resolution of
$\approx$ 100 $ps$ (PbSc) and $\approx$ 300 $ps$ (PbGl) and energy
resolution of $\sigma_E/E$=1.9\%$\oplus$8.2\%/$\sqrt{E(GeV)}$ (PbSc)
and $\sigma_E/E$=0.8\%$\oplus$8.4\%/$\sqrt{E(GeV)}$ (PbGl). In order
to improve the signal/background ratio we require the minimum hit
energy $>0.3$ \gev, a shower profile cut as described in \cite{PHENIX_cal_2003},
and no accompanying hit in the RICH
detector, which serves as a veto for conversion electrons.  A sample of the
invariant mass distribution of photon pairs detected in the EMCal is shown
in \fig{fig:mass}.

Charged particles are reconstructed in each PHENIX central arm using a
drift chamber, followed by two layers of multiwire proportional
chambers with pad readout \cite{PHENIX_NIM_overview}.  Particle
momenta are measured with a resolution $\delta p/p = 0.7\% \oplus
1.1\%p$ (GeV/c). A confirmation hit is required in PC3. We also
require that no signal in the RICH detector is associated with these
tracks. These requirements eliminate charged particles which do not
originate from the event vertex, such as beam albedo and weak decays,
as well as conversion electrons.

High momentum charged pions 
(above the RICH {\v C}erenkov threshold) are identified using
the RICH and
EMCal detectors. Candidate tracks must be associated with a hit in 
the RICH~\cite{PHENIX_CAID_2003}, 
which corresponds to a minimum 
momentum of 18 MeV/$c$ for electrons, 3.5 GeV/$c$ for muons, and
4.9 GeV/$c$ for charged pions. In a previous PHENIX
publication~\cite{PHENIX_supp_hAuAu200}, we have shown that charged
particles with reconstructed \pt\ above 4.9 \gevc, which have
an associated hit in the RICH, are dominantly charged pions and
background electrons from photon conversions albedo. The efficiency for
detecting charged pions rises quickly past 4.9 \gevc, reaching
an efficiency of $>90\%$ at $\pt{}>6$ \gevc.

%%%%%%%%%%%%%%%%%%%%%%%%%%%%%%%%%%%%%%%%%%%%%%%%%%%%%%%%%%%% Fig. 4
\begin{figure}[ht]
\includegraphics[width=1.0\linewidth]{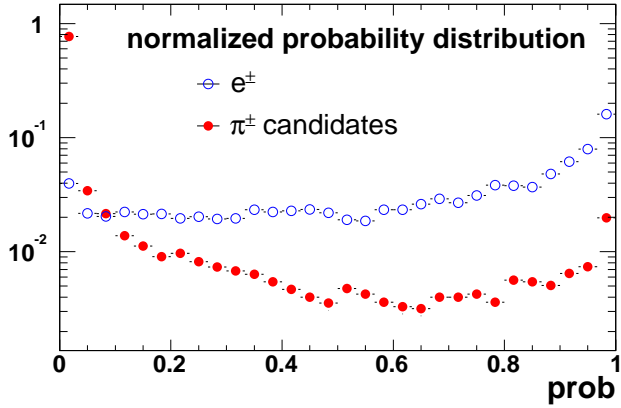}
\caption{\label{fig:prob} (color online)
The probability distribution for charged
pion candidates and electrons derived from the EM shower shape using 
identified electrons and pions. The integrals have been
normalized to one.}
\end{figure}

To reject the electron background in the charged pion candidates, the
shower information at the EMCal is used. Since most of the
background electrons are genuine low $p_T$ particles that were
mis-reconstructed as high \pt\ particles, simply requiring a
large deposition of shower energy in the EMCal is effective in
suppressing the electron background. In this analysis, a momentum-
dependent energy cut on the EMCal is applied
\bge\label{eq:ecut}
E>0.3+0.15 p_T.
\ende
In addition to this energy cut, the shower shape
information~\cite{PHENIX_cal_2003} is used to further separate
the broad hadronic showers from the narrow electromagnetic showers
and hence reduce the conversion backgrounds.
The difference of the EM shower and hadronic shower is
typically characterized by a $\chi^2$
variable~\cite{PHENIX_cal_2003},
\bge\label{eq:chisq}
\chi^2 =\sum_{i}\frac{(E^{meas}_{i}-E^{pred}_{i})^2}{\sigma_{i}^2},
\ende
where $E^{meas}_{i}$ is the energy measured at tower $i$ and
$E^{pred}_{i}$ is the predicted energy for an electromagnetic
particle of total energy $\sum_{i}E^{meas}_{i}$.

In this analysis we use the probability calculated from this
$\chi^2$ value for an EM shower, ranging from 0 to 1 with a 
flat distribution expected for an EM shower,
and a peak around 0 for an hadronic shower.

Figure~\ref{fig:prob} shows the probability distribution for pion and
electron candidates, each normalized to one. The pion candidates were
required to pass the energy cut and the electrons were selected using
particle ID cuts similar to that used in
\cite{Adler:2004electron}. The electron distribution is relatively
flat, while the charged pion distribution peaks at 0. A cut of shower
shape probability~$< 0.2$ selects pions above the energy cut with an
efficiency of $\gtrsim 80$\%. Detailed knowledge of the pion
efficiency is not necessary, since we present in this paper the 
per-trigger pion conditional-yield distributions, for which this
efficiency cancels out.

%%%%%%%%%%%%%%%%%%%%%%%%%%%%%%%%%%%%%%%%%%%%%%%%%%%%%%%%%%%% Fig. 5
\begin{figure}[ht]
\includegraphics[width=1.0\linewidth]{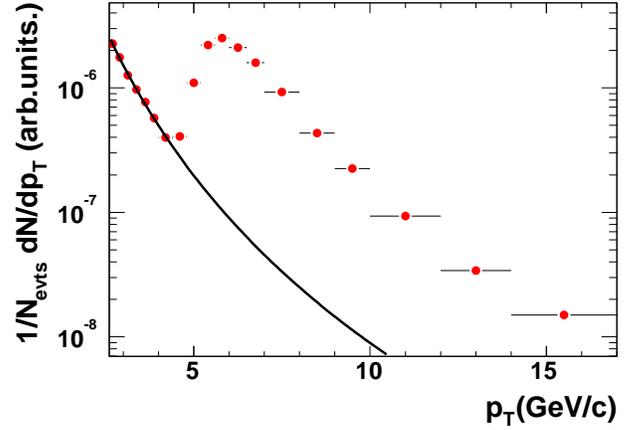}
\caption{\label{fig:pion} (color online)
The raw charged pion transverse momentum
spectrum, with the final cuts applied. The level of the remaining
background is estimated from an extrapolation from low-p$_T$
and is shown as a black line.}
\end{figure}

Since the energy and shower shape cuts are independent of each other,
we can fix one cut and then vary the second to check the remaining
background level from conversions. The energy cut in \eq{eq:ecut}
is chosen such that the raw pion yield is found to be insensitive to
the variation in the shower shape probability. Figure~\ref{fig:pion}
shows the raw pion spectra for EMCal-RICH triggered events as a function
of $p_T$, with the above cuts applied. The pion turn on from $4.9-7$
GeV/$c$ is clearly visible. Below $p_T$ of 5 GeV/$c$, the remaining
background comes mainly from the random association of charged
particles with hits in the RICH detector. The background level is less
than 5\% from $5-16$ GeV/$c$, which is the $p_T$ range for the charged
pion data presented in this paper.

%\clearpage

%%%%%%%%%%%%%%%%%%%%%%%%%%%%%%%%%%%%%%%%%\input{ppg_raw_res}
%%%%%%%%%%%%%%%%%%%%%%%%%%%%%%%%%%%%%%%%%%%%%%%%%%%%%%%%%%%%%%%%%%
\section{Correlation function}
\label{sec:rawResults}
%%%%%%%%%%%%%%%%%%%%%%%%%%%%%%%%%%%%%%%%%%%%%%%%%%%%%%%%%%%%%%%%%%

\begin{figure}[tbh]
\includegraphics[width=0.9\linewidth]{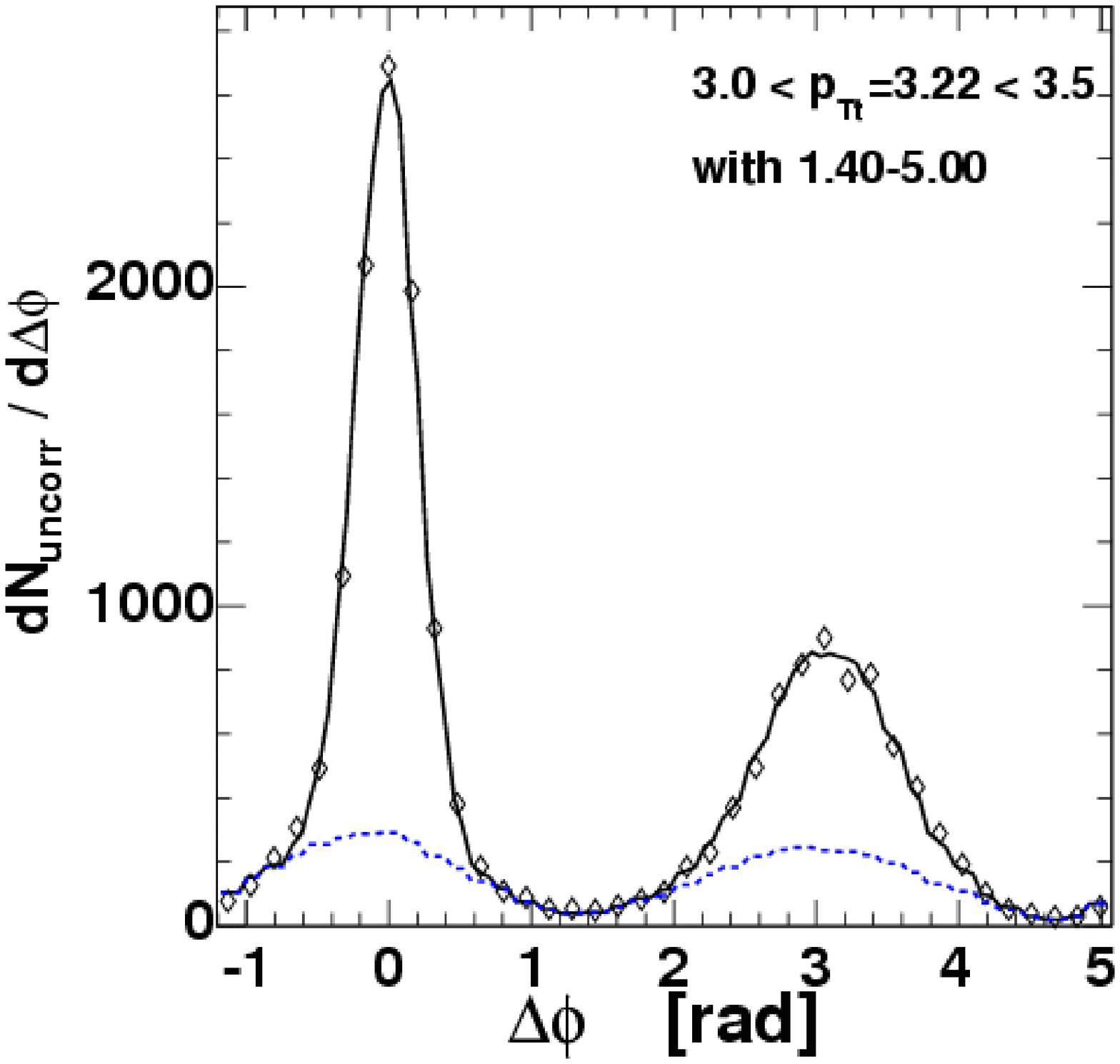}
\includegraphics[width=0.9\linewidth]{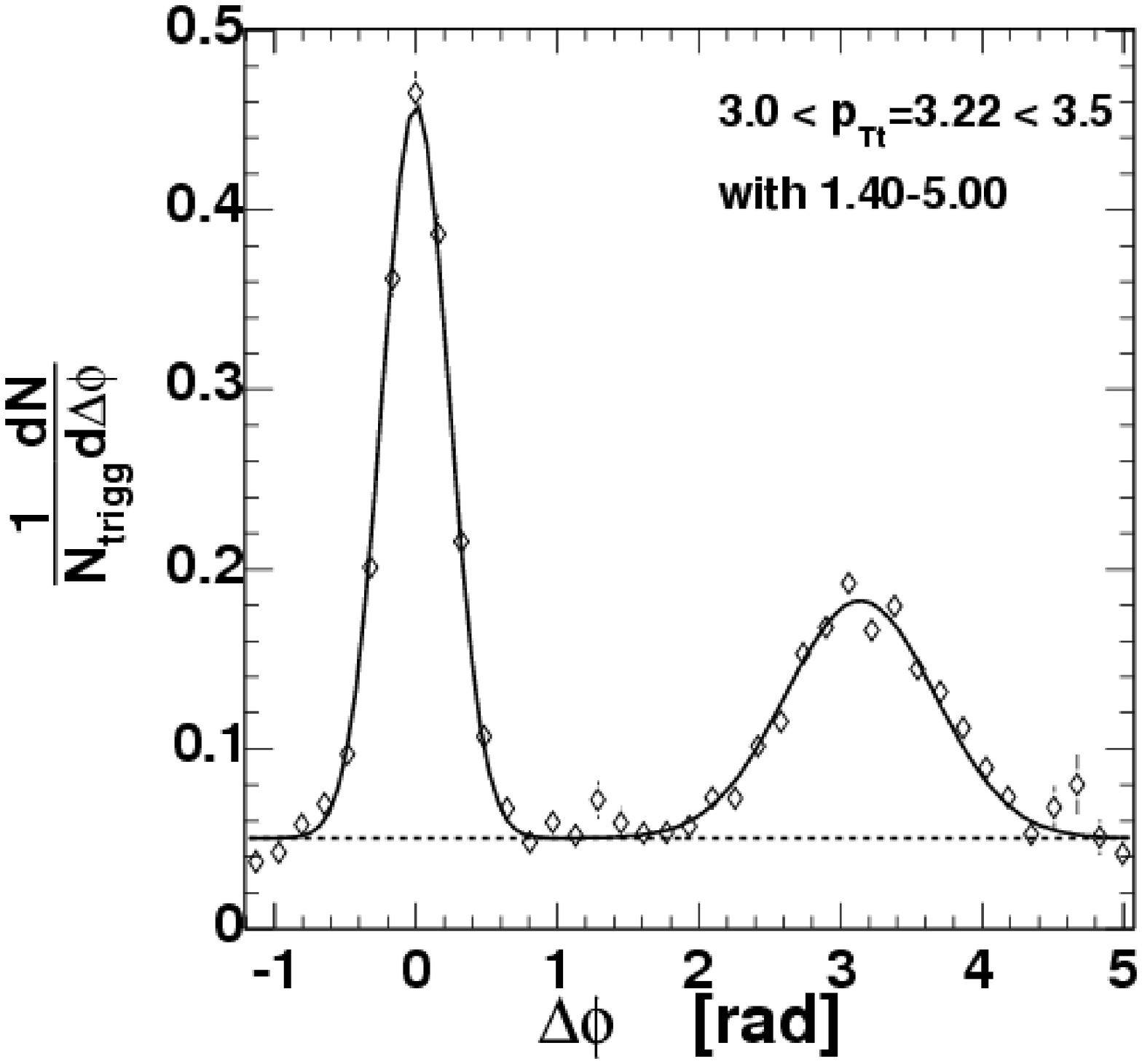}
\caption{\label{fig:CFs} (color online)
An example of the correlation functions for 3$<\ptt<$3.5 \gevc\ and
associated particles in 1.4$<\pta<$5 \gevc. 
(upper) 
Unnormalized pair-yield
distribution plotted with the fit function which is two Gaussians
modulated by the 
pair detection efficiency derived from the mixed
distribution (blue dashed line). 
(lower) Per \piz\ trigger yield distribution
corrected for the pair detection efficiency. Dashed line
represents the constant term in the fit.  }
%\endc
\end{figure}

The analysis uses two-particle azimuthal correlation functions between
a neutral 
pion and an associated charged hadron to measure the distribution of
the azimuthal angle difference $\Delta\phi = \phi_t - \phi_a$ (see
\fig{fig:CFs}).  Whenever a \piz\ was found in an event,  the real,
$dN_{\rm uncorr}/d\Delta\phi$, and mixed, $dN_{\rm mix}/d\Delta\phi$, 
distributions for given \ptt\ (\piz) and \pta\ (charged hadron) were 
accumulated (left panel of \fig{fig:CFs}). Mixed events were obtained by
randomly selecting each member of a particle pair from different
events having similar vertex position. Then the mixed event
distribution was used to correct the correlation function for effects
of the limited PHENIX azimuthal acceptance and for the detection
efficiency, to the extent that it remains constant over the data
sample.

%%%%%%%%%%%%%%%%%%%%%%%%%%%%%%%%%%%%%%%%%%%%%%%%%%%%%%%%%%%% Fig. 6

%%%%%%%%%%%%%%%%%%%%%%%%%%%%%%%%%%%%%%%%%%%%%%%%%%%%%%%%%%%% Fig. 7
\begin{figure}[tbh]%\bgc
\includegraphics[width=0.9\linewidth]{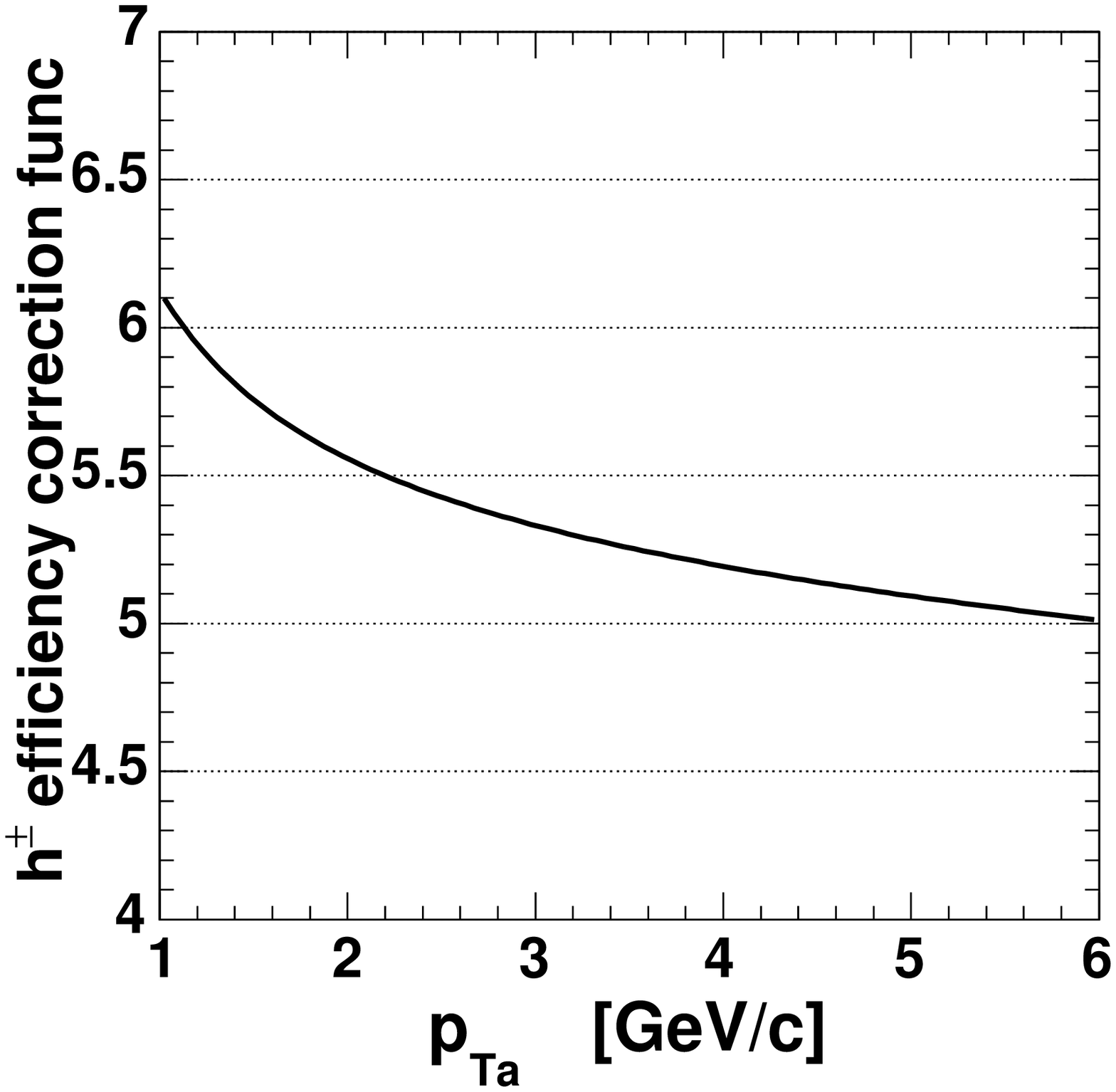}
\includegraphics[width=0.9\linewidth]{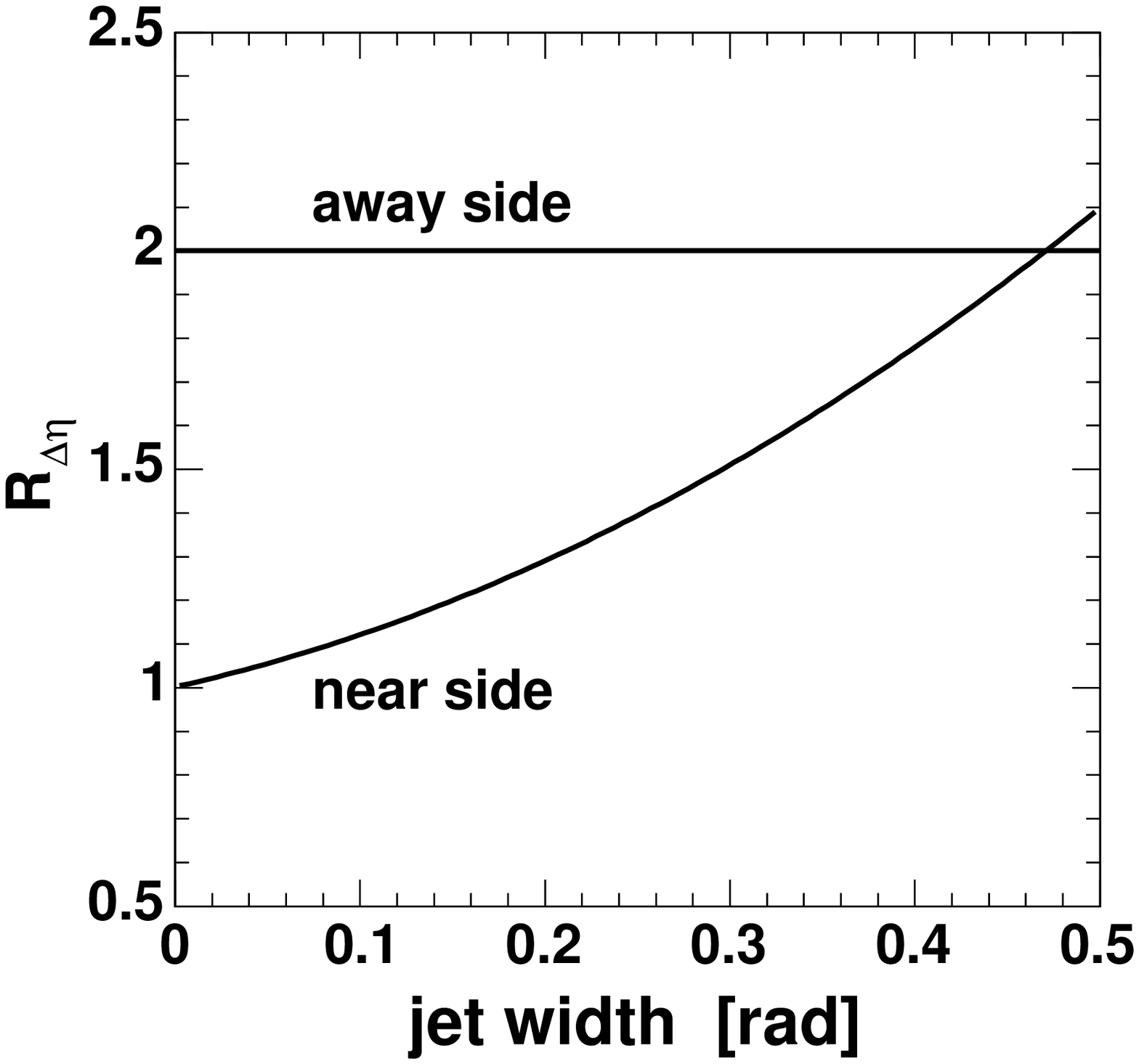}
\caption{\label{fig:effcorr}
(upper) Inclusive charged hadron efficiency correction function.
(lower) $\eta$ acceptance correction
factor for loss of jet pairs outside the limited $\eta$-acceptance
of the PHENIX experiment.}
%\endc
\end{figure}

We fit the raw $dN_{\rm uncorr}/d\Delta\phi$ distribution with the product 
\begin{eqnarray}\label{eq:dnuncor}
  \lefteqn{{dN_{\rm uncorr}\over d\Delta\phi}=}\mbox{\hskip 8cm} \\
  \lefteqn{{1\over\cal{N}}{dN_{\rm mix}\over d\Delta\phi}\cdot
	\left(C	_0+C_1\cdot f_{\rm near}(\Delta\phi)+C_2\cdot f_{\rm away}(\Delta\phi)\right)}\mbox{\hskip 7.6cm}\nonumber
\end{eqnarray}
%\Big|_{\pi/2}^{3\pi/2})
where the mixed event distribution is normalized to $2\pi$
($\mathcal{N}=\sum dN^i_{\rm mix}/d\Delta\phi$ see blue dashed line on
the left panel of \fig{fig:CFs}), $C_{0-2}$ are constant factors to be
determined from the fit, $f_{\rm near}(\Delta\phi)$ and $f_{\rm
away}(\Delta\phi)$ are the near- and away-side peak fit functions
respectively.  Traditionally, the Gaussian functions, around
$\Delta\phi =0$ and around $\Delta\phi =\pi$, are used for $f_{\rm
near}(\Delta\phi)$ and $f_{\rm away}(\Delta\phi)$. This leaves a total
of five free parameters to be determined - the areas and widths of the
above two Gaussians: \yn, \sign\ for the near-angle component and \yf,
\siga\ for the away-angle component and the constant term describing
an uncorrelated distribution of particle pairs which are not
associated with jets. However, the assumption of the Gaussian shape of
the angular correlation induced by jet fragmentation is justified only
in the high-\pt\ region where the relative angles are small.  In order
to access also a lower \pt\ region we used an alternative
parameterization of $f_{\rm near}(\Delta\phi)$ and $f_{\rm
away}(\Delta\phi)$ which will be discussed later in the text.

The normalized correlation function was constructed as a ratio of real
and mixed distributions multiplied by $\eta$-acceptance correction
factor $R_{\Delta\eta}$, divided by \pt{}-dependent efficiency
correction $\epsilon(\pt{})$ (see left panel of \fig{fig:effcorr}) and
divided by the number of \piz\ triggers.
\bge
{1\over N_{\rm trigg}}{dN\over d\Delta\phi} =
{R_{\Delta\eta}\over N_{\rm trigg}\, \epsilon(\pt{})}
{dN_{\rm uncorr}(\Delta\phi)/d\Delta\phi \over dN_{\rm mix}(\Delta\phi)/d\Delta\phi}
\cdot\cal{N}.
\ende

%%%%%%%%%%%%%%%%%%%%%%%%%%%%%%%%%%%%%%%%%%%%%%%%%%%%%%%%%%%% Fig. 8
\begin{figure*}[tbh]
\includegraphics[width=0.9\linewidth]{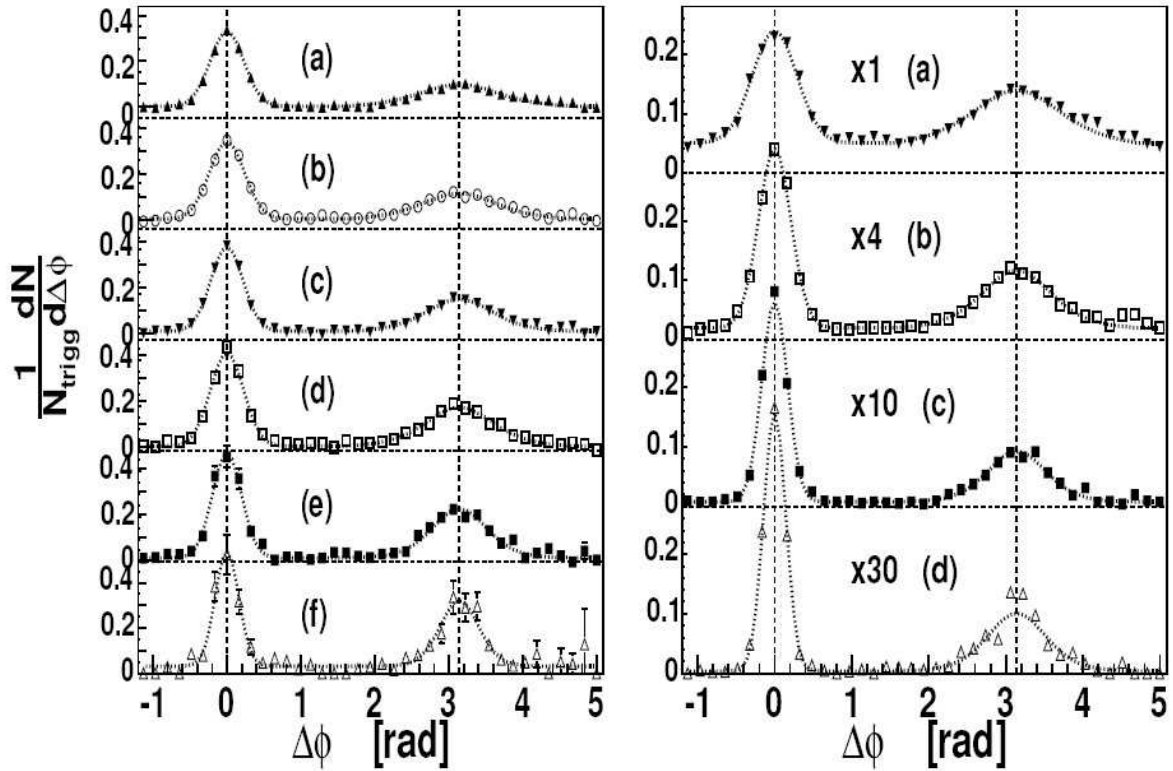}
\caption{\label{fig:correl} 
(left) Measured yield of charged hadrons 
with away-side transverse momentum 
$1.4<\pta<5.0\gevc$ associated with a trigger \piz\ of 
transverse momenta given in in Table~\protect\ref{tab:app_chikv}.
%(a)~$2.5<\ptt<3.0\gevc$,
%(b)~$3.0<\ptt<3.5\gevc$,
%(c)~$3.5<\ptt<4.5\gevc$,
%(d)~$4.5<\ptt<5.5\gevc$,
%(e)~$5.5<\ptt<6.5\gevc$, and
%(f)~$6.5<\ptt<8.0\gevc$.
(right) Measured yield of charged hadrons 
associated with a trigger \piz\ of fixed transverse
momentum $3.0<\ptt<10.0\gevc$ and the  
away-side transverse momenta given in Table~\protect\ref{tab:app_chikv}.
%(a)~$1.0<\pta<2.0\gevc$,
%(b)~$2.0<\pta<3.0\gevc$,
%(c)~$3.0<\pta<4.0\gevc$, and
%(d)~$4.0<\pta<5.0\gevc$.
The dashed lines corresponds to the fit of two Gaussian functions
representing the trigger (t) jet and away-side (a) jet correlation. 
The $\chi^2$(DOF) \sign\ and \sqrtrms{\pout}\ values  extracted 
from these fits are tabulated in Table~\protect\ref{tab:app_chikv}. 
}
\end{figure*}

%%%%%%%%%%%%%%%%%%%%%%%%%%%%%%%%%%%%%%%%%%%%%%%%%%%%%%%%%%%% Table I
\begin{table*}[tbh]
\caption{\label{tab:app_chikv}
The $\chi^2$(DOF) \sign\ and \sqrtrms{\pout}\ values  extracted  for the 
correlation function shown in \fig{fig:correl}. All units in rad and \gevc.  Only the statistical errors are shown.}
\begin{ruledtabular}
\begin{tabular}{cccccccc}
 \multicolumn{4}{c}{$1.4<\pta<5.0\gevc$} &  \multicolumn{4}{c}{$3.0<\ptt<10.0\gevc$}\\
 $\ptt$ \gevc           & $\chi^2$(DOF=34) &  \sign            &  \sqrtrms{\pout}                 &   \pta       & $\chi^2$(DOF=34) &  \sign            &  \sqrtrms{\pout}  \\\hline
(a) $2.5-3.0$  & 69.4  & 0.26 $\pm$ 4E-03 & 1.17 $\pm$ 0.07& (a) $1.0-2.0$ & 188.4 & 0.29 $\pm$ 4E-03 & 0.87 $\pm$ 0.03\\
(b) $3.0-3.5$  & 79.6  & 0.24 $\pm$ 4E-03 & 1.19 $\pm$ 0.05& (b) $2.0-3.0$ & 63.2  & 0.21 $\pm$ 3E-03 & 1.16 $\pm$ 0.04\\
(c) $3.5-4.5$  & 61.2  & 0.22 $\pm$ 3E-03 & 1.04 $\pm$ 0.04& (c) $3.0-4.0$ & 50.3  & 0.16 $\pm$ 4E-03 & 1.36 $\pm$ 0.06\\
(d) $4.5-5.5$  & 52.7  & 0.22 $\pm$ 6E-03 & 1.08 $\pm$ 0.06& (d) $4.0-5.0$ & 63.2  & 0.14 $\pm$ 4E-03 & 1.69 $\pm$ 0.13\\
(e) $5.5-6.5$  & 38.4  & 0.20 $\pm$ 8E-03 & 0.90 $\pm$ 0.06\\
(f) $6.5-8.0$  & 31.6  & 0.16 $\pm$ 1E-02 & 0.64 $\pm$ 0.06\\
\end{tabular}			
\end{ruledtabular}
\end{table*}

The $R_{\Delta\eta}$ correction factor which accounts for limited
$\eta$ acceptance of the PHENIX experiment (see right panel of
\fig{fig:effcorr}) for the the near-side yield, with an assumption
that the angular jet width is the same in $\Delta\eta$ and in $\Delta\phi$,
can be written as
\bge
R_{\Delta\eta}={1\over {1\over\sqrt{2\pi\sign^2}} \int_{-0.7}^{0.7}\exp\left(-{\Delta\eta\over 2\sign^2}\right)
acc(\Delta\eta)d\eta},
\ende
where $acc(\Delta\eta)$ represent the PHENIX pair acceptance
function in $|\Delta\eta|$. It can be obtained by convolving two
flat distributions in $|\Delta\eta|<$0.35, so $acc(\Delta\eta)$ has a simple
triangular shape: $acc(\Delta\eta)=(0.7-|\Delta\eta|)/0.7$.
For the away-side yield the corresponding $R_{\Delta\eta}$ is
\bge
R_{\Delta\eta}={2(0.7) \over {1\over\sqrt{2\pi\sign^2}} \int_{-0.7}^{0.7} acc(\Delta\eta)d\eta}
=2.
\ende
$R_{\Delta\eta}$ equals 2, because the pair efficiency has a triangular
shape in $|\Delta\eta|<$0.7, which results in 50\% average efficiency
when the real jet pair distribution is flat in
$|\Delta\eta|<$0.7. Normalized correlation functions for various
\pt{t}\ and \pt{a}\ are shown in \fig{fig:correl}.

%\bge\label{eq:dndpout}
%{dN_{away}\over d\Delta\phi}%\Big|_{\pi/2}^{3\pi/2}
%={-\pt{a}\cos\Delta\phi\over\sqrt{2\pi\rms{\pout}}{\rm Erf}({\sqrt{2}\pt{a}\over\sqrtrms{\pout}})}
%\exp\left(-{\ptkv{a}\sin^2\Delta\phi\over 2\rms{\pout}}\right)
%\ende
%and $a_0$, $C_1$, $C_2$, \sign\ and \rms{\pout}\ are
%parameters to be determined from the fit.

%%%%%%%%%%%%
% formulae %
%%%%%%%%%%%%
For two particles with transverse momenta \ptt, \pta\ from the same
jet, the width of the near-side correlation distribution can be
related to the RMS value of the \jt{}\ vector component, \jt{y}, perpendicular
to the parton momentum as
\begin{equation}
\sign^2 = \meankv{\Delta\phi} = \mean{\left({\jt{y}\over\pta}\right)^2+\left({\jt{y}\over\ptt}\right)^2},
\end{equation}

\clearpage

\noindent where we assume \rms{\jt{y}}$\ll$\ptkv{t}\ and \ptkv{a}\ and 
thus the arc-sine function
can be approximated by its argument and we can solve for 
\footnote{For relations between $\sqrt{\la X^2\ra}$ and
$\la|X_y|\ra$, where $X$ is any 2D quantity, see App.\ref{app:means}}
\bge
\sqrtrms{\jt{}}=\sqrt{2\,\rms{\jt{y}}}
\simeq\sqrt{2}{\ptt\,\pta\over\sqrt{\ptt^2+\pta^2}}\; \sign \quad.
\label{eq:rmsjt}
\ende

%%%%%%%%%%%%%%%%%
In order to extract \meanabs{\kt{y}}, or \rms{\kt{}}, 
we start with the relation~\cite{Feynman5,CCORjt} 
between the magnitude of \pout,(see \fig{fig:correl-schematic})
\bge\label{eq:def_pout}
\pout = \pt{a}\,\sin\Delta\phi,
\ende
which is the transverse momentum component of the away-side particle \vpt{a}\
perpendicular to trigger particle \vpt{t}\ in the azimuthal plane (see
\fig{fig:correl-schematic}), and \kt{y}: 
\bge\label{eq:CCOR_pout}
\meanabs{\pout}^2 = \xe^2\left[2\meanabs{\kt{y}}^2 + \meanabs{\jt{y}}^2\right] + \meanabs{\jt{y}}^2,
\ende
where
\bge\label{eq:xe}
\xe=-{\vpt{t}\cdot\vpt{a}\over\ptt^2}= -{\pta\,\cos{\Delta\phi}\over \ptt}
\simeq{\za\,\ptq{a}\over\zt\,\ptq{t}}
\ende
represents the fragmentation variable of the away-side jet.~\cite{Darriulat_poutxe,CCHK_jet_structure} We note
however, that \cite{Feynman5} explicitly neglected
$\mzt=\la\ptt/\ptq{t}\ra$ in the formula at ISR energies, where
$\mzt\simeq$~0.85, while it is not negligible at
\s=200~GeV. Furthermore, as mentioned earlier, the average values of
trigger and associated jet momenta are generally not the same. There
is a systematic momentum imbalance due to  %%%mjt101405+020805
\kt{}-smearing of the steeply falling parton momentum distribution. The
event sample with a condition of \ptt$>$\pta\ is dominated by
configurations where the \kt{}-vector is parallel to the trigger jet and
antiparallel to the  associated jet and $\mean{\ptq{t}-\ptq{a}}\ne 0$.
Here we introduce the hadronic %%%mjt021006
variable \xh\ in analogy to the partonic variable \xhq\
\bge\label{eq:xhqdef}
\xh\equiv{\pt{a}\over\pt{t}}\mbox{\hskip 0.5cm} \xhq = \xhq(\rms{\kt{}},\xh)\equiv{\mean{\ptq{a}}\over\mean{\ptq{t}}}
\ende

The detailed discussion on the magnitude of this imbalance is given later.
%sections \ref{sec:kt_smearing} and \ref{sec:pythia}.
In order to derive the relation between the magnitude of \pout\ and
\kt{} let us first consider the simple case where we have neglected
both trigger and associated \mean{\jt{}}\ (see panel (a) on
\fig{fig:correl-schematic}). In this case one can see that
\begin{eqnarray}
\meanabs{\pout}|_{\jt{t}=\jt{a}=0} & \equiv & \meanabs{\pout}_{00} = \nonumber\\
& = & \sqrt{2}\meanabs{\kt{y}}{\pta\over\mean{\ptq{a}}} =\nonumber\\
& = & \sqrt{2}\meanabs{\kt{y}}\mzt{\xh\over\xhq} \; .\nonumber
\end{eqnarray}
Rewriting the formula for \pout\ in terms of RMS we get
$$
\sqrtrms{\pout}_{00}=\mzt\sqrtrms{\kt{}}\,{\xh\over\xhq} \qquad,
$$  %%%mjt101505 add back following comment
where we have taken \rms{k_T}=\rms{2k_{T_y}}\ .

\begin{figure}[tbh]%\bgc 
\includegraphics[width=1.0\linewidth]{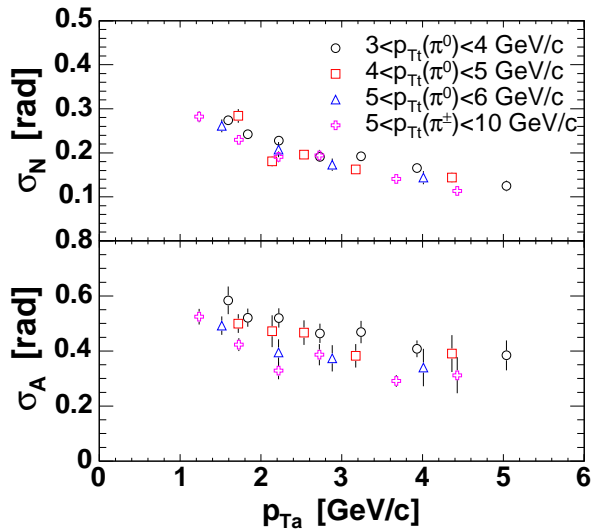}
\caption{\label{fig:sigmas_pta} (color online)
(top)  The width of the near-side peak \sign\ with \pta\ for
various values of \ptt\ as indicated in legend.
(bottom) The width of the far-side peak \siga\ with \pta\ for the same 
\ptt\
selection.}
%\endc
\end{figure}

\begin{figure}[tbh]%\bgc 
\includegraphics[width=1.0\linewidth]{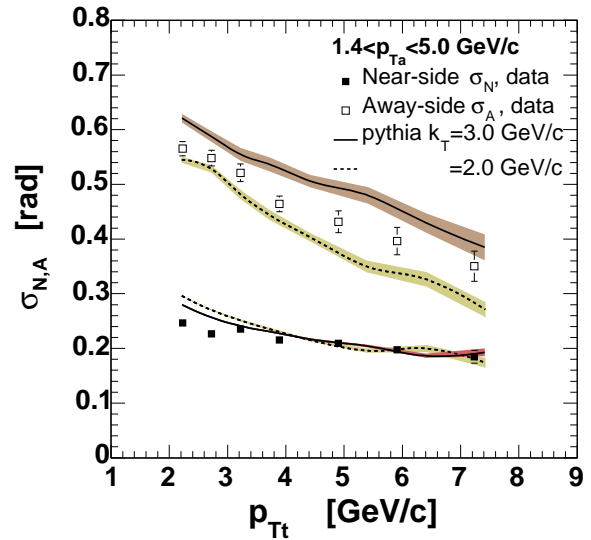}
\caption{\label{fig:sigmas_ptt} (color online)
The near-side (squares) and away-side (circles) width
as a function or trigger-\piz\ \ptt. The associated charged particle
momenta are in the $1.4<\pta<5.0$ \gevc\ region.   The curves are 
from a PYTHIA calculation with the values of $k_T$ indicated.  
The data values are given in Table~\protect\ref{tab:app_sig}.}
%\endc
\end{figure}

	However, the jet fragments are produced with finite jet transverse momentum
\jt{}. The situation when the trigger particle is produced with \jt{ty}$>0$ \gevc\
and the associated particle with \jt{ay}=0 \gevc\ is shown in %%%mjt101405+0206 
\fig{fig:correl-schematic}b. The \pout\ vector picks up an additional component
\begin{eqnarray}
 &   & \rms{\pout}|_{\jt{t}>0,\,\jt{a}=0} =\nonumber\\
 & = & \left[\rms{\pout}_{00}+{\rms{\jt{ty}}\over\ptkv{t}}(\ptkv{a}-\rms{\pout}_{00})\right]
        {\ptkv{t}-\rms{\jt{ty}}\over\ptkv{t}}\nonumber
\end{eqnarray}
With an assumption of $\jt{ty}\ll\pt{t}$ we found that 
$$
\rms{\pout}|_{\jt{t}>0,\,\jt{a}=0}=\xh^2\left[\mzt^2\rms{\kt{}}\,{1\over\xhq^2}+\rms{\jt{ty}}\right]
$$  %%%mjt101405
We include \jt{a}\ in the same approximation, $\jt{ay}\ll\pt{a}$, i.e.  collinearity of \jt{a}\ and \pout\, with result 
\bge\label{eq:pout_jtjt}
\rms{\pout}
%%%mjt101405-I removed this condition, this is pout punkt! %%%|_{\jt{t}>0,\,\jt{a}>0}
=\xh^2\left[\mzt^2\rms{\kt{}}\,{1\over\xhq^2}+\rms{\jt{ty}}\right]+\rms{\jt{ay}}
\ende
and we solve for \zkt/\xhq
$$
{\zkt\over\xhq}={1\over\xh}\sqrt{\rms{\pout}-\rms{\jt{ay}}-\xh^2\rms{\jt{ty}}} \qquad .
$$
If we assume no difference between \jt{t}\ and \jt{a}\ then we have
\bge
\label{eq:zkt}
{\mean{\zt(\kt{},\xh)}\sqrtrms{\kt{}}\over\xhq(\kt{},\xh)} =
{1\over\xh}\sqrt{\rms{\pout}-\rms{\jt{y}}(1+\xh^2)}
\ende
All quantities on the right-hand side of \eq{eq:zkt} can be directly
extracted from the correlation function.  The correlation functions
are measured in the variable $\Delta\phi$ in bins of \ptt\ and \pta\
(\eg\ see \fig{fig:correl}), and the rms of the near and away peaks
\sign\ and \siga\ are extracted. We tabulated \sign\ and \siga\ for
many combinations of \ptt\ and \pta\ (see \fig{fig:sigmas_pta} and
\fig{fig:sigmas_ptt}).  

Initially, we used
the approximation $\sqrtrms{\pout}\sim \pta\sin\siga$ in Eq.~\ref{eq:zkt}. 
However, we have noticed that this approximation and other approximations for
\sqrtrms{\pout}\ proposed \eg\ in reference \cite{Levai_kt_05} (see
appendix \ref{sec:azimuth_correl}) are inadequate for $\siga > 0.4$
radians, so we don't use \siga\ to calculate \kt{}. 

We extract \sqrtrms{\pout}\ directly for all values of \ptt\, \pta\ (even for
wide bins in \pta\, using the \mean{\pta} of the bin) by fitting the
correlation function in the $\pi/2 < \Delta\phi < 3\pi/2$ region by
\begin{eqnarray}
\label{eq:dndpout}
\lefteqn{ {dN_{away}\over d\Delta\phi}\Big|_{\pi/2}^{3\pi/2}
={dN\over d\pout}{d\pout\over d\Delta\phi}= }\mbox{\hskip 8cm} \\
={-\pt{a}\cos\Delta\phi\over\sqrt{2\pi\rms{\pout}}{\rm Erf}({\sqrt{2}\pt{a}\over\sqrtrms{\pout}})}
\exp\left(-{\ptkv{a}\sin^2\Delta\phi\over 2\rms{\pout}}\right)\nonumber
\end{eqnarray}

%\clearpage 

\begin{figure}[tbh]
\includegraphics[width=0.90\linewidth]{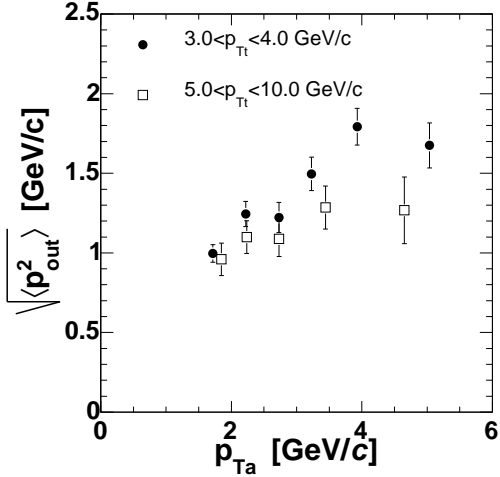}
\caption{\label{fig:pout_pta}
Extracted values of \sqrtrms{\pout}\ for $3.0<\ptt<4.0$
and $5.0<\ptt<10.0$ \gevc\ for various values of \pta\ using the
direct \pout\ extraction method based on fitting the away-side peak by
\eq{eq:dndpout}.
}
\end{figure}

\begin{figure}[tbh]
\includegraphics[width=0.90\linewidth]{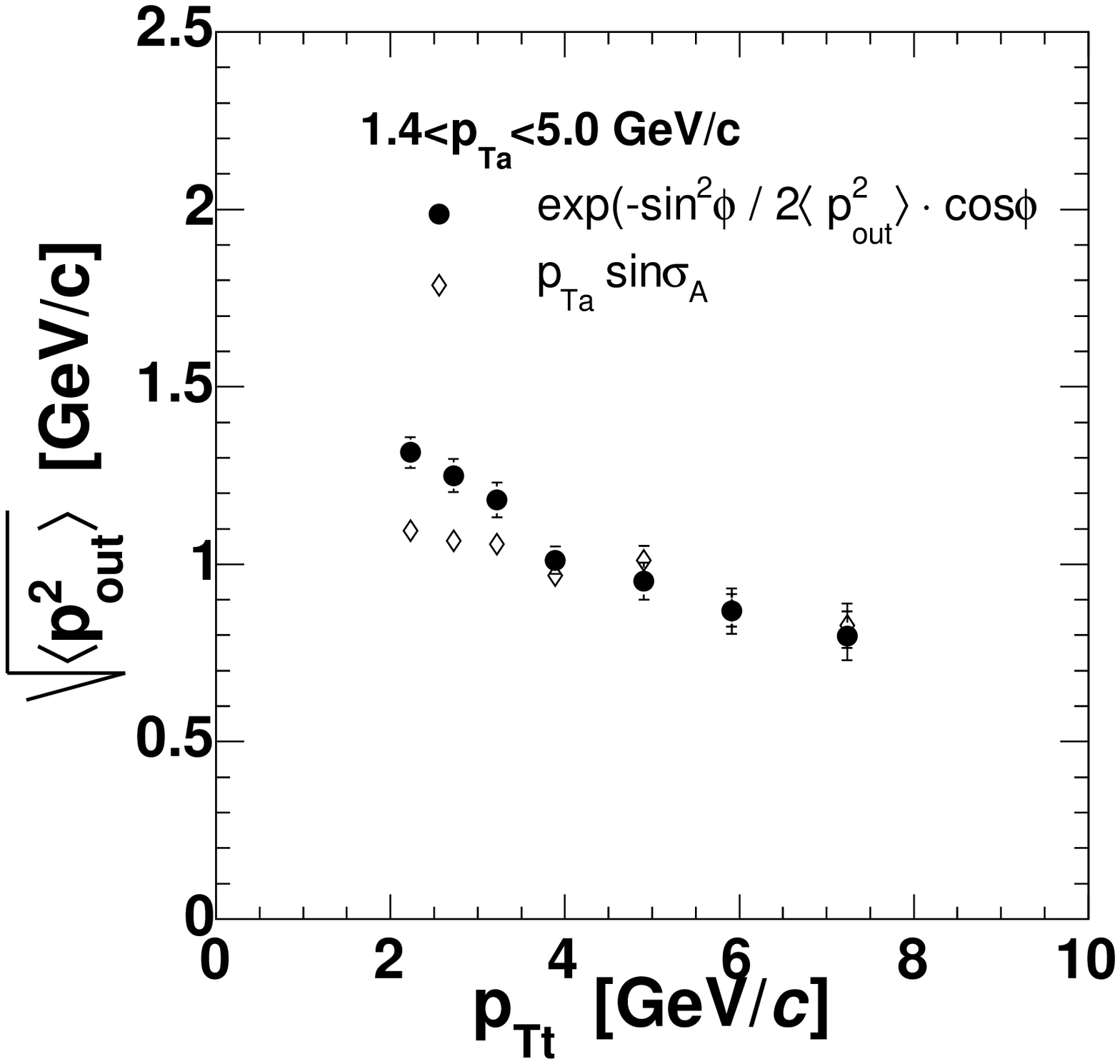}
\caption{\label{fig:pout_ptt}
(solid circles) Extracted values of \sqrtrms{\pout}\
for $1.4<\pta<5.0$ \gevc\ for various values of \ptt\ from
\eq{eq:dndpout}.  (open diamonds) Indirect \pout=\pta sin(\siga) 
values.}
\end{figure}

%%%%%%%%%%%%%%%%%%%%%%%%%%%%%%%%%%%%%%%%%%%%%%%%%%%%%%%%%%%% Table II
\begin{table*}[tbh]
\caption{\label{tab:sigma}
Measured widths of the near- and away-angle \pizh\ correlation peaks for various trigger particle momenta .  Only the statistical errors are shown.
}
\begin{ruledtabular}
\begin{tabular}{cccccccccccc}
 \multicolumn{3}{c}{\ptt=3.39~\gevc} & \multicolumn{3}{c}{\ptt=4.40~\gevc} &
 \multicolumn{3}{c}{\ptt=5.41~\gevc} & \multicolumn{3}{c}{\ptt=6.40~\gevc}\\ \hline
\pta & \sign~rad          &  \siga~rad          & \pta & \sign~rad          &  \siga~rad        & \pta & \sign~rad          & \siga~rad         & \pta & \sign~rad          & \siga~rad \\ \hline
1.59 &  0.27  $\pm$  0.01 &  0.58  $\pm$   0.05 & 1.72 &  0.28  $\pm$  0.02 &  0.50 $\pm$  0.03 & 1.51 &  0.26  $\pm$  0.01 &  0.49 $\pm$  0.03 & 1.34 &  0.40  $\pm$  0.03 &  0.68 $\pm$  0.05 \\
1.84 &  0.24  $\pm$  0.01 &  0.52  $\pm$   0.03 & 2.14 &  0.18  $\pm$  0.01 &  0.47 $\pm$  0.06 & 2.22 &  0.21  $\pm$  0.02 &  0.39 $\pm$  0.05 & 1.64 &  0.30  $\pm$  0.02 &  0.58 $\pm$  0.05 \\
2.22 &  0.23  $\pm$  0.01 &  0.52  $\pm$   0.03 & 2.53 &  0.20  $\pm$  0.01 &  0.47 $\pm$  0.04 & 2.88 &  0.17  $\pm$  0.01 &  0.37 $\pm$  0.05 & 1.94 &  0.23  $\pm$  0.02 &  0.52 $\pm$  0.06 \\
2.73 &  0.19  $\pm$  0.01 &  0.46  $\pm$   0.04 & 3.17 &  0.16  $\pm$  0.01 &  0.38 $\pm$  0.04 & 4.01 &  0.14  $\pm$  0.02 &  0.34 $\pm$  0.07 & 2.29 &  0.23  $\pm$  0.02 &  0.40 $\pm$  0.03 \\
3.24 &  0.19  $\pm$  0.01 &  0.47  $\pm$   0.04 & 4.36 &  0.14  $\pm$  0.01 &  0.39 $\pm$  0.07 &      &                    &                   & 2.74 &  0.17  $\pm$  0.01 &  0.41 $\pm$  0.05 \\
3.93 &  0.17  $\pm$  0.01 &  0.41  $\pm$   0.03 &      &                    &                   &      &                    &                   & 3.36 &  0.17  $\pm$  0.02 &  0.36 $\pm$  0.04 \\
5.04 &  0.12  $\pm$  0.01 &  0.38  $\pm$   0.05 &      &                    &                   &      &                    &                   &      &                    & \\
\end{tabular}
\end{ruledtabular}
\end{table*}

where we assumed a Gaussian distribution in
\pout. We still use a Gaussian function in $\Delta\phi$ in the near
angle peak to extract \sqrtrms{\jt{}}. The \sqrtrms{\pout}\ values
extracted from the fit of the functional form (\ref{eq:dndpout}) 
are shown in \fig{fig:pout_pta} and \fig{fig:pout_ptt}.

%%%%%%%%%%%%%%%%%%%%%%%%%%%%%%%%%%%%%%%%%%%%%%%%%%%%%%%%%%%% Fig. 9

\begin{figure}[tbh]
\includegraphics[width=1.0\linewidth]{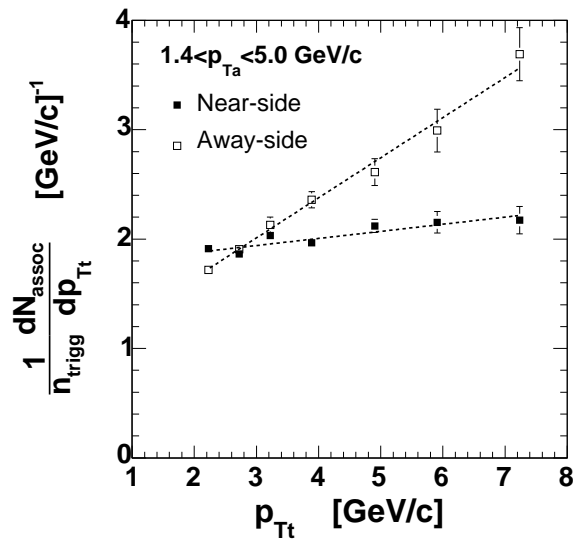}
\caption{\label{fig:assoc_yield} 
Measured yield of charged hadrons associated with one trigger \piz\ 
with transverse momenta indicated in Table~\protect\ref{tab:app_chikv} 
and associated charged hadron with $1.4<\pt<5.0\gevc$. 
%%%mjt020806$1.4<\pt<2.4\gevc$ 
Dashed lines represent the linear fit.
} 
\end{figure}

%%%mjt101405English and other changes below.
The per-trigger yields as a function of \ptt\ for fixed associated $1.4<\pta<5.0$ \gevc\ bin
are shown in \fig{fig:assoc_yield}. There is a distinct behavior of
the near-side yield which varies with trigger \pt{t}\ much less than
the away-side yield.  For the away-side, this reflects the fact that the particle detected
in the fixed associated bin are produced from the lower $z$ region of
the fragmentation function for events with higher trigger \pt{t}.
For the near-side jet, this multiplicity increase is reduced due to the
fact that with increasing \pt{t}\ the near-side jet energy 
increases; however, at the same time the larger fraction of this
energy is taken away by the more energetic trigger particle. Thus the relative change in $z$ is smaller on the near-side. 

%%%%%%%%%%%%%%%%%%%%%%%%%%%%%%%%%%%%%%%%%%%%%%%%%%%%%%%%%%%% Fig. 10

%%%%%%%%%%%%%%%%%%%%%%%%%%%%%%%%%%%%%%%%%%%%%%%%%%%%%%%%%%%% Table III
\begin{table}[bth]
\caption{\label{tab:app_sig}
The \sign\ and \siga\ values shown in \fig{fig:sigmas_ptt}. 
All units in rad and \gevc.  Only the statistical errors are shown.}
%\bgc
\begin{ruledtabular}
\begin{tabular}{ccc}
\hline
\ptt  &  \sign              &  \siga \\\hline
2.23 &  0.247 $\pm$  0.002 &  0.565 $\pm$  0.013\\
2.72 &  0.227 $\pm$  0.003 &  0.548 $\pm$  0.014\\
3.22 &  0.235 $\pm$  0.004 &  0.521 $\pm$  0.016\\
3.89 &  0.215 $\pm$  0.004 &  0.464 $\pm$  0.014\\
4.90 &  0.210 $\pm$  0.006 &  0.431 $\pm$  0.020\\
5.91 &  0.197 $\pm$  0.009 &  0.396 $\pm$  0.025\\
7.23 &  0.185 $\pm$  0.012 &  0.350 $\pm$  0.028\\
\hline
\end{tabular}
\end{ruledtabular}
%\endc
\end{table}

In order to extract \mzt\  and \xhq\ 
knowledge of the fragmentation
function is needed; a detailed discussion is given in following
sections.

\clearpage

%%%%%%%%%%%%%%%%%%%%%%%%%%%%%%%%%%%%%%%%%%%%%%%%%%%%%%%%%%%% Fig. 11

%%%%%%%%%%%%%%%%%%%%%%%%%%%%%%%%%%%%%%%%%%%%%%%%%%%%%%%%%%%% Fig. 12

%%%%%%%%%%%%%%%%%%%%%%%%%%%%%%%%%%%%%%%%%%%%%%%%%%%%%%%%%%%% Table IV
\begin{table}[tbh]
\caption{\label{tab:app_pout}
The \sqrtrms{\pout} values shown in \protect\fig{fig:pout_pta} and 
\protect\fig{fig:pout_ptt}.
All units in \gevc.  Only the statistical errors are shown.
}
%\bgc
\begin{ruledtabular}
\begin{tabular}{cccccc}
\multicolumn{2}{c}{$1.4<\pta<5.0$}&\multicolumn{2}{c}{$3<\ptt<4$}&\multicolumn{2}{c}{$5<\ptt<10$}\\
\ptt  &   \sqrtrms{\pout}   & \pta   &   \sqrtrms{\pout}   & \pta   &   \sqrtrms{\pout} \\\hline
2.23 &  1.315 $\pm$  0.043 &  1.72 &  0.996 $\pm$  0.056 &  1.85 &  0.960 $\pm$  0.102 \\    
2.72 &  1.250 $\pm$  0.046 &  2.22 &  1.244 $\pm$  0.079 &  2.24 &  1.100 $\pm$  0.103 \\
3.22 &  1.182 $\pm$  0.049 &  2.73 &  1.222 $\pm$  0.095 &  2.73 &  1.088 $\pm$  0.110 \\
3.89 &  1.011 $\pm$  0.038 &  3.23 &  1.496 $\pm$  0.105 &  3.44 &  1.285 $\pm$  0.136 \\
4.90 &  0.953 $\pm$  0.052 &  3.93 &  1.793 $\pm$  0.115 &  4.65 &  1.268 $\pm$  0.210 \\
5.91 &  0.868 $\pm$  0.064 &  5.04 &  1.675 $\pm$  0.141 \\
7.24 &  0.798 $\pm$  0.068 \\
\end{tabular}
\end{ruledtabular}
%\endc
\end{table}

%%%%%%%%%%%%%%%%%%%%%%%%%%%%%%%%%%%%%%%%%%%%%%%%%%%%%%%%%%%% Table V
\begin{table}[tbh]
\caption{\label{tab:app_yield}
The near and away side conditional yield per number of triggers for
$1.4<\pta<5.0$ \gevc\ shown in \fig{fig:assoc_yield}.  
All units in rad and \gevc.   Only the statistical errors are shown.
}
%\bgc
\begin{ruledtabular}
\begin{tabular}{ccc}
\ptt &  $Y_N$              &  $Y_A$ \\\hline
2.23 &  1.911 $\pm$  0.018 &   1.717 $\pm$  0.044 \\
2.72 &  1.863 $\pm$  0.022 &   1.908 $\pm$  0.055 \\
3.22 &  2.032 $\pm$  0.032 &   2.130 $\pm$  0.071 \\
3.89 &  1.966 $\pm$  0.033 &   2.360 $\pm$  0.074 \\
4.90 &  2.120 $\pm$  0.061 &   2.611 $\pm$  0.123 \\
5.91 &  2.153 $\pm$  0.098 &   2.992 $\pm$  0.196 \\
7.24 &  2.174 $\pm$  0.125 &   3.690 $\pm$  0.242 \\
\end{tabular}
\end{ruledtabular}
%\endc
\end{table}

%\clearpage

%%%%%%%%%%%%%%%%%%%%%%%%%%%%%%%%%%%%%%%%%%%%%%%%%%%%%%%%%%%% Fig. 13

%%%%%%%%%%%%%%%%%%%%%%%%%%%%%%%%%%%%%%
\subsection{\sqrtrms{\jt{}}\ and \xzkt\ results}
%%%%%%%%%%%%%%%%%%%%%%%%%%%%%%%%%%%%%%

The measurement is performed in two different kinematical
regimes; first the transverse momentum of the trigger particle,
\ptt, is fixed and the peak width is measured for different values
of associated particle transverse momenta \pta\
(\fig{fig:sigmas_pta}).  (Note that in the region 
of overlap, the data are in excellent agreement with a 

\clearpage 

%%%%%%%%%%%%%%%%%%%%%%%%%%%%%%%%%%%%%%%%%%%%%%%%%%%%%%%%%%%% Table VI
\begin{table}[tbh]
\caption{\label{tab:app_jt}
The \sqrtrms{\jt{}}\ values shown in \protect\fig{fig:jt_pta} and 
\protect\fig{fig:jt_ptt}. 
All units in rad and \gevc.  Only the statistical errors are shown.
}
\begin{ruledtabular}
\begin{tabular}{cccc}
          \multicolumn{2}{c}{$1.4<\pta<5.0$}
        & \multicolumn{2}{c}{$3<\ptt<4$}\\
\ptt  &  \sqrtrms{\jt{}} &  \pta &  \sqrtrms{\jt{}}\\\hline
3.22 &  0.587 $\pm$  0.009  & 1.72 &  0.562 $\pm$  0.011 \\
3.89 &  0.577 $\pm$  0.009  & 2.22 &  0.597 $\pm$  0.014 \\
4.90 &  0.600 $\pm$  0.017  & 2.73 &  0.572 $\pm$  0.017 \\
5.91 &  0.596 $\pm$  0.026  & 3.23 &  0.590 $\pm$  0.020 \\
7.24 &  0.597 $\pm$  0.038  & 3.93 &  0.603 $\pm$  0.017 \\
8.34 &  0.632 $\pm$  0.085  & 5.04 &  0.506 $\pm$  0.029 \\\hline
   \multicolumn{2}{c}{$4<\ptt<5$} & \multicolumn{2}{c}{$5<\ptt<6$}\\
\pta &  \sqrtrms{\jt{}} & \pta &  \sqrtrms{\jt{}}\\\hline
1.72 &  0.643 $\pm$  0.036 & 1.52 &  0.529 $\pm$  0.030\\
2.14 &  0.492 $\pm$  0.032 & 2.22 &  0.581 $\pm$  0.049\\
2.53 &  0.608 $\pm$  0.035 & 2.88 &  0.590 $\pm$  0.047\\
3.17 &  0.590 $\pm$  0.032 & 4.01 &  0.603 $\pm$  0.063\\
4.36 &  0.631 $\pm$  0.052\\
\end{tabular}
\end{ruledtabular}
\end{table}
%	  \multicolumn{2}{c}{$1.4<\pta<5.0$}
%	& \multicolumn{2}{c}{$3<\ptt<4$}
%       & \multicolumn{2}{c}{$4<\ptt<5$}
%	& \multicolumn{2}{c}{$5<\ptt<6$}\\
%---\ptt  &  \sqrtrms{\jt{}} &  \pta &  \sqrtrms{\jt{}} &  \pta &  
%---\sqrtrms{\jt{}} &  \pta &  \sqrtrms{\jt{}}  \\ \hline
%3.22 &  0.587 $\pm$  0.009 & 1.72 &  0.562 $\pm$  0.011 & 1.72 &  0.643 $\pm$  0.036 & 1.52 &  0.529 $\pm$  0.030\\
%3.89 &  0.577 $\pm$  0.009 & 2.22 &  0.597 $\pm$  0.014 & 2.14 &  0.492 $\pm$  0.032 & 2.22 &  0.581 $\pm$  0.049\\
%4.90 &  0.600 $\pm$  0.017 & 2.73 &  0.572 $\pm$  0.017 & 2.53 &  0.608 $\pm$  0.035 & 2.88 &  0.590 $\pm$  0.047\\
%5.91 &  0.596 $\pm$  0.026 & 3.23 &  0.590 $\pm$  0.020 & 3.17 &  0.590 $\pm$  0.032 & 4.01 &  0.603 $\pm$  0.063\\
%7.24 &  0.597 $\pm$  0.038 & 3.93 &  0.603 $\pm$  0.017 & 4.36 &  0.631 $\pm$  0.052\\
%8.34 &  0.632 $\pm$  0.085 & 5.04 &  0.506 $\pm$  0.029\\

%%%%%%%%%%%%%%%%%%%%%%%%%%%%%%%%%%%%%%%%%%%%%%%%%%%%%%%%%%%% Fig. 14
\begin{figure}[tbh]%\bgc 
\includegraphics[width=0.90\linewidth]{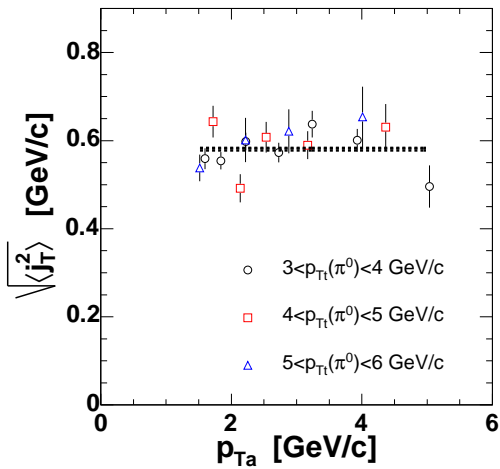}
\caption{\label{fig:jt_pta} (color online)
\rms{\jt{}}\ values calculated according \eq{eq:rmsjt}. The dashed
line represents the 0$^{th}$-order polynomial fit in the $1.5<\pta<5$~\gevc\ 
region.
}
%\endc
\end{figure}

%%%%%%%%%%%%%%%%%%%%%%%%%%%%%%%%%%%%%%%%%%%%%%%%%%%%%%%%%%%% Fig. 15
\begin{figure}[hbt]
%\bgc 
\includegraphics[width=1.0\linewidth]{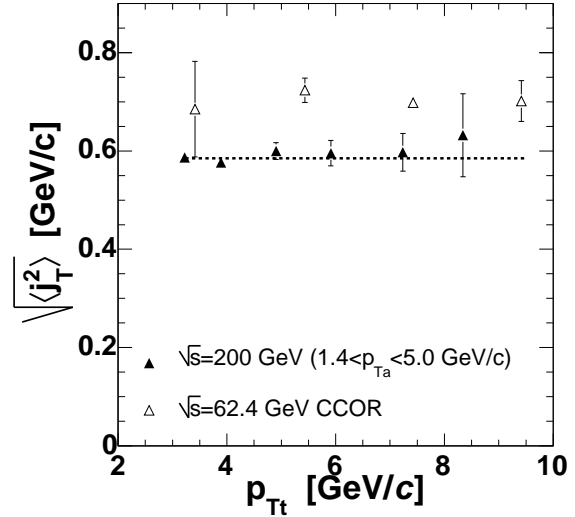}
\caption{
Averaged values of \rms{\jt{}}\ in (1.5$<\pta<$5~\gevc) as a function of the
trigger transverse momentum \ptt\ (solid triangles). The CCOR values measured
at \s=62.4 \gev\ shown by open triangles.}
\label{fig:jt_ptt} 
%\endc
\end{figure}

\noindent previous measurement~\cite{STAR_supp_dAu}.)  
In the second case, particle pairs with a fixed associated bin 
%%%mjt020806$1.4<\pta<2.4$ \gevc\ 
$1.4<\pta<5.0$ \gevc\ and
various \ptt\ are selected (\fig{fig:sigmas_ptt}). It is evident that
both near and away-side correlation peaks in all cases reveal a
decreasing trend with \pta\ and
\ptt.

However, the asymptotic behavior of \sign\ and \siga\ is
different. Whereas the magnitude of \sign, according to \eq{eq:rmsjt},
should vanish for large values of \ptt\ and \pta, the \siga\,
according to \eq{eq:zkt} should be approximately constant around
\xzkt/\ptt\ for large values of \pta. The \mzt\ and \xhq\
quantities are implicitly \pta\ dependent, however, their ratio 
%%mjt020806
is roughly $\sim 1$ so that the asymptotic value of
\siga$\big|_{\pta\rightarrow\infty}\sim$\sqrtrms{\kt{}}/\ptt.
%%mjt021006 delete following incorrect statement
%%With $\ptt\rightarrow\infty$,  \siga\ vanishes regardless of actual \pta\ magnitude.

Extracted values of \rms{\jt{}}\ as a function of \pta\ according to
\eq{eq:rmsjt} are shown in \fig{fig:jt_pta}.  All \rms{\jt{}}\ values are
constant in the explored region (\pta$>$1.5~\gevc). It is expected that \rms{\jt{}}\
can not remain constant for arbitrarily small values of \pta\ because
of the phase space limitation. In the region where \pta$\leq$\sqrtrms{\jt{}}, the
magnitude of the \jt{}-vector is truncated, similar to the ``Seagull
effect'' \cite{Seagull_Votruba_1975}. Since the \sqrtrms{\jt{}}\ values are on the
order of 600~\mevc, we assume that the phase space limitation can be
safely neglected for \pta$>$1.5~\gevc\ and we extract the values of
\sqrtrms{\jt{}}\ averaged over \pt{a}\ and  \pt{t}\ (see \fig{fig:jt_ptt})
\bge\label{res:jt}
  \jtfinal
\ende

%\clearpage

The systematic error originates from the 
%%%mjt042006 from JR  assumption made in
finite momentum resolution and 
\eq{eq:rmsjt} where we assume that the arc-sine function can be
approximated by its argument. For the angular width of the near angle
peak (see \fig{fig:sigmas_pta} and \fig{fig:sigmas_ptt}) it
corresponds to an uncertainty of order of 3\%. 

%%%%%%%%%%%%%%%%%%%%%%%%%%%%%%%%%%%%%%%%%%%%%%%%%%%%%%%%%%%% Fig. 16
\begin{figure}[tbh]
%\bgc 
\includegraphics[width=1.0\linewidth]{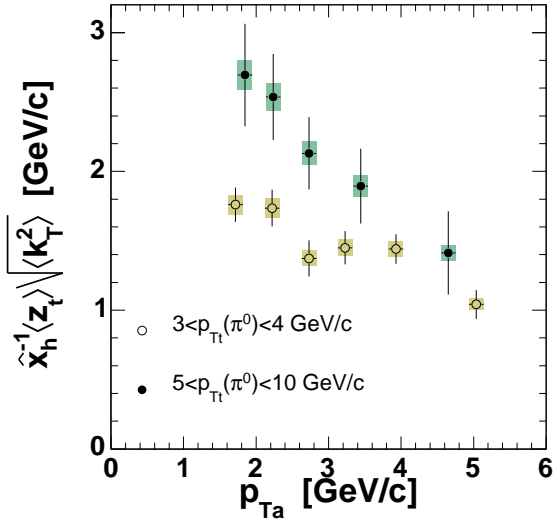}
\caption{\label{fig:zkt_pta} (color online)
\xhq$^{-1}$\zkt\ values calculated according to \eq{eq:zkt} for
trigger \piz\ in $3<\pt{t}<4$~\gevc\ and $5<\pt{t}<6$~\gevc\ as a
function of \pt{a}. The systematic uncertainties are indicated by the shaded
boxes.} 
%\endc
\end{figure}

The independence of \rms{\jt{}}\ on either \ptt\ or $\sqrt{s}$ has been
observed by the CCOR experiment in the range \s=31--62.4 \gev\
\cite{CCORjt}. The \rms{\jt{}}\ values at \s=62.4 \gev\ (open triangles on
\fig{fig:jt_ptt}) are systematically larger then values found in this
analysis. The discrepancy should not be taken as significant, as CCOR
used a slightly different technique than in this paper. CCOR extracted
the \rms{\jt{}}\ values from measurements of \meanabs{\pout}\ for different values
of the $x_E$ variable \eq{eq:xe}. According to \eq{eq:CCOR_pout} the
\meanabs{\pout}$^2$ magnitude should depend linearly on $x^2_E$; and the
\meanabs{\jt{y}}\ value was extracted from the intercept of the \mean{\pout}$^2(x_E$) fit 
at $x_E$=0, rather than from a measurement of \sign.

\begin{figure}[tbh]
%\bgc 
\includegraphics[width=1.0\linewidth]{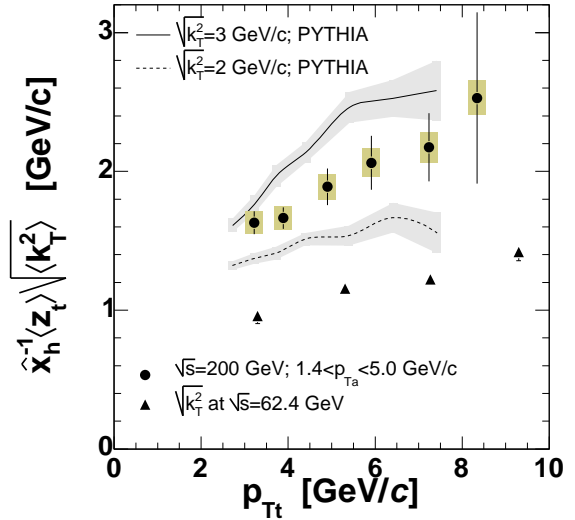}
\caption{\label{fig:zkt_ptt} (color online)
The same calculation according
\eq{eq:zkt} for fixed associated bin $1.4<\pt{a}<5.0$ \gevc\ as a
function of \pt{t}. The systematic errors indicated by colored rectangles.
The \sqrtrms{\kt{}}\ results obtained by CCOR collaboration at \s=62.4 \gev\ \cite{CCORjt} displayed by
solid triangles.}
%\endc
\end{figure}

%%%mjt101405 english below
Knowing the \rms{\jt{}}\ and \pout\ values, we used \eq{eq:zkt} to 
determine 
\xzkt\ (see \fig{fig:zkt_pta} and \fig{fig:zkt_ptt}).
The systematic error was estimated with Monte Carlo simulations to be
on the order of 5\%. The main source of systematic error originates
from the assumption (Eqs. (\ref{eq:rmsjt}) and (\ref{eq:pout_jtjt}))
of the relative smallness of \rms{\jt{}}, collinearity between \pout\
and \jt{ay} and from the limited momentum resolution discussed in
section \ref{sec:experimental}.

%%%%%%%%%%%%%%%%%%%%%%%%%%%%%%%%%%%%%%%%%%%%%%%%%%%%%%%%%%%% Fig. 17

The \pta\ dependence of the extracted \xzkt\ (\fig{fig:zkt_pta})
reveals a strikingly decreasing trend. It was originally expected that
by fixing the value of \ptt, the kinematics of the hard scattering
(i.e. \ptq{t}$\simeq$\ptq{a}) would be fixed, independently of the
value of \pta.  Various values of \pta\ would then sample the \ptq{a}\
fragmentation function, and the value of \xzkt\ was expected to be
constant. It is evident that this assumption is not quite
correct. %%%mjt020806

A similar line of argument applies also for the rising trend when 
\pta\ is fixed and \ptt\ varies (\fig{fig:zkt_ptt}). It is interesting
to note that the CCOR \sqrtrms{\kt{}}\ values measured at \s=62.4
\gev\ (open triangles on \fig{fig:zkt_ptt}) reveal a similar rising
trend. However, the rising trend of \xzkt\ with \ptt\ and falling with
\pta\ suggests that the variation of \sqrtrms{\kt{}}\ with \ptt\
seen by the CCOR collaboration \cite{CCORjt} may be indicative of the
$\langle z_t\rangle \hat{x}_h^{-1}$ variation which was there
neglected.\footnote{Note, however that the method was different. CCOR
determined \jt{}\ and \kt{}\ from the slope and intercept of
Eq.~\ref{eq:pout_jtjt} with respect to \pta\ at each value of \ptt,
with the implicit assumption that $\langle z_t\rangle
\hat{x}_h^{-1}=1$.} In order to understand variation of \mzt\ and
\xhq\ we have to explore the process of dijet fragmentation.

\section{Fragmentation Functions}
\label{sec:inclusFrag}

We have shown in \eq{eq:zkt} that the width of the away side correlation peak

is related to the product of \xzkt. In order to evaluate \mzt, knowledge of the 
scattered parton $\hat{p}_T$ spectrum and fragmentation function is required.

%%%jr and mjt 020806
%The partonic quantities \mzt\ and \xhq\
%are not directly measurable but they can be evaluated, given the
%knowledge of the parton distribution and fragmentation function.
Fragmentation functions from $e^+ e^-$ collisions, weighted
by the appropriate hard-scattering constituent cross-sections and
$Q^2$ evolution could in principle be used. However, it was originally thought
that the shape of the fragmentation function could be deduced from present
measurements using the combined analysis of the inclusive trigger \pt{t}\ and
associated particle \pt{a}\ distributions. Although this idea turned out to be
incorrect, we will follow this line of reasoning for a while as it is
instructive.

Generally, the invariant cross section for inclusive hadron production
from jets can be parametrized in the following way. First, we assume
that the number of parton fragments (consider only pions for
simplicity) at a given \pt{}\ corresponds to the sum over all
contributions from parton momenta, \ptq{}\, from $p_T<\ptq{}<\sqrt{s}/2$.
The joint probability of detecting a pion with \pt{}$=z\ptq{}$ originating 
from a parton with \ptq{}\ can be written as
\begin{eqnarray}
\label{jointprob}
{d^2\sigma_\pi\over\ptq{} d\ptq{} dz} & = & {d\sigma_q\over\ptq{} d\ptq{}}\times \D (z)\nonumber\\
%%%\fkt\-->\fq\ mjt042106
 & = & \fq\times\D(z).
\end{eqnarray}
Here we use \fq\ to represent the final state scattered-parton
invariant spectrum $d\sigma_q/\ptq{} d\ptq{}$ and $\D(z)$ to represent the fragmentation
function.  The first term in \eq{jointprob} can be viewed as a 
probability of finding a parton with transverse momentum \ptq{}\ and the
second term corresponds to the probability that the parton %%%mjt \ptq{}\
fragments into a particle of momentum \pt{}$=z\ptq{}$.  With a simple change of 
variables from \ptq{}\ to \pt{}$=z\ptq{}$, we obtain the joint probability of a pion 
with \pt{}\ which is a fragment with momentum fraction $z$ from a parton with 
\ptq{}=\pt{}$/z$:
\begin{equation}
\label{jointprobpi}
{d^2\sigma_\pi\over\pt{} d\pt{} dz} = f_{q}({p_T\over z})\cdot\D(z)\; {1\over z^{2}}.
\end{equation}
The \pt{}\ and $z$ dependences do not factorize. However, the \pt{}\ spectrum may be 
found by integrating over all values of \ptq{}\ $\geq$ \pt{}\ to \ptq{}$_{max}=\sqrt{s}/2$, 
which corresponds to values of $z$ from $x_T=2 p_T/\sqrt{s}$ to 1.  
\begin{equation}
\label{eq:dsigma_integral}
{1\over \pt{}}{d\sigma_{\pi}\over d\pt{}}  =  
 \int_{x_T}^{1} f_{q}({p_T\over z})\cdot\D(z)\; {dz\over z^{2}} 
\end{equation} 
Alternatively, for any fixed value of \pt{}\ one can 
evaluate the $\la z(p_T)\ra$, integrated over the parton spectrum: 
\bge \label{meanz_inc}
\la z(p_T)\ra = {{\int_{x_T}^{1}z\;\D(z)\; \fq\ (\pt{}/z) {dz\over z^{2}}} \over
{\int_{x_T}^{1}\D(z)\; f_{q}(\pt{}/z) {dz\over z^{2}}}}\qquad .
\ende
 
From the scaling properties of QCD and from the shape of the \piz\
invariant cross section itself, which is a pure power law for $p_T\geq
3$ GeV/c \cite{PHENIX_pi0ppPRL}, one can deduce that \fq\ 
should have a power law shape, $\fq=A \ptq{}^{-n}$. In this case the hadron spectrum 
also has a power law shape because
\begin{eqnarray}
\label{power_law}
{1\over \pt{}}{d\sigma_{\pi}\over d\pt{}} & \approx &
\int_{x_T}^{1} A\, \D(z)\cdot ({p_T\over z})^{-n} {dz\over z^{2}} \nonumber\\
& \approx & {A\over {\pt{}^{n}}} \int_{x_T}^{1} \, \D(z)\cdot {z}^{n-2} dz
\end{eqnarray}
and the last integral depends only weakly on \pt{}\ due to the small value of 
$x_T$. For small parton \ptq{}\ (below 3-4 \gevc) the power law
shape is no longer valid, but the region $p_T < 3$ GeV/c is outside
the scope of this paper.  The \fq\ should also diminish for
very high \ptq{}$\rightarrow\sqrt{s}/2$ where the phase space
available for hard parton production diminishes, again not relevant
for the present purposes.

We used the power law parameterization for the final state scattered-parton invariant spectrum
$\fq \propto \ptq{}^{-n}$ where $n$ is a free parameter 
%following 
%parametrization for the final state scattered-parton 
%invariant spectrum:  
%$$
%\fkt \propto \ptq{}^{-(n1+n2\cdot\log(2\ptq{}/\s))}
%$$
%where $n1$ and $n2$ are a free parameters 
which can be determined from the fit of
\eq{eq:dsigma_integral} to the measured \piz\ cross section. 
There is, however, one more missing piece of information - the shape of the
fragmentation function \D. In 
%%%mjt020806 order 
an attempt to extract this information
from the data, we have analyzed associated $x_E$-distributions, as shown 
in Table~\ref{tab:xe}.

%%%%%%%%%%%%%%%%%%%%%%%%%%%%%%%%%%%%%%%%%%%%%%%%%%%%%%%%%%%% Table VII
\begin{table*}[tbh]
\caption{\label{tab:xe}
Measured $x_E$ distributions associated with various transverse momenta of the trigger \piz.  Only the statistical errors are shown.
} 
\begin{ruledtabular}
\begin{tabular}{llllllllll}
 \multicolumn{2}{c}{\ptt=3.39~\gevc}&\multicolumn{2}{c}{\ptt=4.40~\gevc}&
 \multicolumn{2}{c}{\ptt=5.41~\gevc}&\multicolumn{2}{c}{\ptt=6.40~\gevc}&
 \multicolumn{2}{c}{\ptt=7.39~\gevc} \\ 
$x_E$ & \multicolumn{1}{c}{$dn/dx_E$} & $x_E$ & \multicolumn{1}{c}{$dn/dx_E$} &
$x_E$ & \multicolumn{1}{c}{$dn/dx_E$} & $x_E$ & \multicolumn{1}{c}{$dn/dx_E$} &
$x_E$ & \multicolumn{1}{c}{$dn/dx_E$} \\ \hline
0.32 & 2.7e+00 $\pm$ 4.7e-02  &  0.23 & 6.7e-01 $\pm$ 2.2e-02  & 0.22 & 2.3e-01 $\pm$ 1.2e-02 &  0.18 & 8.0e-02 $\pm$ 6.3e-03  & 0.17 & 1.8e-02 $\pm$ 3.1e-03\\
0.37 & 1.9e+00 $\pm$ 4.0e-02  &  0.27 & 6.8e-01 $\pm$ 2.2e-02  & 0.27 & 1.4e-01 $\pm$ 9.6e-03 &  0.22 & 4.6e-02 $\pm$ 4.7e-03  & 0.24 & 9.0e-03 $\pm$ 1.5e-03\\
0.42 & 1.4e+00 $\pm$ 3.3e-02  &  0.32 & 4.8e-01 $\pm$ 1.9e-02  & 0.32 & 9.4e-02 $\pm$ 7.7e-03 &  0.27 & 3.1e-02 $\pm$ 3.9e-03  & 0.33 & 4.4e-03 $\pm$ 1.0e-03\\
0.47 & 9.6e-01 $\pm$ 2.8e-02  &  0.37 & 2.9e-01 $\pm$ 1.4e-02  & 0.37 & 5.7e-02 $\pm$ 6.0e-03 &  0.35 & 1.7e-02 $\pm$ 2.0e-03  & 0.45 & 2.8e-03 $\pm$ 8.1e-04\\
0.52 & 7.3e-01 $\pm$ 2.4e-02  &  0.42 & 2.2e-01 $\pm$ 1.2e-02  & 0.43 & 4.1e-02 $\pm$ 5.0e-03 &  0.44 & 8.2e-03 $\pm$ 1.4e-03  & 0.55 & 6.9e-04 $\pm$ 4.0e-04\\
0.57 & 5.2e-01 $\pm$ 2.0e-02  &  0.47 & 1.5e-01 $\pm$ 1.0e-02  & 0.47 & 2.8e-02 $\pm$ 4.2e-03 &  0.54 & 3.8e-03 $\pm$ 9.2e-04  & 0.64 & 4.5e-04 $\pm$ 3.2e-04\\
0.62 & 3.8e-01 $\pm$ 1.7e-02  &  0.52 & 8.0e-02 $\pm$ 7.4e-03  & 0.52 & 2.3e-02 $\pm$ 3.8e-03 &  0.64 & 2.4e-03 $\pm$ 7.3e-04  & &\\
0.67 & 3.0e-01 $\pm$ 1.5e-02  &  0.57 & 8.6e-02 $\pm$ 7.7e-03  & 0.57 & 1.9e-02 $\pm$ 3.4e-03 &  0.81 & 9.3e-04 $\pm$ 2.6e-04  & &\\
0.75 & 2.1e-01 $\pm$ 9.0e-03  &  0.62 & 5.8e-02 $\pm$ 6.3e-03  & 0.63 & 1.1e-02 $\pm$ 2.5e-03 &	  & & &\\
0.85 & 1.1e-01 $\pm$ 6.5e-03  &  0.68 & 4.9e-02 $\pm$ 5.7e-03  & 0.67 & 1.1e-02 $\pm$ 2.5e-03 &	  & & &\\
0.95 & 8.2e-02 $\pm$ 5.5e-03  &  0.75 & 3.2e-02 $\pm$ 3.3e-03  & 0.76 & 5.6e-03 $\pm$ 1.3e-03 &	  & & &\\
1.04 & 5.4e-02 $\pm$ 4.5e-03  &  0.85 & 2.0e-02 $\pm$ 2.5e-03  & 0.85 & 2.9e-03 $\pm$ 9.2e-04 &	  & & &\\
1.15 & 3.6e-02 $\pm$ 3.6e-03  &  0.94 & 1.6e-02 $\pm$ 2.2e-03  & 0.97 & 2.3e-03 $\pm$ 8.1e-04 &	  & & &\\
1.25 & 2.8e-02 $\pm$ 3.2e-03  &  1.04 & 7.0e-03 $\pm$ 1.5e-03  & 1.07 & 8.3e-04 $\pm$ 4.8e-04 &	  & & &\\
\end{tabular}
\end{ruledtabular}
\end{table*}

%%%%%%%%%%%%%%%%%%%%%%%%%%%%%%%%%%%%%%%%%\input{ppg_frag_xe}
%%%%%%%%%%%%%%%%%%%%%%%%%%%%%%%%%%%%%%%%%%%%%%%%%%%%%%%%%%%%%%%%%%
\subsection{'Scaling' variable \xe }
\label{sec:frag_xe}
%%%%%%%%%%%%%%%%%%%%%%%%%%%%%%%%%%%%%%%%%%%%%%%%%%%%%%%%%%%%%%%%%%

%%first JR then MJT020806
It was expected~\cite{Darriulat_poutxe} that the \xe\ variable, defined by \eq{eq:xe}, to first order, approximates  the
fragmentation function in the limit of high values of \ptt, where
there is sufficient collinearity between the trigger particle and the
fragmenting parton.  In this case where \jt{}$\ll$\ptt\ and \kt{}$\ll$\ptt\
one can assume that \ptt=\ptq{t}/\zt\ and $\xe\,\zt =
\xhq\,\pt{a}\,\cos{\Delta\phi}/\ptq{a} \simeq \xhq\,\za$, 
and thus the slopes of $D(\za)$ and
\xe\ are related as
\bge\label{eq_xe_z_slopes}
\mza\approx \mean{\xe}\mzt\xhq^{-1}\qquad .
\ende

\begin{figure}[tbh]
%\bgc
\includegraphics[width=1.0\linewidth]{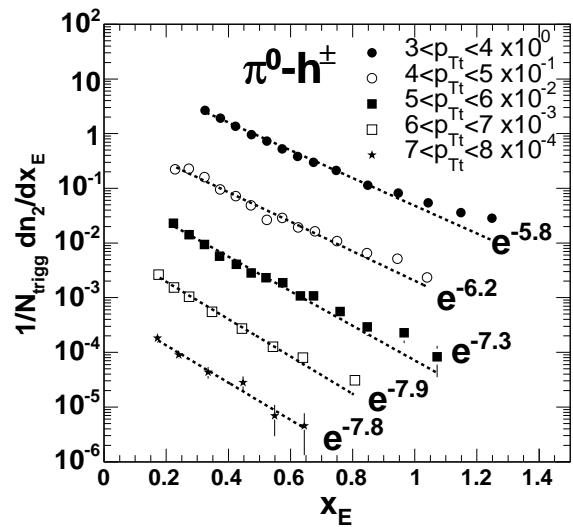}
\caption{\label{fig:xe_expo}
The distribution of associated particles with \xe\ variable
for various trigger particle \ptt\ indicated in the legend.
Exponential fits indicated by dashed lines. 
}
%\endc
\end{figure}

\begin{figure}[tbh]
%\bgc
\includegraphics[width=1.0\linewidth]{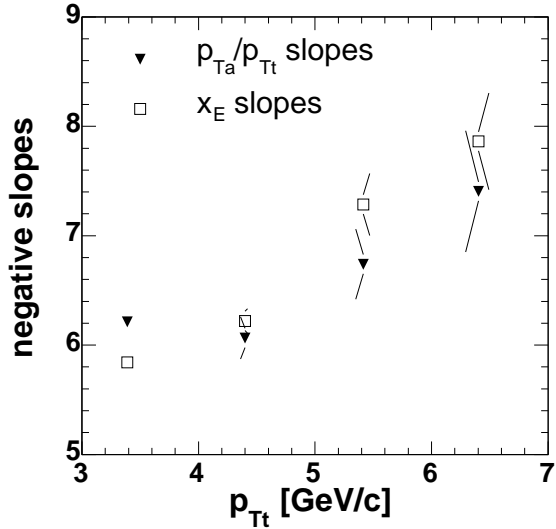}
\caption{\label{fig:slopes}
The negative slope parameters extracted from the fit of a
plain exponential
function into a \xe\ (see \fig{fig:xe_expo}) and \pta/\ptt\
(see \fig{fig:ptaptt_expo}) distributions. 
}
%\endc
\end{figure}

The \xe\ distributions of particles associated with trigger particles
in the 3-8 \gevc\ range of transverse momentum are plotted in
\fig{fig:xe_expo}.  The dashed lines represent exponential
fits. The slopes of these exponentials range from $-5.8$ ($3<\ptt<4$
\gevc) to $-7.8$ (open symbols on \fig{fig:slopes}).
This is qualitatively and quantitatively different from the similar
measurement done by CCOR collaboration at \s=62.4 \gev\ where the
slopes of exponential fits to the \xe\ distributions were found to be
$\approx -5.3$ and independent of the trigger transverse momenta. That
observation also supported the hypothesis of the \xe\ distribution
being a good approximation of the fragmentation function. We also note
that the \xe\ distributions are not quite exponential and at large
values of \xe\ there is a tail similar to the power law tail of the
single inclusive \pt{}\ distribution.

%%%%%%%%%%%%%%%%%%%%%%%%%%%%%%%%%%%%%%%%%%%%%%%%%%%%%%%%%%%% Fig. 18

The reason why the \xe\ distributions do not have the same slope for
different \ptt\ and why there is a ``power law'' tail at large \xe\ is
the same as that which causes \xzkt\ to decrease with the associated
particle transverse momentum. %mjt020806
It turns out that by sampling different regions of \pta\ for fixed \ptt, 
the average momentum of the parton fragmenting into a trigger particle, \mzt,
also changes. This kind of trigger bias causes the hard scattering
kinematics, the value of $\hat{p}_{T}$, to not be fixed for the case
where \ptt\ is fixed but \pta\ varies.

%%%%%%%%%%%%%%%%%%%%%%%%%%%%%%%%%%%%%%%%%%%%%%%%%%%%%%%%%%%% Fig. 19

%%%%%%%%%%%%%%%%%%%%%%%%%%%%%%%%%%%%%%%%%%%%%%%%%%%%%%%%%%%% Fig. 20
\begin{figure}[tbh]
%\bgc
\includegraphics[width=1.0\linewidth]{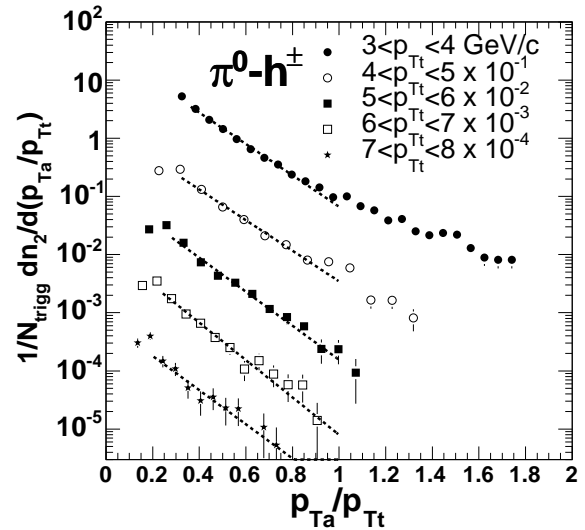}
\caption{\label{fig:ptaptt_expo}
The distribution of associated particles with \pta/\ptt\ variable
for various trigger particle \ptt\ indicated in the legend. The
distribution were fitted in the limited range $0.2<\pta/\ptt<0.4$ by
an exponential function (dashed lines).
}
%\endc
\end{figure}

Taking this into account, one can not treat the associated \xe\
distribution as a rescaled fragmentation function, but rather as a
folding of the two fragmentation processes of trigger and associated
jets. The same line of arguments applies also for other two-particle
variables, \eg \pta/\ptt,\ \cite{Wang_fragModif2} used for an
approximation of the fragmentation variable $z$ (see
\fig{fig:ptaptt_expo}). The negative slopes of an exponential fit in the
$0.2<\pta/\ptt<0.4$ range (solid symbols on \fig{fig:slopes}) are,
within the error bars, the same as for \xe.

In conclusion: the slope parameters extracted from associated \xe\
distributions reveal the rising trend with \ptt\ which reflects the
fact, that the different \pta\ samples not only different \za\ but
also different \zt.

The description of an associated distribution detected under the
condition of the existence of a trigger particle requires an extension
of the formulae discussed in \ref{sec:inclusFrag} and is a subject of
the next section.

%%%%%%%%%%%%%%%%%%%%%%%%%%%%%%%%%%%%%%%%%\input{ppg_frag_cond}
%%%%%%%%%%%%%%%%%%%%%%%%%%%%%%%%%%%%%%%%%%%%%%%%%%%%%%%%%%%%%%%
%%%\subsection{Conditional fragmentation}
\section{Dijet fragmentation}
\label{sec:mz_cond}
%%%%%%%%%%%%%%%%%%%%%%%%%%%%%%%%%%%%%%%%%%%%%%%%%%%%%%%%%%%%%%%

For the description of the detection of a single particle which is the
result of jet fragmentation, recall \eq{jointprob}
\begin{eqnarray}
\label{eq:condjointprob}
{d^2\sigma_\pi\over d\ptq{}d\zt} & = &
  {d\sigma_q \over d\ptq{}}\times \D(\zt) \\
 & = &\ptq{}\fq \times \D(\zt) \equiv \sq \times\D(\zt) \nonumber
\end{eqnarray}
where we have now explicitly labeled the $z$ of the trigger particle
as \zt, and defined
\bge\label{eq:sq_def}
 \sq  \equiv \ptq{}\; \fq  = {d\sigma_q \over d\ptq{}} \qquad .	
\ende
When \kt{}\ smearing is introduced, configurations for which the high
\pt{}\ parton pair is on the average moving towards the trigger
particle are favored due to the steeply falling $\ptq{}$ spectrum,
such that:
\[
\mean{\ptq{t}-\ptq{}}\simeq{1 \over 2}\mean{\ptq{t}-\ptq{a}}\equiv s(\kt{})
\]
with small variance $\sigma^2_s$, and we explicitly introduced
\ptq{t}\ and \ptq{a} to represent the transverse momenta of the
trigger and away partons.  The single inclusive \pt{t}\ spectrum is
now given by
\bge\label{eq:mjt_sig_inclus}
{d^2\sigma_\pi \over d\ptq{t}d\zt} =
\skt \times \D(\zt)
\ende
where the trigger parton \ptq{t}\ spectrum after \kt{} smearing is 
\bge\label{eq:skt_def}
 \skt \equiv \ptq{t}\;\fkt = {d\sigma_q \over d\ptq{t}}\qquad .
\ende
Then, the conditional probability for finding the away side parton
with \ptq{a}\ and \za, given \ptq{t}\ (and \zt), is:
   
\[ \left .  {dP(\ptq{a},\za) \over d\ptq{a}d\za} \right |_{\ptq{t}}
= C(\ptq{a}, \ptq{t}, \kt{}) \D(\za)
\] 
where $C(\ptq{a}, \ptq{t}, \kt{})$ represents the distribution of the
transverse momentum of the away parton \ptq{a}, given
\ptq{t} and \kt{}, which can be written as:

\begin{eqnarray}
\label{eq:mikes_smearing}
\lefteqn{ C(\ptq{a}, \ptq{t}, \kt{}) = } \mbox{\hskip 8cm} \\
\lefteqn{ = {1 \over \sqrt{2\pi\sigma^2_s} }\,
\exp\left(-[\ptq{a} - (\ptq{t} -2s(\kt{}))]^2\over {2\sigma^2_s}\right) }
\mbox{\hskip 6.6cm}\nonumber
\; .
\end{eqnarray}
Then %%%mjtcorrectserror101505 
\[
{d^4\sigma_{\pi} \over d\ptq{t} d\zt d\ptq{a} d\za} =
{d^2\sigma_{\pi} \over d\ptq{t} d\zt } \times
\left .{dP(\ptq{a},\za) \over d\ptq{a}d\za} \right |_{\ptq{t}}\qquad .\] 

In general, $\sigma_s/ s(\kt{})$ is small (see section
\ref{sec:kt_smearing}) so that $C(\ptq{a}, \ptq{t}, \kt{})$
is well approximated by a $\delta$ function and we may take

\[
\ptq{a}=\ptq{t} -2s(\kt{})=\xhq\ptq{t} \qquad,
\]
so that 
\[
{d^3\sigma_{\pi} \over d\ptq{t} d\zt d\za} =
\skt\D(\zt)\D(\za)
\]    
where %%%mjt020806
\[
\za={\pt{a}\over\ptq{a}}={\pt{a}\over\xhq\ptq{t}} = {\zt\pt{a}\over\xhq\pt{t}}\qquad .
\]
Changing variables from $\ptq{t}, \zt$ to $\pt{t}, \zt$ as above, and
similarly from $\za$ to $\pt{a}$, we obtain
\bge\label{eq:mikes_dsig}
{d^3\sigma_\pi \over d\pt{t} d\zt d\pt{a}} =
{1 \over\xhq\pt{t}} \Sigma'_q({\pt{t}\over\zt}) \D(\zt) \D({\zt\pt{a}\over\xhq\pt{t}})
\ende  
where for integrating over $\zt$ or finding $\mean{\zt}$ for fixed
\pt{t}, \pt{a}, the minimum value of $\zt$ is $\zt^{\rm min}=2\pt{t}/\s=x_{Tt}$
and the maximum value is:
\[
\zt^{\rm max}=\xhq {\pt{t}\over\pt{a}}={\xhq\over\xh} \qquad,
\]
where $\xhq(\pt{t},\pt{a})$ is also a function of $k_T$ (\eq{eq:xhqdef}). 
%%%MJT101505 I added this here

     Thus, in order to evaluate $\xhq(\pt{t},\pt{a})$ for use in 
\eq{eq:mikes_dsig}, $k_T$ must be known. We attack this problem by 
successive approximations. First we solve for $k_T$ and \D(z)\ 
assuming $\hat{x}_{h}=1$  as done at the ISR where the smearing correction 
was small. Then we solve for $\xhq(\pt{t},\pt{a})$ with this value of 
$k_T$ and iterate. On the first solution we solve only for \skt\ while 
on the iteration we include the $k_T$ smearing to solve for the unsmeared 
parton spectrum \sq $=\ptq{}\; \fq$ (\eq{eq:sq_def}).

%%%%%%%%%%%%%%%%%%%%%%%%%%%%%%%%%%%%%%%%%\input{ppg_frag_res}
%%%%%%%%%%%%%%%%%%%%%%%%%%%%%%%%%%%%%%%%%%%%%%%%%%%%%%%%%%%%%%%
\subsection{Sensitivity of the associated spectra to the fragmentation function}
\label{sec:no_sensitivity}
%%%%%%%%%%%%%%%%%%%%%%%%%%%%%%%%%%%%%%%%%%%%%%%%%%%%%%%%%%%%%%%
%%%mjt020806

As discussed in section~\ref{sec:frag_xe}, the associated $x_E$ distribution 
was thought to approximate the fragmentation function of the away jet. 
Equation~\ref{eq:mikes_dsig} can be transformed to the $x_E$ distribution at 
fixed \ptt\ with a change of variables from \pta\ to $x_E$ followed by 
integration over \zt :
\begin{eqnarray}\label{eq:dsigma_trig_kt}
%\begin{array}{l}
{d^2\sigma\over d\ptt d\xe} =
{d\pta\over d\xe}\times{d^2\sigma\over d\ptt d\pta}\simeq 
\mbox{\hspace*{1.0in}} \\
{1\over\xhq}
\int_{x_{T\rm t}}^{\xhq{\ptt\over\pta}}
\D(\zt)\D({\zt\pta\over\xhq\ptt})\Sigma'_q({\ptt\over\zt }) %%mjt042206
\,d\zt \nonumber .
%\end{array}
\end{eqnarray}

%mjt020806
%The parton distribution function on the opposite side of the
%trigger particle $\Sigma'$ is smeared by \vkt{}\ and can be
%approximated by a folding of a power law parton distribution with a
%Gaussian function.

    We at first attempted to solve for the fragmentation function by
simultaneous fits of the measured $x_E$ distributions to
Eq.~(\ref{eq:dsigma_trig_kt}) constrained by a fit of the inclusive invariant
$\pi^0$ cross section to Eq.~(\ref{eq:dsigma_integral}). There were
difficulties with convergence.
%%%%%%%%%%%%%%%%%%%%%%%%%%%%%%%%%%%%%%%%%%%%%%%%%%%%%%%%%%%% Fig. 21
\begin{figure}[tbh]
\includegraphics[width=1.0\linewidth]{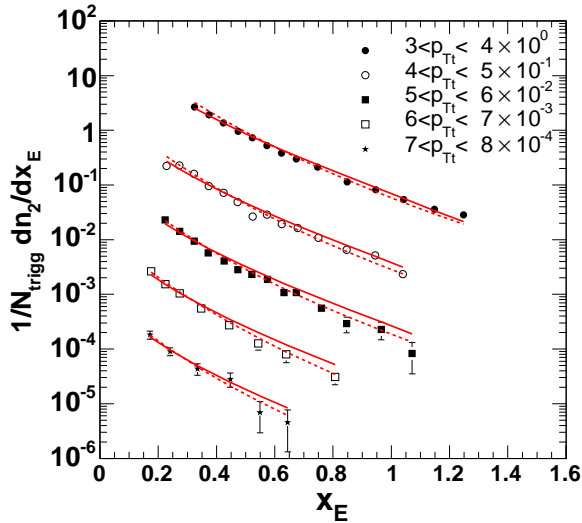}
\caption{\label{fig:xe_LEP} (color online)
The same \xe\ distributions as on \fig{fig:xe_expo} shown with
calculations according to \eq{eq:dsigma_trig_kt} for quark (solid lines)
and for gluon (dashed lines) $D(z)$. An exponential approximation was used and the slopes
for quark and gluon fragmentation function were obtained by fitting to LEP
data \cite{Delphi_Dz_EPJ00} and \cite{Opal_Dz_ZPhys} (see Fig.~\protect\ref{fig:D_LEP}).
}
\end{figure}

%%%%%%%%%%%%%%%%%%%%%%%%%%%%%%%%%%%%%%%%%%%%%%%%%%%%%%%%%%%% Fig. 22
\begin{figure}[tbh]
\includegraphics[width=1.0\linewidth]{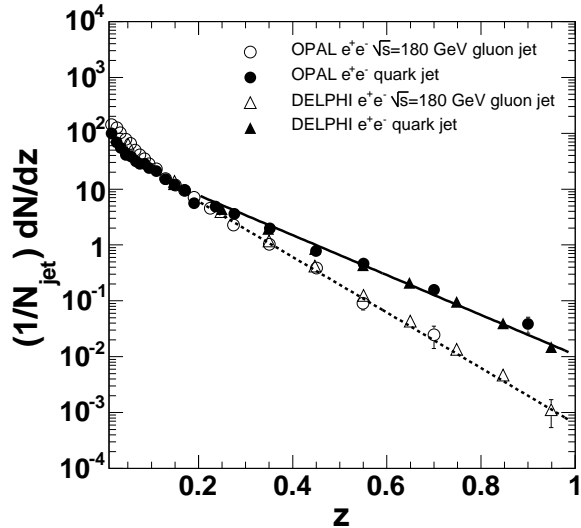}
\caption{\label{fig:D_LEP}
Fragmentation function measured in \ee\ collisions at \s=180 \gev\
for gluon and quark jets. The solid and dashed lines represent the
exponential fit in the $0.2<z<1$ region.
}
\end{figure}

The reason for the lack of convergence became apparent when we calculated $x_E$ distributions according to \eq{eq:dsigma_trig_kt} (Fig.~\ref{fig:xe_LEP}) for two different fragmentation functions corresponding to quark and gluon
jet fragmentation. A simple exponential
parameterization was used and the slopes were obtained from the fit to
the LEP data \cite{Delphi_Dz_EPJ00}, \cite{Opal_Dz_ZPhys} (Fig.~\ref{fig:D_LEP}). For quark and gluon jets, we found
$D_q(z)\approx\exp(-8.2\cdot z)$ and $D_g(z)\approx\exp(-11.4\cdot z)$
respectively. For the parton final state spectrum, we used
$\Sigma'_q\propto\ptq{}^{-8}$. It is evident that the
\xe\ distributions calculated for the quite different quark and gluon
fragmentation functions do not differ significantly (the difference
between solid and dashed lines on \protect\fig{fig:xe_LEP}). Clearly, the $x_E$ distributions are rather 
insensitive to 
the fragmentation functions of the away  jet in contradiction to the previous conventional wisdom. The evidence of this explicit counter example led to attempts to perform the integrals of Eq.~(\ref{eq:mikes_dsig}) and  Eq.~(\ref{eq:mjt_sig_inclus}) analytically which straightforwardly confirmed the observation that the $x_E$ distribution is not sensitive to the fragmentation function. 

     If the smeared trigger parton spectrum is taken as a power law, 
     \[ \Sigma'_q ({\ptt\over\zt })= A({\ptt\over\zt })^{-(n-1)} \] 
     and the fragmentation function as an exponential, $D(z)=B\,\exp (-bz)$, then the integral of \eq{eq:mikes_dsig} over \zt\  becomes:
%%%mjt042105     
\begin{eqnarray}\label{eq:int_mikes_dsig}
\lefteqn{  {d\sigma_\pi \over d\pt{t} d\pt{a}} = } \mbox{\hskip 8cm} \\
\lefteqn{ {B^2\over\xhq} {A\over p_{T_t}^{n}}
\int_{x_{T_t}}^{\xhq {\pt{t}\over\pt{a}}} d\zt \zt^{n-1}
\exp [-b\zt(1+ {\pt{a} \over{\xhq\pt{t}}})] } \mbox{\hskip 7.8cm}\nonumber
\end{eqnarray}
which is an incomplete gamma function. Since $\hat{x}_h\sim 1$, we make the assumption that it is constant. %%%WAZ-mjt050306-next line
Similarly, the integrals of Eqs.~\ref{power_law},~\ref{eq:mjt_sig_inclus} are  also incomplete gamma functions: %%%%%mjt 050306of the same form:
 \begin{equation}\label{int_mjt_sig_inclus}
{d\sigma_\pi \over d\pt{t}} =  {AB\over p_{T_t}^{n-1}}\int_{x_{T_t}}^{1} d\zt \zt^{n-2} \exp [-b\zt] \qquad .
\end{equation}

%%%WAZ--mjt050306
     A reasonable approximation for the inclusive single, and two particle cross sections is obtained by taking the lower limit to zero and the upper limit to infinity, leading to the replacement of the incomplete gamma functions by gamma functions, with the result that:
         \begin{equation}\label{eq:result_int_mikes_dsig}
{d^2\sigma_\pi \over d\pt{t} d\pt{a}} \approx 
{\Gamma(n) \over b^{n}} {B^2\over\xhq} {A\over p_{T_t}^{n}}
{1\over {(1+ {\pt{a} \over{\xhq\pt{t}}})^{n}} }  
 \end{equation}
\begin{equation}
{d\sigma_\pi \over d\pt{t}} \approx {\Gamma(n-1) \over b^{n-1}}  {AB\over p_{T_t}^{n-1}} \qquad,
 \label{eq:result_int_mjt_sig_inclus}
\end{equation}
where $\Gamma(n)=(n-1)\Gamma(n-1)$. 

The conditional probability is just the ratio of the joint probability \eq{eq:result_int_mikes_dsig} to the inclusive probability \eq{eq:result_int_mjt_sig_inclus}, or
   \begin{equation}
\left.{dP_\pi \over d\pt{a}}\right|_{\pt{t}}\approx {{B(n-1)}\over {b\pt{t}}} {1\over\xhq} {1\over {(1+ {\pt{a} \over{\xhq\pt{t}}})^{n}}} 
 \qquad . 
\label{eq:condpta2}
\end{equation}
In the collinear limit, where \pt{a}=\xe\pt{t}\ :
   \begin{equation}
\left.{dP_\pi \over d\xe}\right|_{\pt{t}}\approx {{B(n-1)}\over {b}}{1\over\xhq} {1\over
{(1+ {x_E \over{\xhq}})^{n}}} \, 
\qquad . 
\label{eq:condxe2}
\end{equation}

    The only dependence on the fragmentation function, in this approximation, is in the normalization constant $B/b$ which equals $\mean{m}$, the multiplicity in the away-jet from the integral of the fragmentation function.   
The dominant term in Eq.~(\ref{eq:condxe2}) is the 
Hagedorn function $1/(1+\xe/\xhq )^{n}$, so that at fixed \pt{t}\ the \xe\ 
distribution is predominantly a function only of \xe\  and thus does 
exhibit `\xe ' scaling. Also, the Hagedorn function explains the ``power law'' 
tail at large \xe\ noted in section \ref{sec:frag_xe}.   
The reason that the \xe\ distribution is not very sensitive to the 
fragmentation function is that the integral over \zt\ for fixed \pt{t} 
and \pt{a} (\eq{eq:int_mikes_dsig}) is actually an integral over the 
jet transverse momentum $\hat{p}_{T_t}$. However since both the trigger 
and away jets are always roughly equal and opposite in transverse momentum, 
integrating over $\hat{p}_{T_t}$ simultaneously integrates over 
$\hat{p}_{T_a}$, and thus also integrates over the away jet fragmentation 
function. This can be seen directly by the presence of \zt\ in both the 
same and away fragmentation functions in Eqs.~\ref{eq:mikes_dsig} and 
\ref{eq:dsigma_trig_kt}, so that the integral over \zt\ integrates over 
both fragmentation functions simultaneously.

%%%%%%%%%%%%%%%%%%%%%%%%%%%%%%%%%%%%%%%%%\input{ppg_kt_smearing}

%%%%%%%%%%%%%%%%%%%%%%%%%%%%%%%%%%%%%%%%%%%%%%%%%%%%%%%%%%%%%%%
\subsection{\kt{} smearing}
\label{sec:kt_smearing}
%%%%%%%%%%%%%%%%%%%%%%%%%%%%%%%%%%%%%%%%%%%%%%%%%%%%%%%%%%%%%%%
In order to evaluate \xhq(\ptt,\pta)\ and \kt{}\ must be
known. We attack this problem by successive approximations: first we
solve for $k_T$ assuming $\hat{x}_h=1$ as done at the ISR, where the
smearing correction was small.
%[cite owens if you want].
Then we iterate for finite $k_T$. The Gaussian approximation for the smearing
function \eq{eq:mikes_smearing} does not work so well in the low \ptq{}\
region. The product of the steeply falling parton distribution
function and the fragmentation function is peaked at $z\approx$ 1
preferring ``small'' parton momenta. We have developed more accurate
description of the conditional yields taking into account the \kt{}
smearing.
%%%mjt043006-moved figure here

%%%%%%%%%%%%%%%%%%%%%%%%%%%%%%%%%%%%%%%%%%%%%%%%%%%%%%%%%%%% Fig. 23
\begin{figure}[tbh]
%\bgc 
\includegraphics[width=1.0\linewidth]{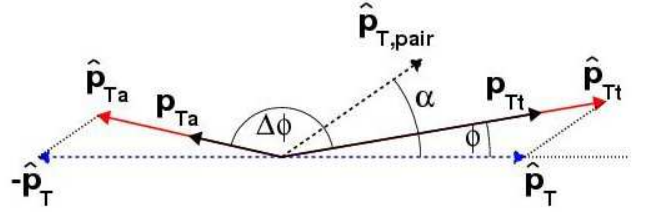}
%\endc
\caption{\label{fig:jetsCondSchem} (color online)
Back-to-back partons in hard scattering rest frame (blue 
dashed arrows) with four-momenta
(\ptq{},0,0,\ptq{}) and (-\ptq{},0,0,\ptq{}) in (-,-,-,+) metrics 
moving along \pn\ ($\pn = \hat{p}_{T\rm pair}$) for an event where
detection of \pt{t}\ and \pt{a}\ is required (the \jt{}\ contribution 
is neglected). The \pt{t}$>$ \pt{a}\ condition implies that the events with
\pn\ pointing more in the direction of \pt{t}\ are selected.
}
\end{figure}

Let us consider the configuration depicted on \fig{fig:jetsCondSchem}. The two
back-to-back partons in $\hat{s}$ frame undergo the Lorentz boost determined by
net pair momentum
\bge\label{eq:pn_def}
\vec{p}_n \equiv \vec{p}_{T\rm pair} \equiv \vptq{t}+\vptq{a} = \vkt{t}+\vkt{a}
\ende

\noindent
If we denote an angle between the unsmeared parton momentum and
\kt{}-vector (or $\vec{p}_n$) as $\alpha$  (see \fig{fig:jetsCondSchem})
then we can write the conditional
probability distribution of trigger parton momenta, \ptq{t}, as
\begin{widetext}
\bge
{d^3\sigma\over d\ptq{t} d\alpha d\ptq{}}\condta
=\ptq{t}\cdot\sq\cdot \pn\cdot G(\pn(\vec{r}_t))
\cdot \D({\pt{t}\over\ptq{t}}){\pt{t}\over\ptqkv{t}}
\cdot \D({\pt{a}\over\ptq{a}(\vec{r}_t)}){\pt{a}\over\ptqkv{a}(\vec{r}_t)}
\ende
\end{widetext}
where $G(\pn)=\exp(-\pnkv/2\rms{\kt{}})$ describes the Gaussian
probability distribution of the net pair momentum magnitude
distribution, \sq\ is the unsmeared parton momentum distribution, \D\
is the fragmentation function and
$\vec{r}_t=(\ptq{t},\phi,\ptq{},\kt{})$ is the phase space
vector. The \ptq{t}\ is chosen to be an integration variable and
\ptq{a}\ is fully determined by given values of \ptq{t}, \ptq{}, angle
$\phi$ and by the requirement of Lorentz invariance.

In order to evaluate
$\mean{\zt(\kt{})}\big|_{\pt{t},\pt{a}}$ and
$\xhq(\kt{})\big|_{\pt{t},\pt{a}}$ we have to evaluate first the
parton distribution for events where given \pt{t}\ and
\pt{a}\ are detected. This conditional cross section can be expressed
as a 
%%WAX-mjt050306 contour 
definite integral over the unobserved variables $\phi$ and \ptq{}\ (see \fig{fig:jetsCondSchem})
\begin{eqnarray}
\label{eq:dsigdqtt}
{d\sigma\over d\ptq{t}}\condta & = &
2\int_0^{\sqrt{s}/2}\int_0^\pi {d^3\sigma\over d\ptq{t}d\ptq{}d\phi}\condta\; d\phi d\ptq{} \nonumber\\
& = & \D({\pt{t}\over\ptq{t}}){2\over\ptq{t}}\int_0^{\sqrt{s}/2}\sq\times  \nonumber
\end{eqnarray}
\bge
\times\int_0^\pi
\pn(\vec{r}_t)\,G(\pn(\vec{r}_t))\cdot\D({\pt{a}\over\ptq{a}(\vec{r}_t)}){1\over\ptqkv{a}(\vec{r}_t)}
\;d\phi\;d\ptq{}   \quad . %%%mjt043006added .
\ende
%%%moved from after {eq:meanxhqnoms} to here
The $d\sigma/d\ptq{a}\big|_{\pt{t},\pt{a}}$ distribution can be derived from \eq{eq:dsigdqtt} just by
rotation $\ptq{t}\rightarrow\ptq{a}$ and $\ptq{a}\rightarrow\ptq{t}$.
The $\mean{\zt(\kt{})}\big|_{\pt{t},\pt{a}}$ and $\xhq(\kt{})\big|_{\pt{t},\pt{a}}$ quantities can then be evaluated as
\bge\label{eq:mzt}
\mean{\zt(\kt{})}\big|_{\pt{t},\pt{a}} = {\mathcal{Z}(1)\over\mathcal{Z}(0)}
\ende
where
\begin{displaymath}
\begin{array}{l}
\mathcal{Z}(n)=
\int_{x_{T\rm t}}^1 \zt^{n-1}\;\D(\zt)
\int_0^{\sqrt{s}/2}\sq\times\\
\\
\times\int_0^\pi
\pn\,G(\pn(\vec{r}_{zt}))\cdot\D({\pt{a}\over\ptq{a}(\vec{r}_{zt})})
{1\over\ptqkv{a}(\vec{r}_{zt})}
\;d\phi\;d\ptq{}\;d\zt
\end{array}
\end{displaymath}
and $\vec{r}_{zt}=(\pt{t}/\zt,\phi,\ptq{},\kt{})$.
The $\xhq(\kt{})\big|_{\pt{t},\pt{a}}$ is evaluated as
\bge\label{eq:meanxhq}
\xhq(\kt{})\big|_{\pt{t},\pt{a}}={\mean{\ptq{a}}\over\mean{\ptq{t}}}\condta=
{\mathcal{X}_a(1)\over\mathcal{X}_a(0)}{\mathcal{X}_t(0)\over\mathcal{X}_t(1)}
\ende
where
\begin{displaymath}
\begin{array}{l}
%\label{eq:meanxhqnoms}
\mathcal{X}_t(n) = 
	\int_{\pt{t}}^{\sqrt{s}/2}\ptq{t}^{n-1} \D({\pt{t}\over\ptq{t}})
	\int_0^{\sqrt{s}/2}\sq\times\\
\\
\times\int_0^\pi\pn(\vec{r}_t)\,G(\pn(\vec{r}_t))\cdot\D({\pt{a}\over\ptq{a}(\vec{r}_t)}){1\over\ptqkv{a}(\vec{r}_t)}
\;d\phi\;d\ptq{}\ptq{t}\\
\\
\mathcal{X}_a(n)  = 
	\int_{\pt{a}}^{\sqrt{s}/2} \ptq{a}^{n-1} \D({\pt{a}\over\ptq{a}})
	\int_0^{\sqrt{s}/2}\sq\times\\
\\
\times\int_0^\pi\pn(\vec{r}_a)\,G(\pn(\vec{r}_a))\cdot\D({\pt{t}\over\ptq{t}(\vec{r}_a)}){1\over\ptqkv{t}(\vec{r}_a)}\nonumber
\;d\phi\;d\ptq{}\ptq{a}
\end{array}
\end{displaymath}

%%%%%%%%%%%%%%%%%%%%%%%%%%%%%%%%%%%%%%%%%%%%%%%%%%%%%%%%%%%% Fig. 24
\begin{figure}[tbh]
%\bgc 
\includegraphics[width=1.0\linewidth]{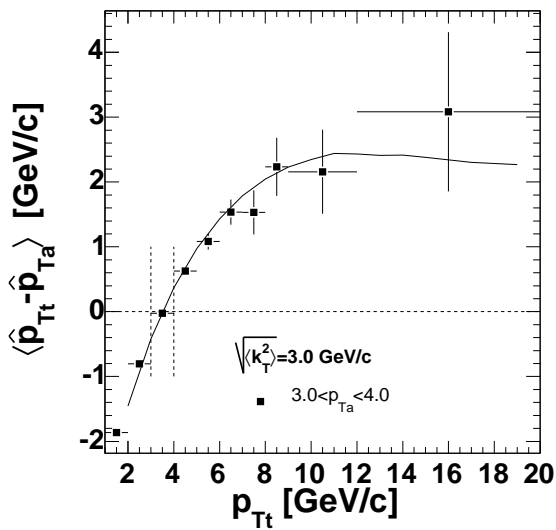}
%\endc
\caption{\label{fig:pow_delta}
PYTHIA simulated average momentum unbalance for the associated particles in
$3.0<\pt{a}<4.0$ \gevc\ bin and calculated according \eq{eq:meanxhq}. The two vertical
dashed line indicates the range where \pt{t}\ and \pt{a}\ bins are equal and the parton momenta
unbalance vanishes (fixed correlations).
}
\end{figure}

We have tested the above formulae on PYTHIA simulation. We have
generated events with \sqrtrms{\kt{}}=3 \gevc\ and evaluated the
partons' momenta unbalance variation with \ptt\ for fixed $3<\pta<4$
\gevc\ bin. The results from the PYTHIA simulation (solid point on
\fig{fig:pow_delta}) are compared to calculation based on
\eq{eq:meanxhq} (solid line on \fig{fig:pow_delta}). The magnitude of
momentum unbalance saturates at \pt{t}$\approx$10\gevc\ around
\sqrtrms{\kt{x}}\ and then starts to decrease. The maximum value
depends on the the \kt{}\ magnitude and on the asymmetry between
\pt{t}\ and \pt{a}. Eventually, the unbalance should vanish at high
\pt{t}\ as a consequence of \sq\ flattening.

%%%%%%%%%%%%%%%%%%%%%%%%%%%%%%%%%%%%%%%%%%%%%%%%%%%%%%%%%%%% Fig. 25
\begin{figure}[tbh]
%\bgc 
\includegraphics[width=1.0\linewidth]{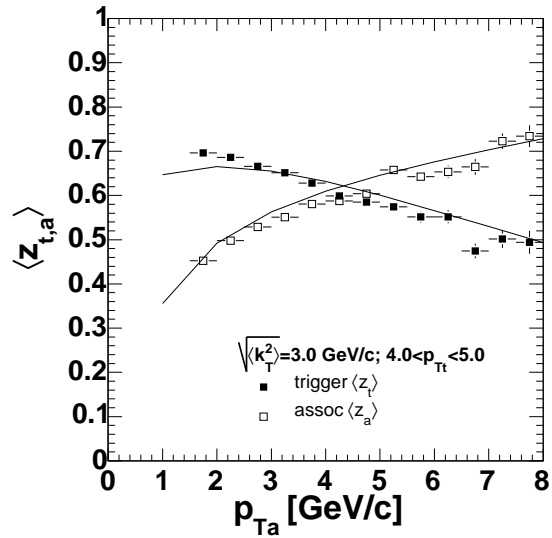}
%\endc
\caption{\label{fig:mztmza}
Average $z$ of a trigger and associated particle as a function
of \pt{a}\ from PYTHIA and according \eq{eq:mzt}.  }
\end{figure}

The comparison of \mzt\ and \mza\ found in PYTHIA and derived
according to \eq{eq:mzt} is shown in \fig{fig:mztmza}. The overall
agreement between the PYTHIA simulations and the calculation is
excellent. The small deviations may be attributed to the fact that in
the PYTHIA simulation, 1\gevc-wide bins were used for trigger and
associated particle identification, whereas the calculation was
performed for fixed values of \pt{t}\ and
\pt{a}.

The last missing piece of information needed before solving
\eq{eq:zkt}\ is the fragmentation function \D\ and unsmeared \sq. The
description of how this knowledge was extracted from the data is a
subject of next section.

\section{Corrected \mean{\kt{}}\ results}
\label{sec:finRes}

The \xzkt\ extracted according to \eq{eq:zkt} for various \ptt\ and \pta\
are shown in \fig{fig:zkt_pta} and \fig{fig:zkt_ptt}. In order to
extract a \sqrtrms{\kt{}}\ values we have solved
\bge\label{eq:kt_root}
\xh^{-1}\sqrt{\rms{\pout}-\rms{\jt{y}}(\xh^2+1)}-\xzkt=0
\ende
for \sqrtrms{\kt{}}\ where the \mzt\ and
\xhq=\mean{\ptq{a}}/\mean{\ptq{t}}\ are evaluated according
\eq{eq:mzt} and \eq{eq:meanxhq} respectively. These two quantities
depend on \sqrtrms{\kt{}}\ so we solved \eq{eq:kt_root} iteratively by
varying a \sqrtrms{\kt{}}\ value and in every step the \mzt\ and \xhq\
were recalculated. To do so the we need to know unsmeared final state
parton spectrum \sq\ and the fragmentation function. For the latter one
we used the LEP data (see \fig{fig:D_LEP}) where the fragmentation
%%WAZ-mjt050306
functions of gluon and quark jets were measured in \ee\ collision at
\s=180 \gev. We have chosen
\bge\label{eq:DzLEP}
 \D  \propto  z^{-\alpha}(1-z)^\beta(1+z)^{-\gamma}
\ende
form used \eg\ in \cite{Delphi_Dz_EPJ00} and extracted $\alpha$,
$\beta$ and $\gamma$ parameters from the fit to distributions shown in
\fig{fig:D_LEP} (see \tab{tab:D_LEP_pars}).

%%%%%%%%%%%%%%%%%%%%%%%%%%%%%%%%%%%%%%%%%%%%%%%%%%%%%%%%%%%% Table VIII
\begin{table}[tbh]
\caption{\label{tab:D_LEP_pars}
Extracted values of $D(z)$ parameters according \eq{eq:DzLEP} from the fit
to the LEP data and power $n$ of the unsmeared final state parton spectra
\sq\ extracted from the fit to the single inclusive \piz\ invariant cross
section \cite{PHENIX_pi0ppPRL} for corresponding fragmentation function
and fixed value of \sqrtrms{\kt{}}=2.5 \gevc.
}
\begin{ruledtabular}
\begin{tabular}{lrrc}
	  & gluon  & quark & (gluon+quark)/2 \\\hline
$\alpha$  & 0.16   & 0.49  & \ 0.32  \\
$\beta$   & 0.88   & 0.57  & \ 0.72  \\
$\gamma$  & 13.29  & 8.00  & 10.65 \\
$n$	  & 7.53   & 7.28  & \ 7.40  \\
\end{tabular}
\end{ruledtabular}
\end{table}

%%%%%%%%%%%%%%%%%%%%%%%%%%%%%%%%%%%%%%%%%%%%%%%%%%%%%%%%%%%% Fig. 26
\begin{figure}[tbh]
%\bgc 
%\subfigure{
\includegraphics[width=1.0\linewidth]{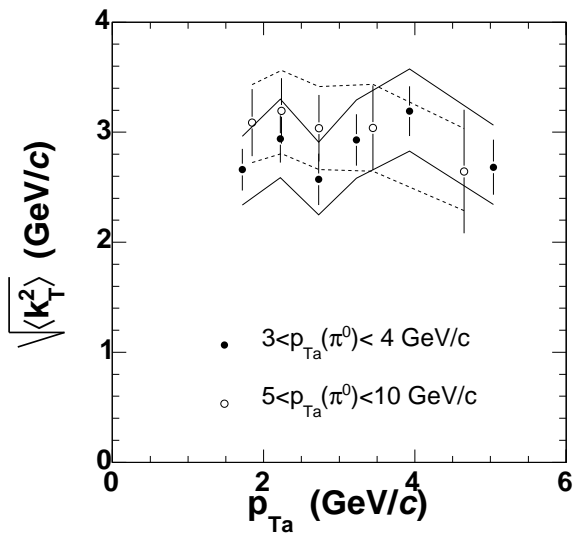}
\caption{\label{fig:rmskt_pta}
\sqrtrms{\kt{}}\ values corresponding to \fig{fig:zkt_pta}  as a
solution to \eq{eq:kt_root} for trigger \piz\ in $3<\ptt<4$\gevc\
(solid symbols) and $5<\ptt<10$\gevc\ (open symbols) range. The solid
and dashed lines bracket the systematic uncertainty due to the unknown ratio
of quark and gluon jets, for the solid and open symbols, respectively.  }
%\endc
\end{figure}

For a given set of parameters $\alpha$, $\beta$ and $\gamma$ the power
of the unsmeared final state parton spectra \sq\ was evaluated from the
fit formula \eq{eq:dsigma_integral} to the single inclusive \piz\
invariant cross section \cite{PHENIX_pi0ppPRL}. Here we used the simplified
\kt{}\ smearing
%%mjt021306--drop the 2 from the gaussian
$$
\fkt={1\over\ptq{t}}\skt={1\over\ptq{}}\sq\otimes\exp{-(\ptq{}-\ptq{t})^2\over 
%%2
\rms{\kt{x}}}
$$ and for the fixed value of \sqrtrms{\kt{}} =
$\sqrt{2}$\sqrtrms{\kt{x}} = 2.5 \gevc\ the power $n$ of
\sq\ distribution was determined.

%%%%%%%%%%%%%%%%%%%%%%%%%%%%%%%%%%%%%%%%%%%%%%%%%%%%%%%%%%%% Fig. 27
\begin{figure}[tbh]
\includegraphics[width=1.0\linewidth]{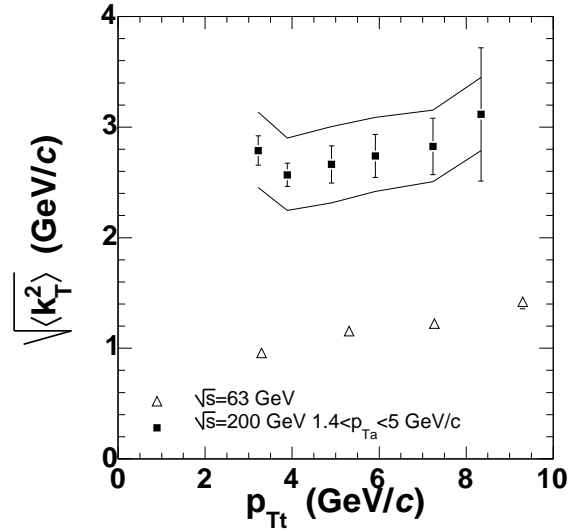}
\caption{\label{fig:rmskt_ptt}
\sqrtrms{\kt{}}\ values corresponding to \fig{fig:zkt_ptt}  as a
solution to \eq{eq:kt_root} for associated particles in
$1.4<\pta<5$\gevc\ region (solid symbols). The solid lines bracket the
systematic error due to the unknown ratio of quark and gluon jets.
The CCOR measurement at \s=62.4 \gev\ \cite{CCORjt} (empty
triangles).}
\end{figure}

%%%WAZ-mjt050306
The measurement of the fragmentation functions at LEP was done
separately for quark and gluon jets and the slopes of these two $D(z)$
distributions are different. Quark jets produce a significantly
harder spectrum than gluon jets (see \fig{fig:D_LEP}).  Since the
relative abundance of quark and gluon jets at \s=200 \gev\ is not
known, for the final results we assumed that the numbers of quark
and gluon jets are equal; the final $D(z)$ uses the averaged 
parameter values between quark and gluon and the difference was
used as a measure of the systematic uncertainty.

%%%%%%%%%%%%%%%%%%%%%%%%%%%%%%%%%%%%%%%%%%%%%%%%%%%%%%%%%%%% Fig. 28
\begin{figure}[tbh]
%\bgc
\includegraphics[width=1.0\linewidth]{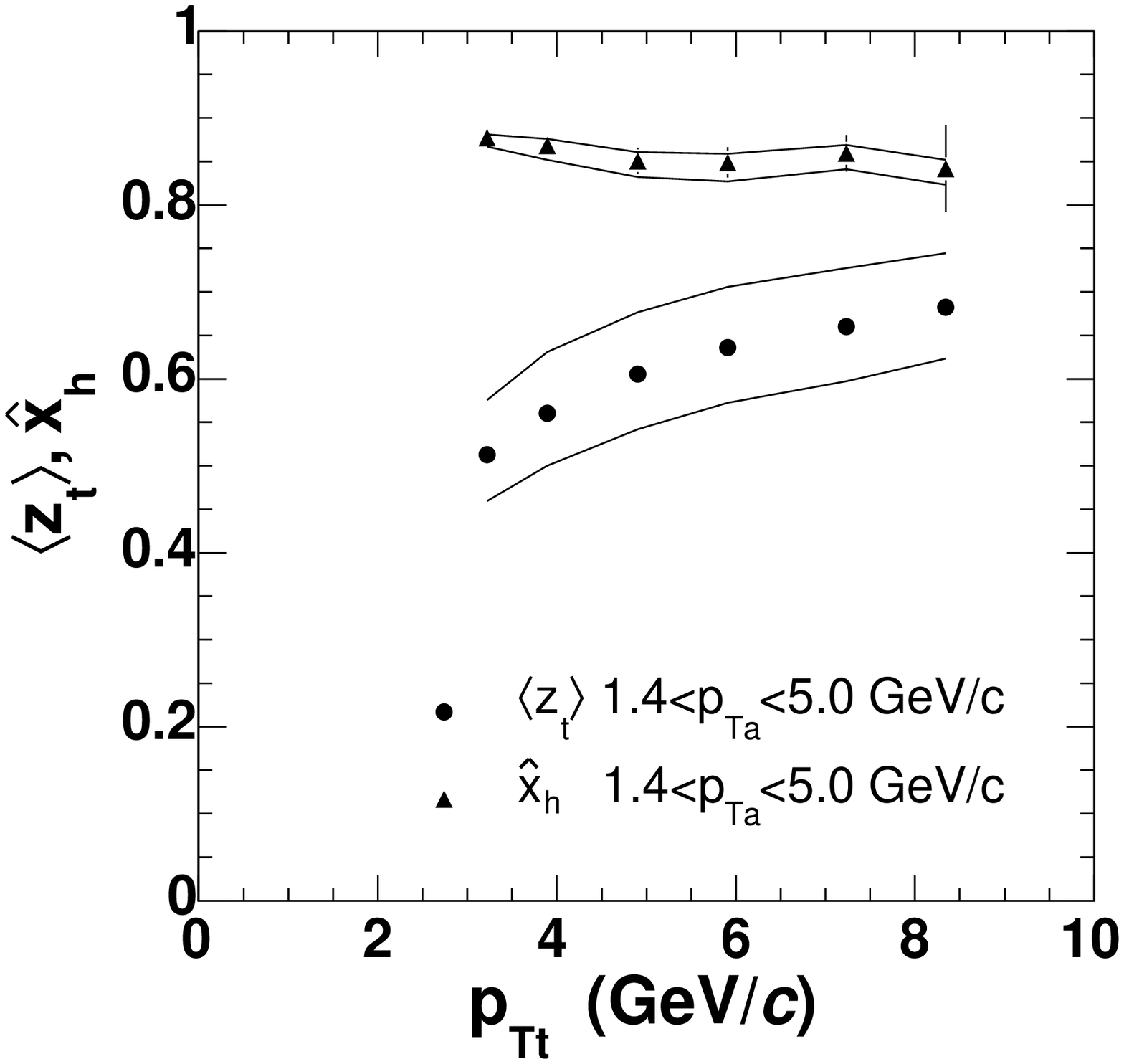}
\caption{\label{fig:zt_xhq_ptt}
\mzt\ and \xhq\ as a function of \ptt\ for the $1.4<\pta<5.0$ \gevc\ associated region. %%%mjt021306
%(solid circles) and \mzt\ with \pta\ for various trigger \ptt\
%as indicated in the legend (open symbols).  
%The \xhq\ values (see definition \ref{eq:xhqdef}) as solution of
%\eq{eq:meanxhq} for $3<\ptt<4$\gevc\ and $5<\ptt<10$\gevc\ as a
%function of \pta\ (left panel) and for $1.4<\pta<5$\gevc\ as a
%function of \ptt.
}
%\endc
\end{figure}

Resulting \sqrtrms{\kt{}} values for $3<\ptt<4$\gevc\ and
$5<\ptt<10$\gevc\ as a function of \pta\ are shown in
\fig{fig:rmskt_pta} (compare to uncorrected values
\fig{fig:zkt_ptt}). The solid and dashed lines bracket the systematic
error due to the unknown ratio of quark and gluon jets. These data
points correspond to the uncorrected \xzkt\ values shown in
\fig{fig:zkt_pta}.  The \sqrtrms{\kt{}} values for varying \ptt\
corresponding to the data shown of \fig{fig:zkt_ptt} are shown in
\fig{fig:rmskt_ptt}.  Also here the solid lines bracket the systematic
error due to the unknown ratio of quark and gluon jets. It is evident
that unfolded \sqrtrms{\kt{}}\ values reveal, within the error bars,
no dependence neither on \pta\ nor on \ptt. The tabulated data are given 
in Table~\ref{tab:kt_ptt}.

%%%%%%%%%%%%%%%%%%%%%%%%%%%%%%%%%%%%%%%%%%%%%%%%%%%%%%%%%%%% Table IX
\begin{table}[tbh]
\caption{\label{tab:kt_ptt}
Values of \xzkt\ and \sqrtrms{\kt{}}\ for various trigger
particle \ptt\ and associated momenta in the $1.4<\pta<5.0$~\gevc\
region shown in \fig{fig:zkt_ptt} and \fig{fig:rmskt_ptt} }
\begin{ruledtabular}
\begin{tabular}{ccc}
\ptt &   \xzkt        &  \sqrtrms{\kt{}}   \\
\gevc&   \gevc        &  \gevc             \\\hline
3.22 &  1.63 $\pm$  0.08 &  2.79 $\pm$  0.13 $\pm$ 0.35\\
3.89 &  1.66 $\pm$  0.08 &  2.57 $\pm$  0.11 $\pm$ 0.33\\
4.90 &  1.89 $\pm$  0.13 &  2.66 $\pm$  0.17 $\pm$ 0.35\\
5.91 &  2.06 $\pm$  0.19 &  2.74 $\pm$  0.20 $\pm$ 0.34\\
7.24 &  2.17 $\pm$  0.25 &  2.83 $\pm$  0.25 $\pm$ 0.32\\
8.34 &  2.53 $\pm$  0.62 &  3.11 $\pm$  0.60 $\pm$ 0.33\\
\end{tabular}
\end{ruledtabular}
\end{table}

We compared the \sqrtrms{\kt{}}\ data obtained in this analysis to
\sqrtrms{\kt{}}\ values found by the CCOR collaboration at \s=62.4
\gev\ \cite{CCORjt} (empty triangles on \fig{fig:rmskt_ptt}). Although
the trend with \ptt\ seems to be similar the overall magnitude at
\s=200 \gev\ is significantly higher. 

%%%%%%%%%%%%%%%%%%%%%%%%%%%%%%%%%%%%%%%%%%%%%%%%%%%%%%%%%%%% Fig. 29
\begin{figure}[tbh]%\bgc 
\includegraphics[width=1.0\linewidth]{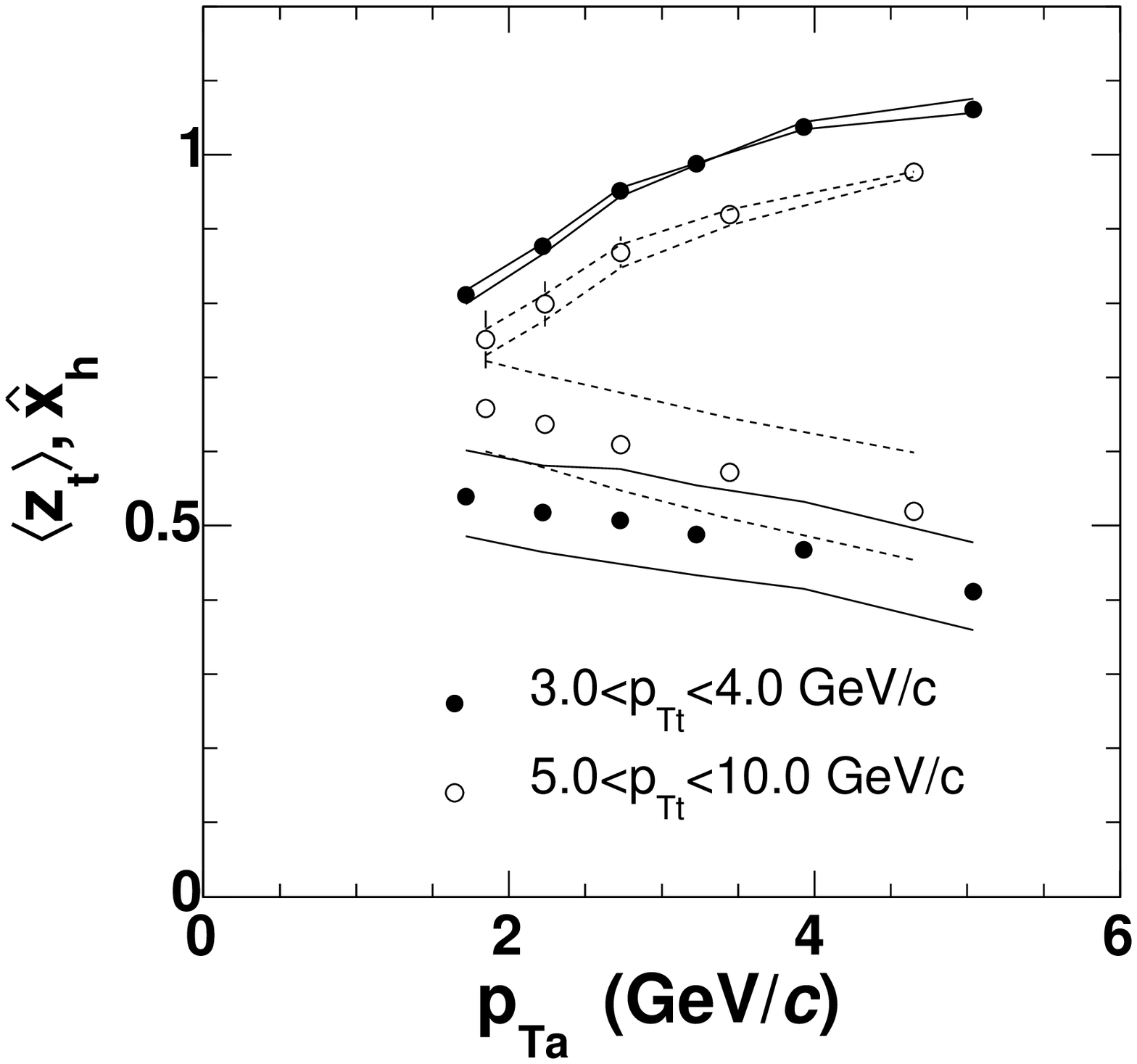}
\caption{\label{fig:zt_xhq_pta}
 The \mzt\ and \xhq\ values (see Eq.~\ref{eq:xhqdef}) as solution of
\eq{eq:meanxhq} for $3<\ptt<4$\gevc\ and $5<\ptt<10$\gevc\ as a
function of \pta\ .
%%mjt021306(left panel) and for $1.4<\pta<5$\gevc\ as a
%%function of \ptt.\mzt\ as a function of \ptt\ for the $1.4<\pta<2.4$
%%\gevc\ associated region (solid circles) and \mzt\ with \pta\ for
%%various trigger \ptt\ as indicated in the legend (open symbols).
}
%\endc
\end{figure}

%%%%%%%%%%%%%%%%%%%%%%%%%%%%%%%%%%%%%%%%%%%%%%%%%%%%%%%%%%%% Table X
\begin{table*}[tbh]
\caption{\label{tab:kt_pta}
The \xzkt\ \sqrtrms{\kt{}}\ values as a function of \pta\ for two different
trigger \piz\ transverse momentum bins shown in \fig{fig:zkt_pta} and 
\fig{fig:rmskt_pta}.  All units in rad and \gevc.
}
%\bgc
\begin{ruledtabular}
\begin{tabular}{cccccc}
 	  \multicolumn{3}{c}{$3<\ptt<4$}
	& \multicolumn{3}{c}{$5<\ptt<10$}\\
\pta&  $\mzt\sqrtrms{\kt{}}\over\xhq$  & \sqrtrms{\kt{}}   &\pta &  $\mzt\sqrtrms{\kt{}}\over\xhq$            & \sqrtrms{\kt{}} \\\hline
1.7 &  1.76 $\pm$  0.12 &  2.66 $\pm$  0.19 & 1.9 &  2.69 $\pm$  0.37 &  3.09 $\pm$  0.30 \\
2.2 &  1.74 $\pm$  0.13 &  2.94 $\pm$  0.22 & 2.2 &  2.54 $\pm$  0.31 &  3.19 $\pm$  0.30 \\
2.7 &  1.37 $\pm$  0.13 &  2.57 $\pm$  0.23 & 2.7 &  2.13 $\pm$  0.26 &  3.04 $\pm$  0.30 \\
3.2 &  1.45 $\pm$  0.12 &  2.93 $\pm$  0.23 & 3.4 &  1.89 $\pm$  0.27 &  3.04 $\pm$  0.38 \\
3.9 &  1.44 $\pm$  0.11 &  3.19 $\pm$  0.23 & 4.7 &  1.41 $\pm$  0.30 &  2.64 $\pm$  0.56 \\
5.0 &  1.04 $\pm$  0.10 &  2.68 $\pm$  0.25 \\
\end{tabular}
\end{ruledtabular}
%\endc
\end{table*}

The \mzt\ and \xhq\ values from the iterative solution of
\eq{eq:kt_root} as a function of the \piz trigger momenta \ptt\ and
associated momenta \pta\ are shown in \fig{fig:zt_xhq_ptt} and
\fig{fig:zt_xhq_pta}. There is an opposite trend; whereas the \mzt\ rises
with \ptt\ it is falling with \pta. It is an interesting consequence
of two effects: competition between steeply falling final state parton
spectra and rising fragmentation function with parton
momentum. Secondly, the detection of trigger particle biases the
\vkt{}\ vector in the direction of the trigger jet as discussed in
section \ref{sec:kt_smearing}.

%%%%%%%%%%%%%%%%%%%%%%%%%%%%%%%%%%%%%%%%%%%%%%%%%%%%%%%%%%%% Table XI
\begin{table}[tbh]
\caption{\label{tab:zt_xhq_ptt}
The \mzt\ and \xhq\ values with \ptt\ shown in \fig{fig:zt_xhq_ptt}.
}
%\bgc
\begin{ruledtabular}
\begin{tabular}{ccc}
\ptt\ (\gevc)&      $\mzt$                       & \xhq  \\\hline
3.22 &  0.51 $\pm$  4.10$^{-3}$ $\pm$ 0.06 &  0.88 $\pm$  0.01\\
3.89 &  0.56 $\pm$  2.10$^{-3}$ $\pm$ 0.07	&  0.87 $\pm$  0.01\\
4.90 &  0.61 $\pm$  1.10$^{-3}$ $\pm$ 0.07	&  0.85 $\pm$  0.01\\
5.91 &  0.64 $\pm$  1.10$^{-4}$ $\pm$ 0.07	&  0.85 $\pm$  0.02\\
7.24 &  0.66 $\pm$  1.10$^{-3}$ $\pm$ 0.07	&  0.86 $\pm$  0.02\\
8.34 &  0.68 $\pm$  5.10$^{-3}$ $\pm$ 0.06	&  0.84 $\pm$  0.05\\
\end{tabular}
\end{ruledtabular}
%\endc
\end{table}

%%%%%%%%%%%%%%%%%%%%%%%%%%%%%%%%%%%%%%%%%%%%%%%%%%%%%%%%%%%% Table XII
\begin{table}[tbh]
\caption{\label{tab:zt_xhq_pta_34}
The \mzt\ and \xhq\ values with \pta\ for two trigger \piz\ momenta bins as shown
on \fig{fig:zt_xhq_pta}.
}
%\bgc
\begin{ruledtabular}
\begin{tabular}{ccc}
 	  \multicolumn{3}{c}{$3<\ptt<4$ \gevc} \\
\pta &  \mzt                               &    \xhq  \\\hline
1.72 &  0.54 $\pm$  8.10$^{-3}$ $\pm$ 0.06 &  0.81 $\pm$  0.01 \\ 
2.22 &  0.52 $\pm$  6.10$^{-3}$ $\pm$ 0.06 &  0.88 $\pm$  0.01 \\
2.73 &  0.51 $\pm$  1.10$^{-3}$ $\pm$ 0.07 &  0.95 $\pm$  0.01 \\
3.23 &  0.49 $\pm$  1.10$^{-3}$ $\pm$ 0.06 &  0.99 $\pm$  0.01 \\
3.93 &  0.47 $\pm$  5.10$^{-3}$ $\pm$ 0.06 &  1.04 $\pm$  0.01 \\
5.04 &  0.41 $\pm$  6.10$^{-3}$ $\pm$ 0.06 &  1.06 $\pm$  0.01 \\\hline
	 \multicolumn{3}{c}{$5<\ptt<10$ \gevc}\\
\pta &  \mzt                               & \xhq  \\\hline
1.85 &  0.66 $\pm$  4.10$^{-3}$ $\pm$ 0.06 & 0.75 $\pm$  0.04\\ 
2.24 &  0.64 $\pm$  1.10$^{-3}$ $\pm$ 0.06 & 0.80 $\pm$  0.03\\
2.73 &  0.61 $\pm$  2.10$^{-3}$ $\pm$ 0.07 & 0.87 $\pm$  0.02\\
3.44 &  0.57 $\pm$  2.10$^{-3}$ $\pm$ 0.07 & 0.92 $\pm$  0.02\\
4.65 &  0.52 $\pm$  5.10$^{-3}$ $\pm$ 0.08 & 0.98 $\pm$  0.01\\
\end{tabular}
\end{ruledtabular}
%\endc
\end{table}
 
\begin{figure}[tbh]
%\bgc
\includegraphics[width=1.0\linewidth]{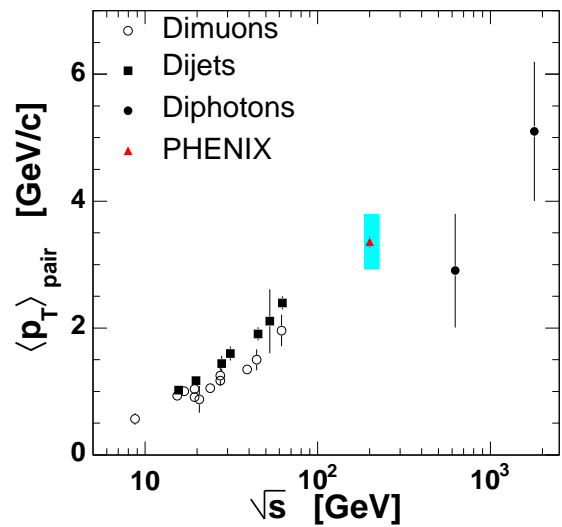}
\caption{\label{fig:apanakT} (color online)
Compilation of mean pair \pt{}\ measurements \cite{Apanasevich_kt_E609} and
comparisons to the \mean{\pt{}}$_{pair}$ measured in this analysis.
}
%\endc
\end{figure}

The \ptt\ averaged value of \sqrtrms{\kt{}}\ (\fig{fig:rmskt_ptt}) is
compared to the average parton pair momentum, \mean{\pn} =\mean{\pt{}}$_{pair}$, presented in
\cite{Apanasevich_kt_E609} (see \fig{fig:apanakT}). The value of
\mean{\pt{}}$_{pair}$ is determined as a sum of the two partons'
\mean{\kt{}}. In the present analysis the \sqrtrms{\kt{}}\ is
determined and thus the value of \mean{\pt{}}$_{pair}$ is evaluated as
$\mean{\pt{}}_{pair} = \sqrt{2}\times\mean{\kt{}} = \sqrt{\pi/2}\times\sqrtrms{\kt{}}.$

%%%%%%%%%%%%%%%%%%%%%%%%%%%%%%%%%%%%%%%%%%%%%%%%%%%%%%%%%%%% Fig. 30

The present value of \mean{\pt{}}$_{pair}$
$$ \ptnfinal $$
appears to be in a good agreement with the lower energy dijet and
dilepton measurements or the higher energy measurement in diphoton
production~\cite{CDF_prompt_phot_1992}.  A UA2 measurement of
\mean{\pt{}} of $Z^0$ production at $\s\sim 600$
\gev\ gives $8.6\pm1.5$ GeV/c~\cite{UA2_Zprod_kT_1987,UA2_dir_phot_kT_1988}.

\section{Summary}  %%mjt021306
\label{sec:summary}

We have made the first measurement of jet \jt{}\ and \kt{}\ for \pp\
collisions at \s~=~200~\gev\ using the method of two-particle
correlations.  Analysis of the angular widths of the near-side peak in
the correlation function has determined that the jet fragmentation
transverse momentum \jt{}\ is constant with trigger particle \ptt\ and
the extracted value \jtfinal\ is comparable with previous lower \s\
measurements. The width of the away-side peak is shown to be a measure
of the convolution of \jt{}\ with the jet momentum fraction $z$ and
the partonic transverse momentum \kt{}.  \mzt\ is determined through a
combined analysis of the measured \piz\ inclusive and associated
spectra %%by determining 
using the jet fragmentation functions from $e^+ e^-$ measurements. 
%The derived
%fragmentation function is found to be steeper than found at ISR
%\cite{CCOR_mz}, indicative of gluonic jet domination in the explored
%\pt{}\ range.  
The average of \mzt\ from the gluon and quark fragmentation functions is used and the difference is taken as the measure of the systematic error. 
The final extracted values of
\kt{}\ are then determined to be also independent of the transverse
momentum of the trigger \piz, in the range measured, with values of
\ktfinal.

%%%%%%%%%%%%%%%%%%%%%%%%%%%%%%%%%%%%%%%%% Acknowledgments
\begin{acknowledgments} 
%%%%%%%%%%%%%%%%%%%%%%%%%%%%%%%%%%%%%%%%%\input{03run_prc}
%\section{Acknowledgements}   % Run-3 long from for PRC, PLB, etc.

We thank the staff of the Collider-Accelerator and Physics
Departments at Brookhaven National Laboratory and the staff of
the other PHENIX participating institutions for their vital
contributions.  We acknowledge support from the Department of
Energy, Office of Science, Office of Nuclear Physics, the
National Science Foundation, Abilene Christian University
Research Council, Research Foundation of SUNY, and Dean of the
College of Arts and Sciences, Vanderbilt University (U.S.A),
Ministry of Education, Culture, Sports, Science, and Technology
and the Japan Society for the Promotion of Science (Japan),
Conselho Nacional de Desenvolvimento Cient\'{\i}fico e
Tecnol{\'o}gico and Funda\c c{\~a}o de Amparo {\`a} Pesquisa do
Estado de S{\~a}o Paulo (Brazil),
Natural Science Foundation of China (People's Republic of China),
Centre National de la Recherche Scientifique, Commissariat
{\`a} l'{\'E}nergie Atomique, and Institut National de Physique
Nucl{\'e}aire et de Physique des Particules (France),
Ministry of Industry, Science and Tekhnologies,
Bundesministerium f\"ur Bildung und Forschung, Deutscher
Akademischer Austausch Dienst, and Alexander von Humboldt Stiftung (Germany),
Hungarian National Science Fund, OTKA (Hungary), 
Department of Atomic Energy (India), 
Israel Science Foundation (Israel), 
Korea Research Foundation, Center for High
Energy Physics, and Korea Science and Engineering Foundation (Korea),
Ministry of Education and Science, Rassia Academy of Sciences,
Federal Agency of Atomic Energy (Russia),
VR and the Wallenberg Foundation (Sweden), 
the U.S. Civilian Research and Development Foundation for the
Independent States of the Former Soviet Union, the US-Hungarian
NSF-OTKA-MTA, and the US-Israel Binational Science Foundation.

\end{acknowledgments}

\appendix
\section{}
%%%%%%%%%%%%%%%%%%%%%%%%%%%%%%%%%%%%%%%%%%%%%%%%%%%%%%%%%%%%%%%%%%
\subsection[A-]{First and second moments of normally distributed 
quantities}
\label{app:means}
%%%%%%%%%%%%%%%%%%%%%%%%%%%%%%%%%%%%%%%%%%%%%%%%%%%%%%%%%%%%%%%%%%

Let $x$ be a 1D variable with normal (Gaussian)
distribution and $r=\sqrt{x^2+y^2}$ is a 2D variable with $x$ and $y$
of normal distribution then the following relations can be easily derived
%\bgc
\begin{ruledtabular}
\begin{tabular}{lcllcl}
\mean{x}    & = & 0                                     & \mean{r}    & = & $\sqrt{\pi\over 2}\sigma_1$ \\
\meanabs{x} & = & $\sqrt{2\over\pi}\sigma_1$ \hskip 1cm & \meanabs{r} & = & $\la r\ra$ \\
\rms{x}     & = & $\sigma^2_1$                          & \rms{r}     & = & $2\sigma^2_1\equiv\sigma^2_2$ \\
\end{tabular}
\end{ruledtabular}
%\endc

Both \vjt\ and \vkt{}\ are two dimensional vectors. We assume Gaussian
distributed $x$ and $y$ components and thus the mean value
\mean{\kt{}x}\ and \mean{\kt{y}}\ is equal to zero. The non-zero moments of 2D Gaussian
distribution are \eg\ the root mean squares $\sqrt{\la j_T^2\ra}$,
$\sqrt{\la k_T^2\ra}$ or the mean absolute values of the \vjt, \vkt{}\
projections into the perpendicular plane to the jet axes \meanabs{\jt{y}}\ and
\meanabs{\kt{y}}. Note that there are a trivial correspondences
\bge
\sqrtrms{\kt{}}  = {2\over\sqrt{\pi}} \mean{\kt{}} =  \sqrt{\pi}\meanabs{\kt{y}}\\
\ende

%%%%%%%%%%%%%%%%%%%%%%%%%%%%%%%%%%%%%%%%%%%%%%%%%%%%%%%%%%%%%%%%%%
\subsection[A-0]{The correct way to analyze the azimuthal correlation 
function.}
\label{sec:azimuth_correl}
%%%%%%%%%%%%%%%%%%%%%%%%%%%%%%%%%%%%%%%%%%%%%%%%%%%%%%%%%%%%%%%%%%

Construction and fitting of the two-particle azimuthal correlation
function is discussed in section \ref{sec:rawResults}. Traditionally
the correlation function is fitted by two Gaussian functions - one
for intra-jet correlation (near peak) and one for the inter-jet
correlations (away-side peak). From the extracted variances of the
Gaussian functions the \jt{}\ and \pout\ magnitudes are
extracted. 

%%%%%%%%%%%%%%%%%%%%%%%%%%%%%%%%%%%%%%%%%%%%%%%%%%%%%%%%%%%% Fig. 31
\begin{figure}[tbh]
\includegraphics[width=1.0\linewidth]{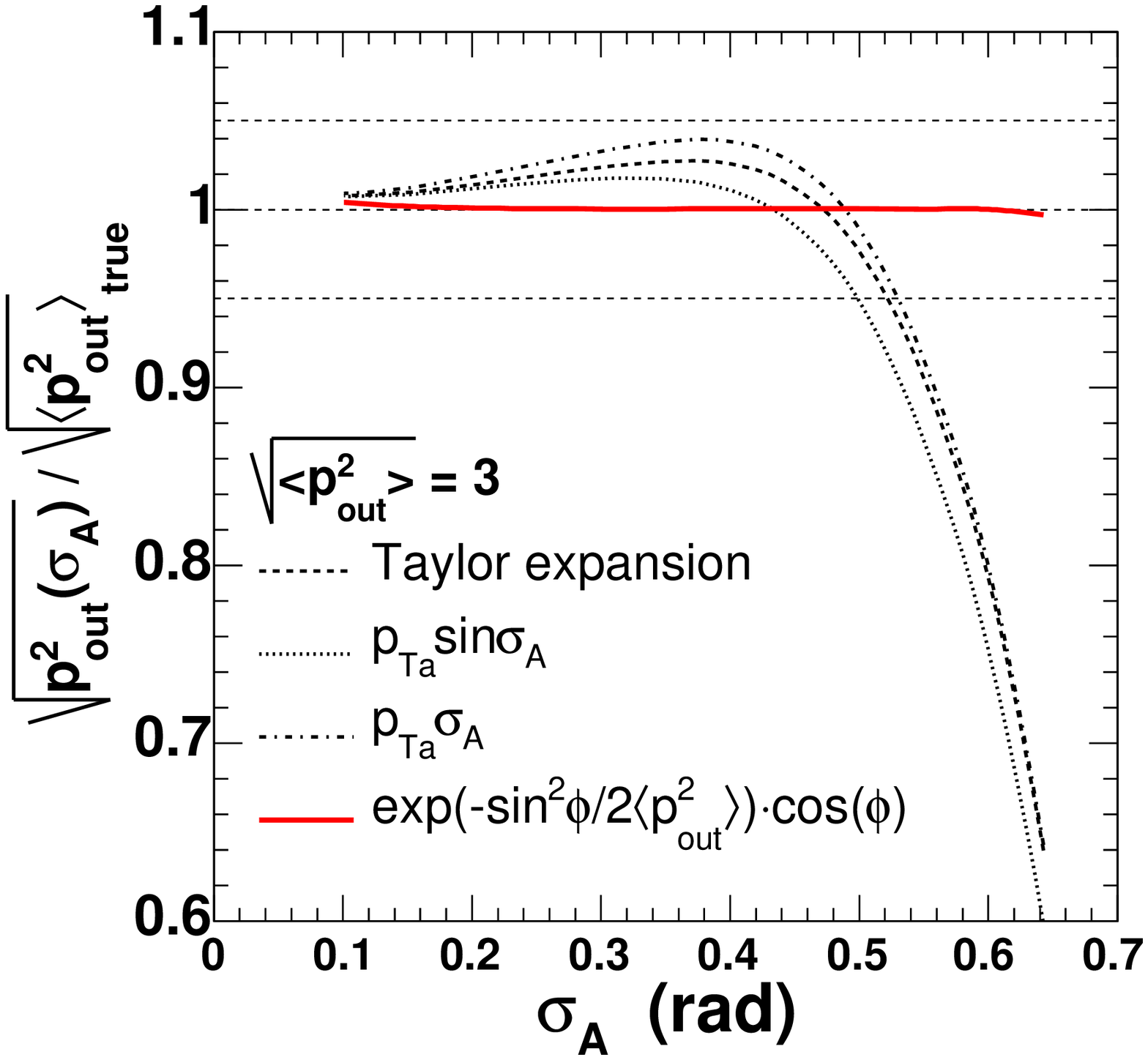}
\caption{\label{fig:pout_approx} (color online)
The relative error on \pout\ determination from the azimuthal correlation function
based on the Taylor expansion of $\la\sin\Delta\phi^2\ra$ (dashed line),
with an assumption of \sqrtrms{\pout}=\pt{a}~sin\siga\ (dotted line) and
\sqrtrms{\pout}=\pt{a}\siga\ (dotted-dashed line). The solid red line corresponds to
\sqrtrms{\pout}\ from \eq{eq:dndpout}.
}
\end{figure}

There is, however, a fundamental problem with this approach.
The \pout-vector defined in \eq{eq:def_pout} is equal to
$\pta\sin\Delta\phi$ event by event. However, we measure the width of the
correlation peak and this corresponds to $\sqrtrms{\Delta\phi}=\siga$. 
The relation
\sqrtrms{\pout}$\approx\pt{a}\sin\siga$ is not a good approximation
for \siga$>$0.4 rad (see \fig{fig:pout_approx}). The assumption that
the away-side correlation has a Gaussian shape is also good only for 
small values of \siga\ (see \fig{fig:pout_approx}).

One way of relating \sqrtrms{\pout}\ and \siga\ was proposed \eg\ by
Peter Levai \cite{Levai_kt_05} and used in several other analyzes.
Since \sqrtrms{\pout}=$\pt{a}\sqrt{\la\sin^2\Delta\phi\ra}$ 
one possibility how to relate \pout\
and \siga\ is to expand
\begin{eqnarray}
\la\sin^2\Delta\phi\ra & = &
\la \Delta\phi^2-
{1\over 3}\Delta\phi^4+{2\over 45}\Delta\phi^6\dots \ra \nonumber\\
& = &  \siga^2-\siga^4+{2\over 3}\siga^6\dots\nonumber
\end{eqnarray}
where we assumed a Gaussian distribution of $\Delta\phi$.
The comparison of $\pt{a}\cdot(\siga^2-\siga^4+{2\over
3}\siga^6\dots)$ with the true \pout\ magnitude (simple monte carlo) for
various \siga\ values is shown in \fig{fig:pout_approx}. It is obvious
that there is only a little difference between \sqrtrms{\pout}=\pt{a}~sin\siga,
\sqrtrms{\pout}=\pt{a}\siga\ and the Taylor series.  In the region where
\siga$>$0.4 rad, all approximations seems to be equally bad. 

However, \pout, the only quantity with a truly Gaussian
distribution (if we neglect the radiative corrections responsible for
non-Gaussian tails in the \pout\ distribution which are anyway not
relevant for the \kt{}\ analysis) can be directly extracted from the
correlation function. With the assumption of Gaussian distribution in
\pout, we can write the away-side $\Delta\phi$-distribution (normalized to unity) as
\begin{eqnarray}
%\label{eq:dndpout}
\lefteqn{ {dN_{away}\over d\Delta\phi}\Big|_{\pi/2}^{3\pi/2}
={dN\over d\pout}{d\pout\over d\Delta\phi}= }\mbox{\hskip 8cm} \nonumber\\
={-\pt{a}\cos\Delta\phi\over\sqrt{2\pi\rms{\pout}}{\rm Erf}({\sqrt{2}\pt{a}\over\sqrtrms{\pout}})}
\exp\left(-{\ptkv{a}\sin^2\Delta\phi\over 2\rms{\pout}}\right)\nonumber
\end{eqnarray}
This is the correct way of extracting a dimensional quantity from the
azimuthal correlation function in the case of narrow associated bin.
Similar line of arguments can be drawn also in the case of near peak.
However, given the narrowness of the near angle peak, the simple Gaussian
approximation is good enough.
\ 
%\newpage
%\pagebreak
%%%%%%%%%%%%%%%%%%%%%%%%%%%%%%%%%%%%%%%%%%%%%%%%%%%%%%%%%%%%%%%%%%
%\section{Data tables}
%\label{app:data}         All tables moved near their figures
%%%%%%%%%%%%%%%%%%%%%%%%%%%%%%%%%%%%%%%%%%%%%%%%%%%%%%%%%%%%%%%%%%

%%% \bibliographystyle{prsty} %h-physrev3  unsrt phaip 
%%% No!  must use \bibliographystyle{apsrev} -- called in header

%%\bibliography{ppg029}

\begin{thebibliography}{50}
\expandafter\ifx\csname natexlab\endcsname\relax\def\natexlab#1{#1}\fi
\expandafter\ifx\csname bibnamefont\endcsname\relax
  \def\bibnamefont#1{#1}\fi
\expandafter\ifx\csname bibfnamefont\endcsname\relax
  \def\bibfnamefont#1{#1}\fi
\expandafter\ifx\csname citenamefont\endcsname\relax
  \def\citenamefont#1{#1}\fi
\expandafter\ifx\csname url\endcsname\relax
  \def\url#1{\texttt{#1}}\fi
\expandafter\ifx\csname urlprefix\endcsname\relax\def\urlprefix{URL }\fi
\providecommand{\bibinfo}[2]{#2}
\providecommand{\eprint}[2][]{\url{#2}}

\bibitem[{\citenamefont{Angelis et~al.}(1980)}]{CCORjt}
\bibinfo{author}{\bibfnamefont{A.~L.~S.} \bibnamefont{Angelis}}
  \bibnamefont{et~al.}
  (\bibinfo{collaboration}{CERN-Columbia-Oxford-Rockefeller}),
  \bibinfo{journal}{Phys. Lett.} \textbf{\bibinfo{volume}{B97}},
  \bibinfo{pages}{163} (\bibinfo{year}{1980}).

\bibitem[{\citenamefont{Darriulat et~al.}(1976)}]{Darriulat_poutxe}
\bibinfo{author}{\bibfnamefont{P.}~\bibnamefont{Darriulat}}
  \bibnamefont{et~al.}, \bibinfo{journal}{Nucl. Phys.}
  \textbf{\bibinfo{volume}{B107}}, \bibinfo{pages}{429} (\bibinfo{year}{1976}).

\bibitem[{\citenamefont{Della~Negra et~al.}(1977)}]{CCHK_jet_structure}
\bibinfo{author}{\bibfnamefont{M.}~\bibnamefont{Della~Negra}}
  \bibnamefont{et~al.} (\bibinfo{collaboration}{CERN-College de
  France-Heidelberg-Karlsruhe}), \bibinfo{journal}{Nucl. Phys.}
  \textbf{\bibinfo{volume}{B127}}, \bibinfo{pages}{1} (\bibinfo{year}{1977}).

\bibitem[{\citenamefont{Adcox et~al.}(2002)}]{PHENIX_supp_130}
\bibinfo{author}{\bibfnamefont{K.}~\bibnamefont{Adcox}} \bibnamefont{et~al.}
  (\bibinfo{collaboration}{PHENIX}), \bibinfo{journal}{Phys. Rev. Lett.}
  \textbf{\bibinfo{volume}{88}}, \bibinfo{pages}{022301}
  (\bibinfo{year}{2002}), \eprint{nucl-ex/0109003}.

\bibitem[{\citenamefont{Adler et~al.}(2003{\natexlab{a}})}]{PHENIX_supp_dAu}
\bibinfo{author}{\bibfnamefont{S.~S.} \bibnamefont{Adler}} \bibnamefont{et~al.}
  (\bibinfo{collaboration}{PHENIX}), \bibinfo{journal}{Phys. Rev. Lett.}
  \textbf{\bibinfo{volume}{91}}, \bibinfo{pages}{072303}
  (\bibinfo{year}{2003}{\natexlab{a}}), \eprint{nucl-ex/0306021}.

\bibitem[{\citenamefont{Adler et~al.}(2003{\natexlab{b}})}]{PHENIX_v2}
\bibinfo{author}{\bibfnamefont{S.~S.} \bibnamefont{Adler}} \bibnamefont{et~al.}
  (\bibinfo{collaboration}{PHENIX}), \bibinfo{journal}{Phys. Rev. Lett.}
  \textbf{\bibinfo{volume}{91}}, \bibinfo{pages}{182301}
  (\bibinfo{year}{2003}{\natexlab{b}}), \eprint{nucl-ex/0305013}.

\bibitem[{\citenamefont{Adler et~al.}(2003{\natexlab{c}})}]{STAR_v2}
\bibinfo{author}{\bibfnamefont{C.}~\bibnamefont{Adler}} \bibnamefont{et~al.}
  (\bibinfo{collaboration}{STAR}), \bibinfo{journal}{Phys. Rev. Lett.}
  \textbf{\bibinfo{volume}{90}}, \bibinfo{pages}{032301}
  (\bibinfo{year}{2003}{\natexlab{c}}), \eprint{nucl-ex/0206006}.

\bibitem[{\citenamefont{Adler
  et~al.}(2003{\natexlab{d}})}]{STAR_b2b_suppression}
\bibinfo{author}{\bibfnamefont{C.}~\bibnamefont{Adler}} \bibnamefont{et~al.}
  (\bibinfo{collaboration}{STAR}), \bibinfo{journal}{Phys. Rev. Lett.}
  \textbf{\bibinfo{volume}{90}}, \bibinfo{pages}{082302}
  (\bibinfo{year}{2003}{\natexlab{d}}), \eprint{nucl-ex/0210033}.

\bibitem[{\citenamefont{Migdal}(1956)}]{LPM_1956}
\bibinfo{author}{\bibfnamefont{A.~B.} \bibnamefont{Migdal}},
  \bibinfo{journal}{Phys. Rev.} \textbf{\bibinfo{volume}{103}},
  \bibinfo{pages}{1811} (\bibinfo{year}{1956}).

\bibitem[{\citenamefont{Wang and Gyulassy}(1992)}]{quenching_WangMiklos}
\bibinfo{author}{\bibfnamefont{X.-N.} \bibnamefont{Wang}} \bibnamefont{and}
  \bibinfo{author}{\bibfnamefont{M.}~\bibnamefont{Gyulassy}},
  \bibinfo{journal}{Phys. Rev. Lett.} \textbf{\bibinfo{volume}{68}},
  \bibinfo{pages}{1480} (\bibinfo{year}{1992}).

\bibitem[{\citenamefont{Wang}(1998)}]{Wang_jetQuenching98_D(z)pars}
\bibinfo{author}{\bibfnamefont{X.-N.} \bibnamefont{Wang}},
  \bibinfo{journal}{Phys. Rev.} \textbf{\bibinfo{volume}{C58}},
  \bibinfo{pages}{2321} (\bibinfo{year}{1998}), \eprint{hep-ph/9804357}.

\bibitem[{\citenamefont{Salgado and Wiedemann}(2004)}]{Urs_ktBroadening}
\bibinfo{author}{\bibfnamefont{C.~A.} \bibnamefont{Salgado}} \bibnamefont{and}
  \bibinfo{author}{\bibfnamefont{U.~A.} \bibnamefont{Wiedemann}},
  \bibinfo{journal}{Phys. Rev. Lett.} \textbf{\bibinfo{volume}{93}},
  \bibinfo{pages}{042301} (\bibinfo{year}{2004}), \eprint{hep-ph/0310079}.

\bibitem[{\citenamefont{Qiu and Vitev}(2003)}]{Ivan_dAu}
\bibinfo{author}{\bibfnamefont{J.-w.} \bibnamefont{Qiu}} \bibnamefont{and}
  \bibinfo{author}{\bibfnamefont{I.}~\bibnamefont{Vitev}},
  \bibinfo{journal}{Phys. Lett.} \textbf{\bibinfo{volume}{B570}},
  \bibinfo{pages}{161} (\bibinfo{year}{2003}), \eprint{nucl-th/0306039}.

\bibitem[{\citenamefont{Wang}(2002)}]{Wang_fragModif1}
\bibinfo{author}{\bibfnamefont{X.-N.} \bibnamefont{Wang}},
  \bibinfo{journal}{Nucl. Phys.} \textbf{\bibinfo{volume}{A702}},
  \bibinfo{pages}{238} (\bibinfo{year}{2002}), \eprint{hep-ph/0208094}.

\bibitem[{\citenamefont{Berman et~al.}(1971{\natexlab{a}})\citenamefont{Berman,
  Bjorken, and Kogut}}]{Feynman1}
\bibinfo{author}{\bibfnamefont{S.~M.} \bibnamefont{Berman}},
  \bibinfo{author}{\bibfnamefont{J.~D.} \bibnamefont{Bjorken}},
  \bibnamefont{and} \bibinfo{author}{\bibfnamefont{J.~B.} \bibnamefont{Kogut}},
  \bibinfo{journal}{Phys. Rev.} \textbf{\bibinfo{volume}{D4}},
  \bibinfo{pages}{3388} (\bibinfo{year}{1971}{\natexlab{a}}).

\bibitem[{\citenamefont{Owens and Kimel}(1978)}]{Feynman2}
\bibinfo{author}{\bibfnamefont{J.~F.} \bibnamefont{Owens}} \bibnamefont{and}
  \bibinfo{author}{\bibfnamefont{J.~D.} \bibnamefont{Kimel}},
  \bibinfo{journal}{Phys. Rev.} \textbf{\bibinfo{volume}{D18}},
  \bibinfo{pages}{3313} (\bibinfo{year}{1978}).

\bibitem[{\citenamefont{Owens et~al.}(1978)\citenamefont{Owens, Reya, and
  Gluck}}]{Feynman3}
\bibinfo{author}{\bibfnamefont{J.~F.} \bibnamefont{Owens}},
  \bibinfo{author}{\bibfnamefont{E.}~\bibnamefont{Reya}}, \bibnamefont{and}
  \bibinfo{author}{\bibfnamefont{M.}~\bibnamefont{Gluck}},
  \bibinfo{journal}{Phys. Rev.} \textbf{\bibinfo{volume}{D18}},
  \bibinfo{pages}{1501} (\bibinfo{year}{1978}).

\bibitem[{\citenamefont{Feynman et~al.}(1978)\citenamefont{Feynman, Field, and
  Fox}}]{Feynman4}
\bibinfo{author}{\bibfnamefont{R.~P.} \bibnamefont{Feynman}},
  \bibinfo{author}{\bibfnamefont{R.~D.} \bibnamefont{Field}}, \bibnamefont{and}
  \bibinfo{author}{\bibfnamefont{G.~C.} \bibnamefont{Fox}},
  \bibinfo{journal}{Phys. Rev.} \textbf{\bibinfo{volume}{D18}},
  \bibinfo{pages}{3320} (\bibinfo{year}{1978}).

\bibitem[{\citenamefont{Owens}(1987)}]{Owens:1987mp}
\bibinfo{author}{\bibfnamefont{J.~F.} \bibnamefont{Owens}},
  \bibinfo{journal}{Rev. Mod. Phys.} \textbf{\bibinfo{volume}{59}},
  \bibinfo{pages}{465} (\bibinfo{year}{1987}).

\bibitem[{\citenamefont{Bunce et~al.}(2000)\citenamefont{Bunce, Saito, Soffer,
  and Vogelsang}}]{Bunce:2000uv}
\bibinfo{author}{\bibfnamefont{G.}~\bibnamefont{Bunce}},
  \bibinfo{author}{\bibfnamefont{N.}~\bibnamefont{Saito}},
  \bibinfo{author}{\bibfnamefont{J.}~\bibnamefont{Soffer}}, \bibnamefont{and}
  \bibinfo{author}{\bibfnamefont{W.}~\bibnamefont{Vogelsang}},
  \bibinfo{journal}{Ann. Rev. Nucl. Part. Sci.} \textbf{\bibinfo{volume}{50}},
  \bibinfo{pages}{525} (\bibinfo{year}{2000}), \eprint{hep-ph/0007218}.

\bibitem[{\citenamefont{Cutler and Sivers}(1978)}]{Cutler:1978qm}
\bibinfo{author}{\bibfnamefont{R.}~\bibnamefont{Cutler}} \bibnamefont{and}
  \bibinfo{author}{\bibfnamefont{D.~W.} \bibnamefont{Sivers}},
  \bibinfo{journal}{Phys. Rev.} \textbf{\bibinfo{volume}{D17}},
  \bibinfo{pages}{196} (\bibinfo{year}{1978}).

\bibitem[{\citenamefont{Cutler and Sivers}(1977)}]{Cutler:1977mw}
\bibinfo{author}{\bibfnamefont{R.}~\bibnamefont{Cutler}} \bibnamefont{and}
  \bibinfo{author}{\bibfnamefont{D.~W.} \bibnamefont{Sivers}},
  \bibinfo{journal}{Phys. Rev.} \textbf{\bibinfo{volume}{D16}},
  \bibinfo{pages}{679} (\bibinfo{year}{1977}).

\bibitem[{\citenamefont{Combridge et~al.}(1977)\citenamefont{Combridge,
  Kripfganz, and Ranft}}]{Combridge:1977dm}
\bibinfo{author}{\bibfnamefont{B.~L.} \bibnamefont{Combridge}},
  \bibinfo{author}{\bibfnamefont{J.}~\bibnamefont{Kripfganz}},
  \bibnamefont{and} \bibinfo{author}{\bibfnamefont{J.}~\bibnamefont{Ranft}},
  \bibinfo{journal}{Phys. Lett.} \textbf{\bibinfo{volume}{B70}},
  \bibinfo{pages}{234} (\bibinfo{year}{1977}).

\bibitem[{\citenamefont{Feynman et~al.}(1977)\citenamefont{Feynman, Field, and
  Fox}}]{Feynman5}
\bibinfo{author}{\bibfnamefont{R.~P.} \bibnamefont{Feynman}},
  \bibinfo{author}{\bibfnamefont{R.~D.} \bibnamefont{Field}}, \bibnamefont{and}
  \bibinfo{author}{\bibfnamefont{G.~C.} \bibnamefont{Fox}},
  \bibinfo{journal}{Nucl. Phys.} \textbf{\bibinfo{volume}{B128}},
  \bibinfo{pages}{1} (\bibinfo{year}{1977}).

\bibitem[{\citenamefont{Dokshitzer et~al.}()\citenamefont{Dokshitzer, Khoze,
  Mueller, and Troian}}]{Dokshitzer_basics_of_pQCD}
\bibinfo{author}{\bibfnamefont{Y.~L.} \bibnamefont{Dokshitzer}},
  \bibinfo{author}{\bibfnamefont{V.~A.} \bibnamefont{Khoze}},
  \bibinfo{author}{\bibfnamefont{A.~H.} \bibnamefont{Mueller}},
  \bibnamefont{and} \bibinfo{author}{\bibfnamefont{S.~I.}
  \bibnamefont{Troian}}, \emph{\bibinfo{title}{Basics of perturbative qcd}},
  \bibinfo{note}{gif-sur-Yvette, France: Ed. Frontieres (1991) 274 p. (Basics
  of)}.

\bibitem[{\citenamefont{Apanasevich et~al.}(1999)}]{Apanasevich_kt_E609}
\bibinfo{author}{\bibfnamefont{L.}~\bibnamefont{Apanasevich}}
  \bibnamefont{et~al.}, \bibinfo{journal}{Phys. Rev.}
  \textbf{\bibinfo{volume}{D59}}, \bibinfo{pages}{074007}
  (\bibinfo{year}{1999}), \eprint{hep-ph/9808467}.

\bibitem[{\citenamefont{Kulesza et~al.}(2003)\citenamefont{Kulesza, Sterman,
  and Vogelsang}}]{Werner_resummation}
\bibinfo{author}{\bibfnamefont{A.}~\bibnamefont{Kulesza}},
  \bibinfo{author}{\bibfnamefont{G.}~\bibnamefont{Sterman}}, \bibnamefont{and}
  \bibinfo{author}{\bibfnamefont{W.}~\bibnamefont{Vogelsang}},
  \bibinfo{journal}{Nucl. Phys.} \textbf{\bibinfo{volume}{A721}},
  \bibinfo{pages}{591} (\bibinfo{year}{2003}), \eprint{hep-ph/0302121}.

\bibitem[{\citenamefont{Busser et~al.}(1973)}]{Busser:1973hs}
\bibinfo{author}{\bibfnamefont{F.~W.} \bibnamefont{Busser}}
  \bibnamefont{et~al.}, \bibinfo{journal}{Phys. Lett.}
  \textbf{\bibinfo{volume}{B46}}, \bibinfo{pages}{471} (\bibinfo{year}{1973}).

\bibitem[{\citenamefont{Berman et~al.}(1971{\natexlab{b}})\citenamefont{Berman,
  Bjorken, and Kogut}}]{Berman:1971xz}
\bibinfo{author}{\bibfnamefont{S.~M.} \bibnamefont{Berman}},
  \bibinfo{author}{\bibfnamefont{J.~D.} \bibnamefont{Bjorken}},
  \bibnamefont{and} \bibinfo{author}{\bibfnamefont{J.~B.} \bibnamefont{Kogut}},
  \bibinfo{journal}{Phys. Rev.} \textbf{\bibinfo{volume}{D4}},
  \bibinfo{pages}{3388} (\bibinfo{year}{1971}{\natexlab{b}}).

\bibitem[{\citenamefont{Blankenbecler et~al.}(1972)\citenamefont{Blankenbecler,
  Brodsky, and Gunion}}]{Blankenbecler:1972cd}
\bibinfo{author}{\bibfnamefont{R.}~\bibnamefont{Blankenbecler}},
  \bibinfo{author}{\bibfnamefont{S.~J.} \bibnamefont{Brodsky}},
  \bibnamefont{and} \bibinfo{author}{\bibfnamefont{J.~F.}
  \bibnamefont{Gunion}}, \bibinfo{journal}{Phys. Lett.}
  \textbf{\bibinfo{volume}{B42}}, \bibinfo{pages}{461} (\bibinfo{year}{1972}).

\bibitem[{\citenamefont{Antreasyan et~al.}(1979)}]{Cronin}
\bibinfo{author}{\bibfnamefont{D.}~\bibnamefont{Antreasyan}}
  \bibnamefont{et~al.}, \bibinfo{journal}{Phys. Rev.}
  \textbf{\bibinfo{volume}{D19}}, \bibinfo{pages}{764} (\bibinfo{year}{1979}).

\bibitem[{\citenamefont{Darriulat}(1980)}]{Darriulat:1980nk}
\bibinfo{author}{\bibfnamefont{P.}~\bibnamefont{Darriulat}},
  \bibinfo{journal}{Ann. Rev. Nucl. Part. Sci.} \textbf{\bibinfo{volume}{30}},
  \bibinfo{pages}{159} (\bibinfo{year}{1980}).

\bibitem[{\citenamefont{Adachi et~al.}(1999)}]{e+e-jet_width}
\bibinfo{author}{\bibfnamefont{K.}~\bibnamefont{Adachi}} \bibnamefont{et~al.}
  (\bibinfo{collaboration}{TOPAZ}), \bibinfo{journal}{Phys. Lett.}
  \textbf{\bibinfo{volume}{B451}}, \bibinfo{pages}{256} (\bibinfo{year}{1999}),
  \eprint{hep-ex/9901036}.

\bibitem[{\citenamefont{Adcox
  et~al.}(2003{\natexlab{a}})}]{PHENIX_NIM_overview}
\bibinfo{author}{\bibfnamefont{K.}~\bibnamefont{Adcox}} \bibnamefont{et~al.}
  (\bibinfo{collaboration}{PHENIX}), \bibinfo{journal}{Nucl. Instrum. Meth.}
  \textbf{\bibinfo{volume}{A499}}, \bibinfo{pages}{469}
  (\bibinfo{year}{2003}{\natexlab{a}}).

\bibitem[{\citenamefont{Aphecetche
  et~al.}(2003{\natexlab{a}})}]{PHENIX_NIM_EMC}
\bibinfo{author}{\bibfnamefont{L.}~\bibnamefont{Aphecetche}}
  \bibnamefont{et~al.} (\bibinfo{collaboration}{PHENIX}),
  \bibinfo{journal}{Nucl. Instrum. Meth.} \textbf{\bibinfo{volume}{A499}},
  \bibinfo{pages}{521} (\bibinfo{year}{2003}{\natexlab{a}}).

\bibitem[{\citenamefont{Adcox et~al.}(2003{\natexlab{b}})}]{PHENIX_NIM_CA}
\bibinfo{author}{\bibfnamefont{K.}~\bibnamefont{Adcox}} \bibnamefont{et~al.}
  (\bibinfo{collaboration}{PHENIX}), \bibinfo{journal}{Nucl. Instrum. Meth.}
  \textbf{\bibinfo{volume}{A499}}, \bibinfo{pages}{489}
  (\bibinfo{year}{2003}{\natexlab{b}}).

\bibitem[{\citenamefont{Adler et~al.}(2003{\natexlab{e}})}]{PHENIX_pi0ppPRL}
\bibinfo{author}{\bibfnamefont{S.~S.} \bibnamefont{Adler}} \bibnamefont{et~al.}
  (\bibinfo{collaboration}{PHENIX}), \bibinfo{journal}{Phys. Rev. Lett.}
  \textbf{\bibinfo{volume}{91}}, \bibinfo{pages}{241803}
  (\bibinfo{year}{2003}{\natexlab{e}}), \eprint{hep-ex/0304038}.

\bibitem[{\citenamefont{Aphecetche
  et~al.}(2003{\natexlab{b}})}]{PHENIX_cal_2003}
\bibinfo{author}{\bibfnamefont{L.}~\bibnamefont{Aphecetche}}
  \bibnamefont{et~al.} (\bibinfo{collaboration}{PHENIX}),
  \bibinfo{journal}{Nucl. Instrum. Meth.} \textbf{\bibinfo{volume}{A499}},
  \bibinfo{pages}{521} (\bibinfo{year}{2003}{\natexlab{b}}).

\bibitem[{\citenamefont{Aizawa et~al.}(2003)}]{PHENIX_CAID_2003}
\bibinfo{author}{\bibfnamefont{M.}~\bibnamefont{Aizawa}} \bibnamefont{et~al.}
  (\bibinfo{collaboration}{PHENIX}), \bibinfo{journal}{Nucl. Instrum. Meth.}
  \textbf{\bibinfo{volume}{A499}}, \bibinfo{pages}{508} (\bibinfo{year}{2003}).

\bibitem[{\citenamefont{Adler et~al.}(2004)}]{PHENIX_supp_hAuAu200}
\bibinfo{author}{\bibfnamefont{S.~S.} \bibnamefont{Adler}} \bibnamefont{et~al.}
  (\bibinfo{collaboration}{PHENIX}), \bibinfo{journal}{Phys. Rev.}
  \textbf{\bibinfo{volume}{C69}}, \bibinfo{pages}{034910}
  (\bibinfo{year}{2004}), \eprint{nucl-ex/0308006}.

\bibitem[{\citenamefont{Adler et~al.}(2005)}]{Adler:2004electron}
\bibinfo{author}{\bibfnamefont{S.~S.} \bibnamefont{Adler}} \bibnamefont{et~al.}
  (\bibinfo{collaboration}{PHENIX}), \bibinfo{journal}{Phys. Rev. Lett.}
  \textbf{\bibinfo{volume}{94}}, \bibinfo{pages}{082301}
  (\bibinfo{year}{2005}), \eprint{nucl-ex/0409028}.

\bibitem[{\citenamefont{Levai et~al.}()\citenamefont{Levai, Fai, and
  Papp}}]{Levai_kt_05}
\bibinfo{author}{\bibfnamefont{P.}~\bibnamefont{Levai}},
  \bibinfo{author}{\bibfnamefont{G.}~\bibnamefont{Fai}}, \bibnamefont{and}
  \bibinfo{author}{\bibfnamefont{G.}~\bibnamefont{Papp}},
  \emph{\bibinfo{title}{Dijet correlations at isr and rhic energies}},
  \eprint{hep-ph/0502238}.

\bibitem[{\citenamefont{Adams et~al.}(2003)}]{STAR_supp_dAu}
\bibinfo{author}{\bibfnamefont{J.}~\bibnamefont{Adams}} \bibnamefont{et~al.}
  (\bibinfo{collaboration}{STAR}), \bibinfo{journal}{Phys. Rev. Lett.}
  \textbf{\bibinfo{volume}{91}}, \bibinfo{pages}{072304}
  (\bibinfo{year}{2003}), \eprint{nucl-ex/0306024}.

\bibitem[{\citenamefont{van Apeldoorn et~al.}(1975)}]{Seagull_Votruba_1975}
\bibinfo{author}{\bibfnamefont{G.~W.} \bibnamefont{van Apeldoorn}}
  \bibnamefont{et~al.}, \bibinfo{journal}{Nucl. Phys.}
  \textbf{\bibinfo{volume}{B91}}, \bibinfo{pages}{1} (\bibinfo{year}{1975}).

\bibitem[{\citenamefont{Wang}(2004)}]{Wang_fragModif2}
\bibinfo{author}{\bibfnamefont{X.-N.} \bibnamefont{Wang}},
  \bibinfo{journal}{Phys. Lett.} \textbf{\bibinfo{volume}{B595}},
  \bibinfo{pages}{165} (\bibinfo{year}{2004}), \eprint{nucl-th/0305010}.

\bibitem[{\citenamefont{Abreu et~al.}(2000)}]{Delphi_Dz_EPJ00}
\bibinfo{author}{\bibfnamefont{P.}~\bibnamefont{Abreu}} \bibnamefont{et~al.}
  (\bibinfo{collaboration}{DELPHI}), \bibinfo{journal}{Eur. Phys. J.}
  \textbf{\bibinfo{volume}{C13}}, \bibinfo{pages}{573} (\bibinfo{year}{2000}).

\bibitem[{\citenamefont{Alexander et~al.}(1996)}]{Opal_Dz_ZPhys}
\bibinfo{author}{\bibfnamefont{G.}~\bibnamefont{Alexander}}
  \bibnamefont{et~al.} (\bibinfo{collaboration}{OPAL}), \bibinfo{journal}{Z.
  Phys.} \textbf{\bibinfo{volume}{C69}}, \bibinfo{pages}{543}
  (\bibinfo{year}{1996}).

\bibitem[{\citenamefont{Abe et~al.}(1993)}]{CDF_prompt_phot_1992}
\bibinfo{author}{\bibfnamefont{F.}~\bibnamefont{Abe}} \bibnamefont{et~al.}
  (\bibinfo{collaboration}{CDF}), \bibinfo{journal}{Phys. Rev. Lett.}
  \textbf{\bibinfo{volume}{70}}, \bibinfo{pages}{2232} (\bibinfo{year}{1993}).

\bibitem[{\citenamefont{Ansari et~al.}(1987)}]{UA2_Zprod_kT_1987}
\bibinfo{author}{\bibfnamefont{R.}~\bibnamefont{Ansari}} \bibnamefont{et~al.}
  (\bibinfo{collaboration}{UA2}), \bibinfo{journal}{Phys. Lett.}
  \textbf{\bibinfo{volume}{B194}}, \bibinfo{pages}{158} (\bibinfo{year}{1987}).

\bibitem[{\citenamefont{Ansari et~al.}(1988)}]{UA2_dir_phot_kT_1988}
\bibinfo{author}{\bibfnamefont{R.}~\bibnamefont{Ansari}} \bibnamefont{et~al.}
  (\bibinfo{collaboration}{UA2}), \bibinfo{journal}{Z. Phys.}
  \textbf{\bibinfo{volume}{C41}}, \bibinfo{pages}{395} (\bibinfo{year}{1988}).

\end{thebibliography}

%\vfill\eject
\end{document}